\documentclass[a4paper,fleqn,usenatbib]{mnras}

\usepackage[T1]{fontenc}
\usepackage{ae,aecompl}

\usepackage{graphicx}	
\usepackage{amsmath}	
\usepackage{amssymb}	
\usepackage{natbib}
 \usepackage{booktabs}
 \usepackage{multi row}
 
\usepackage{epsfig}
\usepackage{comment}
\usepackage{url}

\title[morpho SMFs CANDELS]{Mass assembly and morphological transformations since $z\sim3$ from CANDELS}

\author[M. Huertas-Company et al.]{
M. Huertas-Company$^{1},$\thanks{E-mail: marc.huertas@obspm.fr}
M. Bernardi$^{2}$,
P.~G. P\'erez-Gonz\'alez$^{3}$,
M.~L.~N. Ashby$^{4}$,
\newauthor
G. Barro$^{5}$,
C. Conselice$^{6}$,
E. Daddi$^{7}$,
A. Dekel$^{8}$,
P. Dimauro$^{1}$,
S.~M. Faber$^{9}$,
\newauthor
N.~A. Grogin$^{10}$,
J.~S. Kartaltepe$^{11}$,
D.~D. Kocevski$^{12}$,
A.~M. Koekemoer$^{10}$,
\newauthor
D.~C. Koo$^{9}$,
S. Mei$^{1,14}$,
F. Shankar$^{13}$
\\
$^{1}$GEPI, Observatoire de Paris, CNRS, Universit\'e Paris  Diderot, 61, Avenue de l'Observatoire 75014, Paris  France\\
$^{2}$Department of Physics and Astronomy, University of Pennsylvania, Philadelphia, PA 19104, USA\\
$^{3}$Departamento de Astrof\'isica, Facultad de CC. F\'isicas, Universidad Complutense de Madrid, E-28040 Madrid, Spain\\
$^{4}$Harvard-Smithsonian Center for Astrophysics, 60 Garden St., Cambridge, MA 02138, USA\\
$^5$ Department of Astrophysics, University of California, Berkeley, USA\\
$^6$ School of Physics \& Astronomy, University of Nottingham, Nottingham NG7 2RD, UK\\
$^7$ Laboratoire AIM, CEA/DSM-CNRS-Universit\'e Paris Diderot, Irfu/Service d'Astrophysique, CEA Saclay, Orme des Merisiers, 91191 Gif-sur-Yvette Cedex, France\\
$^8$ Center for Astrophysics and Planetary Science, Racah Institute of Physics, The Hebrew University, Jerusalem 91904, Israel\\
$^9$ UCO/Lick Observatory, Department of Astronomy and Astrophysics, University of California, Santa Cruz, CA 95064\\
$^{10}$  Space Telescope Science Institute, Baltimore, MD, USA\\
$^{11}$  School of Physics and Astronomy, Rochester Institute of Technology, Rochester, NY, USA\\
$^{12}$ Department of Physics and Astronomy, Colby College, Waterville, ME 04961, USA\\
$^{13}$ Department of Physics and Astronomy, University of Southampton, Southampton SO17 1BJ, UK\\
$^{14}$ California Institute of Technology, Pasadena, CA 91125, USA
}

\date{Accepted XXX. Received YYY; in original form ZZZ}

\pubyear{2016}

\begin{document}
\label{firstpage}
\pagerange{\pageref{firstpage}--\pageref{lastpage}}
\maketitle

\begin{abstract}
We quantify the evolution of the stellar mass functions (SMFs) of star-forming and quiescent galaxies as a function of morphology from $z\sim 3$ to the present.  Our sample consists of $\sim 50,000$ galaxies in the CANDELS fields ($\sim880$ $arcmin^2$), which we divide into four main morphological types, i.e. pure bulge dominated systems, pure spiral disk dominated, intermediate 2-component bulge+disk systems and irregular disturbed galaxies. Our main results are: \\
{\bf 1) Star-formation: }At $z\sim 2$, 80\% of the stellar mass density of star-forming galaxies is in irregular systems.  However, by $z\sim 0.5$, irregular objects only dominate at stellar masses below $10^9M\odot$.  A majority of the star-forming irregulars present at $z\sim 2$ undergo a gradual transformation from disturbed to normal spiral disk morphologies by $z\sim 1$ without significant interruption to their star-formation.  Rejuvenation after a quenching event does not seem to be common except perhaps for the most massive objects, because the fraction of bulge dominated star-forming galaxies with $M_*/M_\odot>10^{10.7}$ reaches 40\% at $z<1$. \\
{\bf 2) Quenching: } We confirm that galaxies reaching a stellar mass of $M_*\sim10^{10.8}M_\odot$ ($M^*$) tend to quench. Also, quenching implies the presence of a bulge: the abundance of massive red disks is negligible at all redshifts over 2~dex in stellar mass.  However the dominant quenching mechanism evolves. At $z>2$, the SMF of quiescent galaxies above $M^*$ is dominated by compact spheroids. Quenching at this early epoch destroys the disk and produces a compact remnant unless the star-forming progenitors at even higher redshifts are significantly more dense.  
    At $1<z<2$, the majority of newly quenched galaxies are disks with a significant central bulge. 
This suggests that mass-quenching at this epoch starts from the inner parts and preserves the disk. At $z<1$, the high mass end of the passive SMF is globally in place and the evolution mostly happens at stellar masses below $10^{10}M_\odot$. These low-mass galaxies are compact, bulge dominated systems, which were environmentally-quenched:  destruction of the disk through ram-pressure stripping is the likely process. 

\end{abstract}




\section{Introduction}

Lying at the centers of dark matter potential wells, galaxies are the building blocks of our universe. How they assemble their mass and acquire their morphology are two central open questions today.  The answer requires a complete understanding of the complex baryonic physics which dominate at these scales. At first order, however, a galaxy is a system that transforms gas into stars. The life of a galaxy is therefore a balance between processes that trigger star formation by accelerating gas cooling and others which tend to prevent star formation by expelling or heating gas (e.g.~\citealp{2013ApJ...772..119L}). Stellar mass functions (SMFs) are a key first-order observable which allow one to statistically trace back the formation of stars in the universe. Comparison with predicted SMFs constrains the mechanisms which trigger, enhance or inhibit star formation.

Deep NIR surveys over large areas undertaken in the last years probe the evolution of the stellar-mass functions from $z\sim 4$ (e.g. \citealp{2008ApJ...675..234P, 2013A&A...556A..55I, 2013ApJ...777...18M}). They have shown that some form of feedback, to avoid the over-formation of stars both at the high mass and low mass ends, is necessary. Another key result is that the abundance of passive galaxies steadily increases from $z\sim 4$ to $z\sim 0$. The quenching of star formation is therefore a key process in the evolution of baryons. It causes a bimodal color distribution at least from $z\sim 3$ (e.g.~\citealp{2011ApJ...735...86W}) and is probably the main explanation for the decrease of the star-formation rate density in the universe (e.g.~\citealp{2014ARA&A..52..415M}). 

What makes a galaxy quench is still an open and extensively debated question. The evolution of the SMFs of passive and star-forming galaxies suggests that stellar mass (or more generally halo mass) is a fundamental property tightly linked to the star-formation activity. 
The $z\sim 0$ SMF has a \emph{knee} ($M^{*}$) around $\sim10^{10.7}M_\odot$, and this mass scale seems to be independent of redshift.  Galaxies tend to quench when they reach that characteristic mass (e.g.~\citealp{2010ApJ...721..193P, 2013A&A...556A..55I, 2016arXiv160205917M}). This \emph{mass-quenching process} primarily takes place at $z>1$ because, at later times, there are more passive than star-forming galaxies with this mass, so mass-quenching becomes less relevant.
Therefore, at late times, most of the quenching activity happens below $\sim10^{10.7}M_\odot$, and this tends to flatten the low mass end of the passive galaxy SMF (e.g.~\citealp{2016arXiv160205917M}). Since most of these galaxies are satellites, this quenching is generally referred to as \emph{environmental quenching}. Even though this empirical description of quenching has been extremely successful in explaining the global trends, the actual physical mechanisms behind quenching are still largely unconstrained. 

Stellar mass functions alone do not provide  information on how the formation of stars affects galaxy structure. It is however well established that star formation activity is strongly correlated with morphology. Galaxies which live on the main sequence of star-formation tend to have a disk-like morphology with low Sersic indices, while passive galaxies tend to have early-type morphologies and Sersic indices larger than 2 (e.g.~\citealp{2011ApJ...742...96W}). Whether this is a cause or a consequence is not yet known \citep{2016arXiv160406459L}. Several studies claim that the observed relation between structure and star-formation is in fact a consequence of very dissipative quenching processes. A large amount of gas would be driven into the central parts of the galaxies producing a central burst of star-formation and therefore a bulge with high central stellar mass density (e.g. \citealp{2013ApJ...765..104B, 2015arXiv150900469B}). However, recent evidence suggests that the dominant quenching mechanism at intermediate stellar masses might be simply a shutting off of the gas supply through strangulation (e.g.~\citealp{2015Natur.521..192P}) without significant morphological transformations even at very high redshifts (e.g. \citealp{2016MNRAS.458L..14F}). The observed correlation between central stellar mass density and star-formation rate could be mostly explained by the fading of the disk after the strangulation event (e.g. \citealp{2014arXiv1402.1172C}). This would also explain the relative large abundance of fast-rotating passive galaxies in the local universe (see~\citealp{2016arXiv160204267C} for a review). 

{Properly quantifying how the joint distribution of morphology and mass evolves might shed new light on which are the main quenching processes. It also provides a new element of comparison with recent numerical and empirical simulations which now predict morphologies and structure (e.g~\citealp{2014MNRAS.444.1518V}). However, there is currently no benchmark measurement of this type.
{Large surveys such as SDSS \citep[$z\le 0.25$][]{2013MNRAS.436..697B} and more recently GAMA \citep[$z\le 0.06$][]{2016MNRAS.457.1308M} have enabled a good quantification of the morphological dependence of the SMF at low redshift (the larger volume of the SDSS means it is able to probe rarer higher masses than GAMA). Pushing to higher redshift requires better angular resolution over large areas.  As a result, there are very few complete studies of the morphological dependence of the SMF at high redshift. \cite{2005ApJ...625..621B} made a first attempt but stopped at $z\sim 0.8$. Most of the studies at higher redshift are based on smaller subsets of objects (e.g.~\citealp{2005ApJ...620..564C, 2011arXiv1111.6993B,2011MNRAS.413.2845M}) or broad morphologies \citep{2015MNRAS.447....2M}. Future space based wide surveys, such as EUCLID and WFIRST-AFTA, will clearly be a major step forward. In the meanwhile, the largest area observed by the Hubble Space Telescope both in the optical and infrared is the CANDELS survey~\citep{2011ApJS..197...35G, 2011ApJS..197...36K}. Even though it does not reach the same coverage as ground based surveys such as UltraVista~\citep{2012A&A...544A.156M}, it probes at high angular resolution the rest-frame optical morphologies of galaxies from $z\sim3$ with a similar depth to the deepest NIR ground-based surveys. In this sense, it is currently the best available dataset for establishing robust constraints on the abundance of different morphologies in the early universe. This is the main purpose of the present work.  

In \cite{2015ApJS..221....8H}, we used new deep-learning techniques to estimate the morphologies of all galaxies with $H<24.5$  in the five CANDELS fields with unprecedented accuracy\footnote{The catalog is available at \url{http://rainbowx.fis.ucm.es/Rainbow\_navigator\_public/}}. We now use these morphologies, together with robust stellar mass estimates from extensive multi-band imaging, to study the evolution of the SMFs of quiescent and star-forming galaxies of different morphologies from $z\sim3$, for the first time. We then discuss the implications for the dominant quenching processes and morphological transformations. The data on the mass functions are made public so that they can be directly compared with the predictions of different models.

The paper is organized as follows. In~\S~\ref{sec:dataset}, we describe the dataset used as well as the main physical parameters we measure (morphologies, structural parameters, stellar masses etc.). In~\S~\ref{sec:MFs} we describe the methodology used to derive the stellar mass functions. \S~\ref{sec:evol_MF} discusses their evolution. Finally in ~\S~\ref{sec:discussion}, we discuss the implications for the star formation histories of the different morphologies and the evolution of the quenching mechanisms at different cosmic epochs.  

Throughout the paper, we assume a flat cosmology with $\Omega_M=0.3$, $\Omega_\Lambda=0.7$ and $H_0= 70$ $km.s^{-1}.Mpc^{-1}$ and we use magnitudes in the AB system. All stellar masses were scaled to a \cite{2003PASP..115..763C} IMF.

\section{Dataset}
\label{sec:dataset}

\subsection{Parent sample}
Galaxies in the 5 CANDELS fields (UDS, COSMOS, EGS, GOODS-S, GOODS-N) are selected in the F160W by applying a magnitude cut F160W$<$24.5~mag (AB). The total area is $\sim 880$ $arcmin^2$. We use the CANDELS public photometric catalogs for UDS \citep{2013ApJS..206...10G} and GOODS-S \citep{2013ApJS..207...24G} and soon-to-be published CANDELS catalogs for COSMOS, EGS (Stefanon et al. 2016) and GOODS-N (Barro et al. 2016). The magnitude cut is required to ensure the availability of morphologies \citep{2015ApJS..221....8H} a key quantity for the analysis presented in this work. The stellar mass completeness resulting from this magnitude cut is extensively discussed in section~\ref{sec:compl} given its importance to derive reliable stellar mass functions.

\subsection{Structural properties}
\label{sec:struct}
We use the publicly available 2D single Sersic fits from \cite{2012ApJS..203...24V} to estimate basic structural parameters (radii, Sersic indeces, axis ratios). The fist were done using  galfit~\citep{2002AJ....124..266P} on the three NIR images (F105W,F125W,F160W). The expected uncertainty on the main parameters is less than $20\%$ for the magnitude cut applied in this work as widely discussed in~\cite{2012ApJS..203...24V, 2014ApJ...788...28V}. 

\subsection{Morphological classification}
\label{sec:morphos}
We use the deep-learning morphology catalog described in \cite{2015ApJS..221....8H}. In brief, the ConvNets-based algorithm is trained with visual morphologies available in GOODS-S and then applied to the remaining 4 fields. Following the CANDELS classification scheme, we assign 5 numbers to each galaxy: $f_{sph}$, $f_{disk}$, $f_{irr}$, $f_{PS}$, $f_{Unc}$.  These measure the frequency with which hypothetical classifiers would have flagged the galaxy as having a spheroid, a disk, presenting an irregularity, being compact (or a point source), and being unclassifiable/unclear.  For a given image, ConvNets are able to predict the various $f_{type}$ values with negligible bias on average, scatter of $\sim 10\%-15\%$, and fewer than $1\%$ mis-classifications \citep{2015ApJS..221....8H}. 


In what follows, we primarily use the H band (F160W) since our sample is dominated by galaxies at $z>1$, where NIR filters probe the optical rest-frame. For $z<1$ galaxies, we also explored the I band filters (814W, 850LP) but because the classes we define below are quite broad, the classifications do not change significantly (also see Kartaltepe et al. 2015). In addition, as we show below, at low redshifts our classifications match those in the SDSS rather well:  morphological k-corrections do not have a big impact on our results.

In this work we distinguish 4 main morphological types defined as follows:

\begin{itemize}
 \item \emph{spheroids [SPH]:}~~$f_{sph}>2/3$ AND $f_{disk}<2/3$ AND $f_{irr}<0.1$
 \item \emph{late-type disks [DISK]:}~~$f_{sph}<2/3$ AND $f_{disk}>2/3$ AND $f_{irr}<0.1$
 \item \emph{early-type disks [DISKSPH]:}~~$f_{sph}>2/3$ AND $f_{disk}>2/3$ AND $f_{irr}<0.1$
 \item \emph{irregulars [IRR]:}~~$f_{sph}<2/3$ AND $f_{irr}>0.1$
\end{itemize}

The thresholds above are somewhat arbitrary but have been calibrated through visual inspection first to make sure that they indeed result in distinct morphological classes (see also~\citealp{2014arXiv1401.2455K}).

In appendix~\ref{app:morph_stamps} we show some randomly selected postage stamps of the different morphological classes in the COSMOS/CANDELS field sorted by stellar mass and redshift. Slight changes to the thresholds used to define these classes do not affect our main results. The SPH class contains bulge dominated galaxies with little or no disk:  it should be close to the classical Elliptical classification used in the local universe. The DISK class is made of galaxies in which the disk component dominates over the bulge (typically Sb-c galaxies). Between both classes, lies the DISKSPH class in which there is no clear dominant component:  it should include typical S0 galaxies and early-type spirals (Sa). We also distinguish galaxies with clear asymmetry in their light profiles.  This category should capture the variety of irregular systems usually observed in the high redshift universe (e..g clumpy, chain, tadpole etc.). This irregular class might contain a wide variety of galaxies with different physical properties since the classification is based on the irregularity of the light profile. Notice that \emph{IRR} is defined with no condition on $f_{disk}$; therefore, this class can include many late-type disks at low redshifts (i.e. Sds). 

We have verified that the different classes have distinct structural properties. Spheroids are more compact, rounder ($b/a\sim0.8$) and have larger Sersic indices ($n\sim4-5$) than all other morphologies at all stellar masses and at all redshifts. On the other extreme, disks are larger, more elongated ($b/a\sim0.5$) and have Sersic indices close to 1, as expected. Disk+sperhoids systems lie somewhat in between: they have Sersic indices $\sim 2$, but are less compact than the spheroids and have similar axis ratios to disks (in agreement with the visual classification; also see Huertas-Company et al. 2015a). Although a detailed analysis of the structural properties of the different morphologies will be presented elsewhere, appendix~\ref{app:struct_props_morph} shows that the different morphologies also have different stellar mass bulge-to-total mass ratios.  

 \subsection{Stellar masses and completeness}
\label{sec:compl}

SED fitting is used to estimate photometric redshifts and stellar masses used in this work. The detailed methodology is described in \cite{2011ApJ...742...96W,2012ApJ...753..114W} and \cite{2013ApJ...765..104B, 2014ApJ...791...52B}. Therefore only the main points are discussed here. Photometric redshifts are the result of combining different codes to improve the individual performance. The technique is fully described in \cite{2013ApJ...775...93D}.  Based on the best available redshifts (spectroscopic or photometric) we then estimate stellar mass-to-light ratios from the PEGASE01 models \citep{1999astro.ph.12179F}. For these, we assume solar metallicity, exponentially declining star formation histories, a \cite{2000ApJ...533..682C} extinction law and a Salpeter IMF (1955). The $M_*/L$ values are then converted to a Chabrier IMF by applying a constant 0.22 dex shift.  The stellar mass is estimated by multiplying the $M_*/L$ value by the Sersic-based $L$ (from galfit 2D fits - see section~\ref{sec:struct}).  See also Bernardi et al. 2013 and 2016 for extensive discussion of the systematics associated with all these choices.

The stellar mass completeness of the sample is estimated following the methodology of \cite{2010A&A...523A..13P} and \cite{2013A&A...556A..55I}  separately for star-forming and quiescent galaxies. We first compute the lowest stellar mass ($M_{lim}^*$) which could be observed for each galaxy of magnitude $H$ given the applied magnitude cut ($H<24.5$): $log(M^*_{lim})=log(M_*)+0.4(H-24.5)$. We then estimate the completeness as the 90th percentile of the distribution of $M_{lim}$, i.e. the stellar mass for which $90\%$ of the galaxies have lower limiting stellar masses. By adopting this threshold, we make sure that at most $10\%$ of the low mass galaxies are lost in each redshift bin. Figure~\ref{fig:mass_comp} shows the distribution of galaxies in our sample in the mass-redshift plane and the adopted stellar mass completeness as a function of redshift for passive and all galaxies. The sample is roughly complete for galaxies above $10^{10}$ solar masses at $z\sim3$ and goes down to $10^9$ at $z\sim0.5$ (see also table~\ref{tbl:fits_all}).  As a sanity check, we use an alternative estimate of the stellar mass completeness by taking advantage of the fact that the CANDELS data are significantly deeper than the H-band selected sample used here ($H<24.5$). We therefore compute in bins of redshift, the stellar mass at which 90\% of the galaxies in the full CANDELS catalog are also included in our bright selection. The resulting stellar mass completeness is over-plotted in figure~\ref{fig:mass_comp}. It agrees reasonably well with the one estimated independently using the methodology by \cite{2010A&A...523A..13P}. The largest differences are observed at the high mass-end. It can be a consequence of low statistics in these stellar mass bins. In the following we will adopt therefore  the first estimate, keeping in mind however that at high redshift we might under-estimate the completeness.
 
  \begin{figure*}
\begin{center}
$\begin{array}{c c c}
\includegraphics[width=0.45\textwidth]{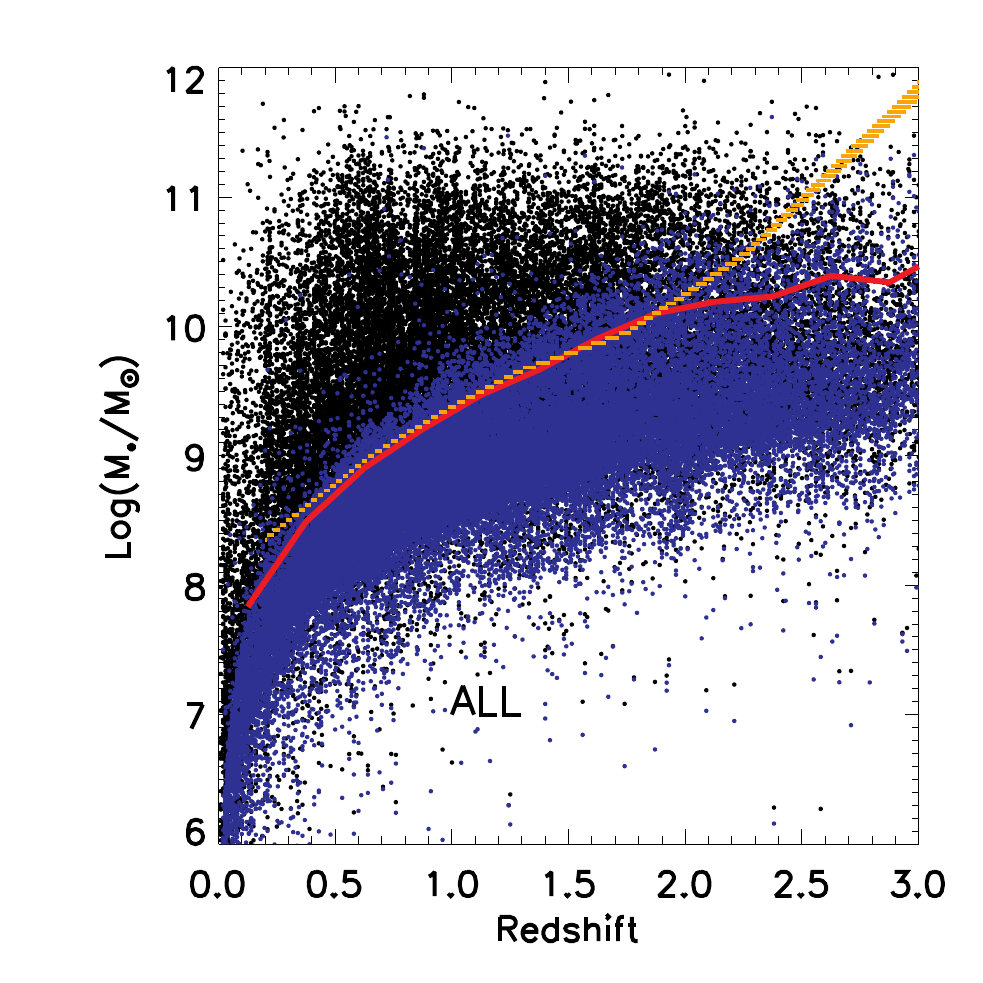} & \includegraphics[width=0.45\textwidth]{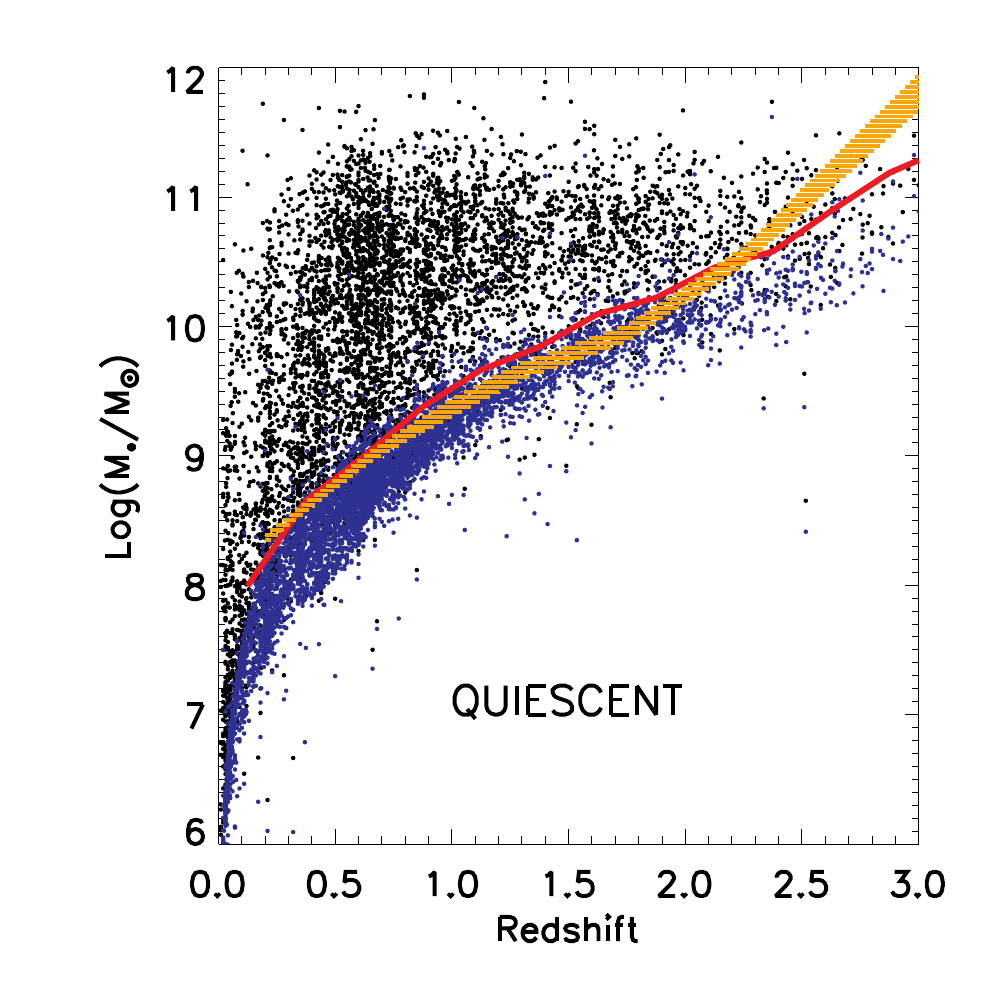} \\
 \end{array}$
\caption{Stellar mass as a function of redshift for all galaxies (left panel)  and quiescent galaxies (right panel) for our $H<24.5$ selected sample. Blue points show the minimum stellar mass which can be observed for a given galaxy computed as explained in the main text. The red line shows the 90th percentile of the distribution of $M_{lim}$ which is the adopted mass completeness in this work (see text for details). The orange line shows the mass completeness estimated using the full depth CANDELS catalog (see text for details)}
\label{fig:mass_comp}
 \end{center}
 \end{figure*}

 \subsection{Quiescent / star-forming separation}
 
Rest-frame magnitudes (U, V and J) are computed based on the best-fit redshifts and stellar templates (see section~\ref{sec:compl})  and are then used to separate the passive and star-forming populations as widely used in the previous literature \citep{2012ApJ...754L..29W}. This color-color separation has the advantage of properly distinguishing galaxies reddened by dust from real passive galaxies with old stellar populations. 
 
  \section{Estimation of morphological stellar mass functions}
\label{sec:MFs}

We use the $V_{max}$ estimator \citep{1968ApJ...151..393S} to derive the stellar mass functions in this work. It has the advantage of being very simple but can easily diverge when the incompleteness becomes too important. For this reason, we restrict our analysis to stellar masses above the thresholds derived in section~\ref{sec:compl} and quoted in table~\ref{tbl:fits_all}. Recent works have shown that above the completeness limits, very consistent results are obtained with maximum-likelihood methods (e.g.~\citealp{2013A&A...556A..55I}). For simplicity, we restrict our analysis to  one single estimator throughout this work. 

\subsection{Uncertainties}

We consider 3 sources of errors which contribute to the uncertainties on the SMFs. Namely Poisson errors ($\sigma_P$), cosmic variance ($\sigma_{CV}$) and errors associated with the estimation of stellar masses and photometric redshifts ($\sigma_{T}$).  Poisson errors reflect exclusively statistical uncertainties due to the limited number of galaxies in each bin. They are proportional to the the square root of the number of objects. Cosmic variance errors are related to the fact that we observe a small area in the sky so our measurements can be affected by statistical fluctuations in the number of galaxies due to the underlying large scale density fluctuations. Cosmic variance can be computed from the galaxy bias and the dark matter cosmic variance assuming a CDM model. We use the tool of \cite{2011ApJ...731..113M} to estimate the fractional error in density given the size of the CANDELS fields and also their spatial distribution.  Finally, uncertainties in stellar mass and redshifts do have an impact on the density of galaxies. Stellar masses are obtained through SED fitting assuming a photometric redshift (spectroscopic redshifts are available for a minority of sources). There are therefore systematic (e.g. template errors, IMF assumptions, SFHs) and statistical errors associated with this methodology. 

To estimate this uncertainty, we take advantage of the various measurements of stellar masses and redshifts existing in CANDELS.  E.g., the 3D-HST team has computed an independent set of photometric redshifts and derived stellar masses using the FAST and EAZY codes~\citep{2014ApJS..214...24S}. They used BC03 models and a Chabrier IMF.  A comparison of the two should provide an estimate of the errors induced in the SMFs due to errors in redshifts and stellar masses. We therefore generated a set of $50$ catalogs by randomly combining stellar masses and photometric redshifts from the CANDELS and 3D-HST catalogs and recomputed the SMFs for each of them. We then measured the scatter in the final 50 SMFs in bins of redshift and stellar mass.  This scatter combines the statistical error associated with estimating $M_*$ from fitting noisy photometry to a given set of templates, with the systematic error associated with the fact that the templates used have built-in assumptions about the star formation history (bursty or not? dusty or not? etc.). We note however that this approach certainly under estimates the errors. There is in fact a large overlap in assumptions made, notably the exponentially declining tau models and
Calzetti reddening law. Additionally, although the photometric extractions by the 3D-HST and CANDELS teams were done independently, the actual data on which the photometry
is based are nearly identical. This is clearly not ideal.  Nevertheless, we lump these together and add in quadrature to the other two terms.  Hence, to each bin we assign an uncertainty
 $$\sigma=\sqrt{\sigma_{P}^2+\sigma_{CV}^2+\sigma_T^2}$$.

Figure~\ref{fig:errors} shows the different fractional errors on the number density of galaxies as a function of stellar mass and redshift for the total sample. Cosmic variance dominates the error budget for stellar masses below $\sim10^{11}$. It is in fact always greater than $10\%$ while Poisson and template fitting errors are generally below $\sim5\%$. At larger stellar masses, the small number statistics generate an increase in the Poisson and template errors, which can exceed $50\%$ at the very highest masses. In the morphology divided samples, the number of objects is obviously reduced and therefore the statistical errors dominate over comic variance effects at all stellar masses. Similar trends are observed when the objects are separated into star-forming and quiescent samples. 

  \begin{figure*}
\begin{center}
$\begin{array}{c c c}
\includegraphics[width=0.330\textwidth]{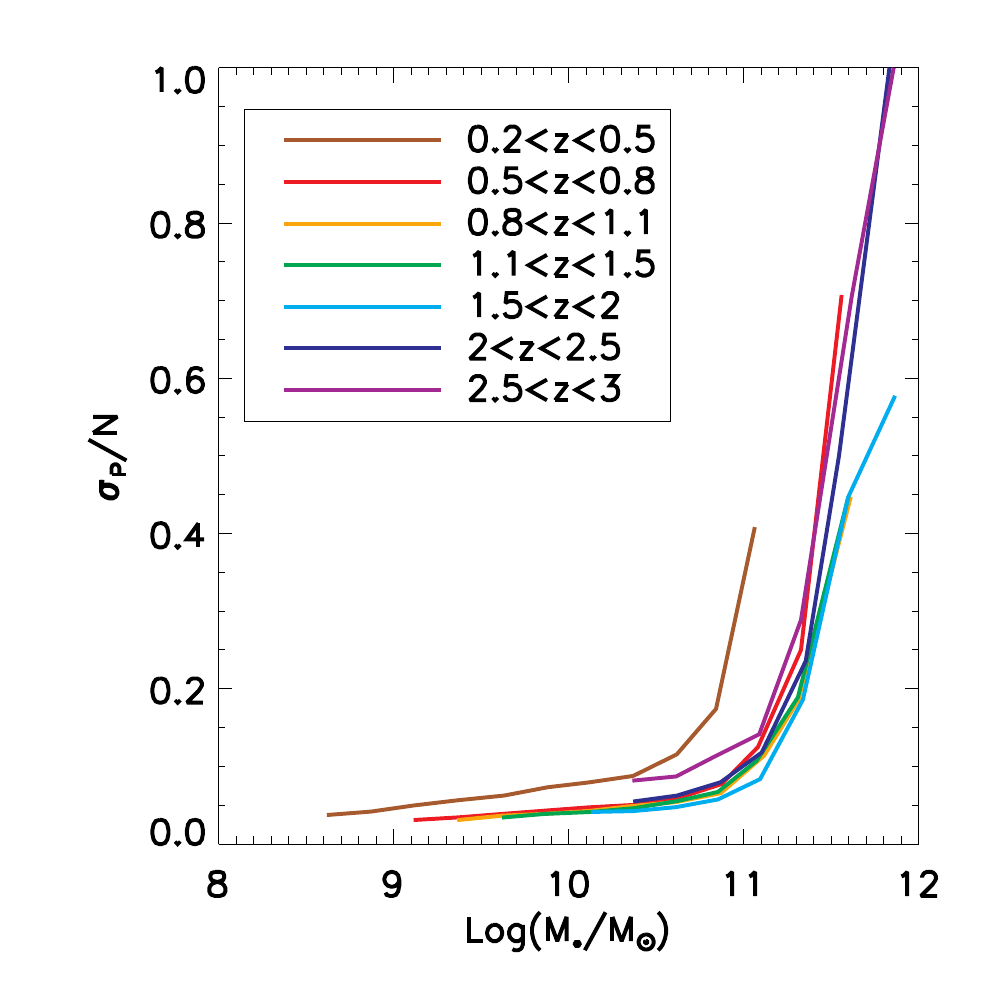} & \includegraphics[width=0.33\textwidth]{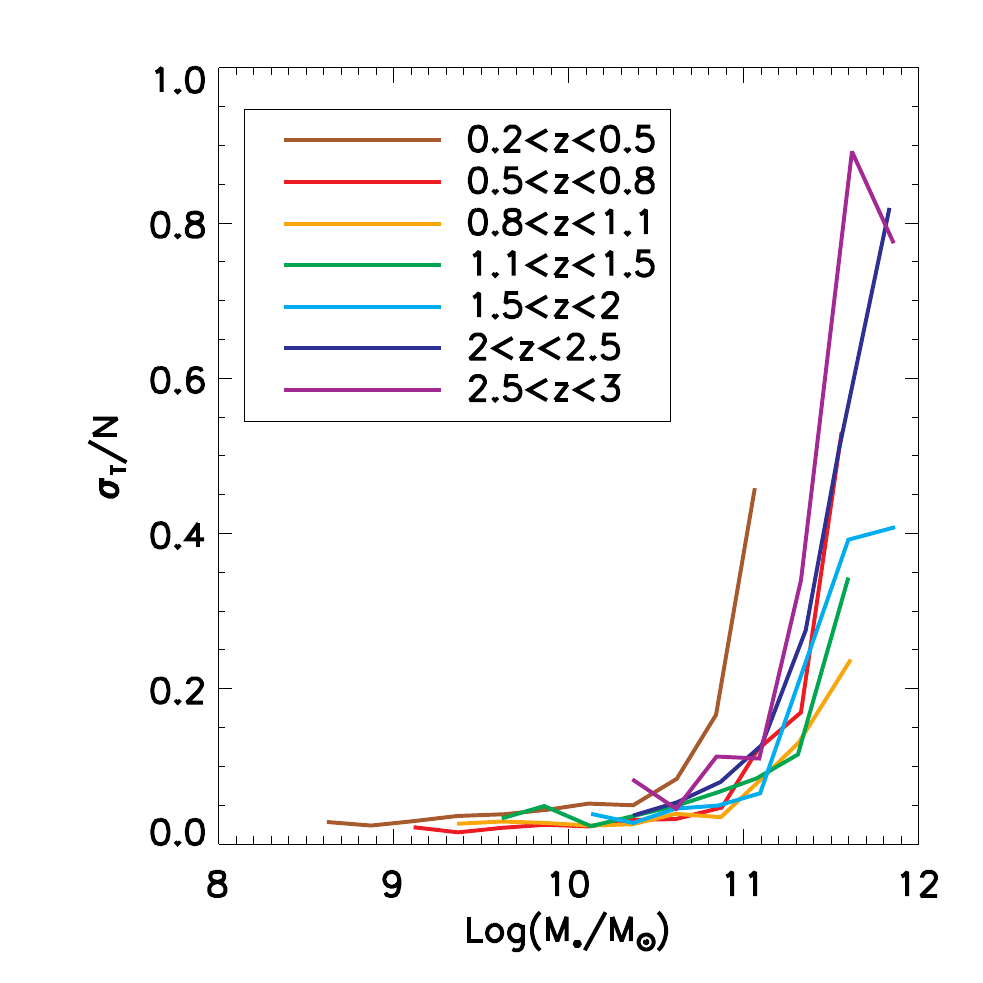} & \includegraphics[width=0.33\textwidth]{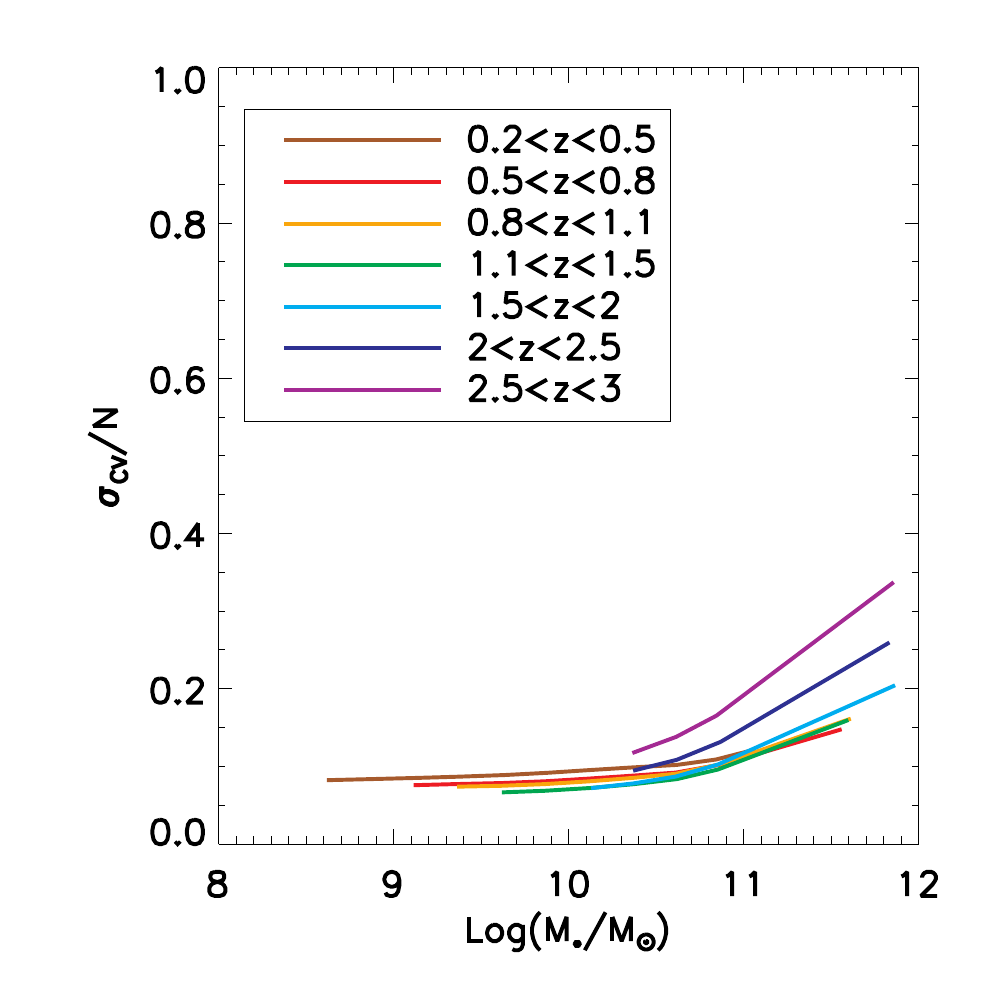}\\
 \end{array}$
\caption{Fractional errors on the number densities of galaxies as a function of stellar mass and redshift for the total sample used in this work. The left, middle and right-hand columns show poisson errors, template fitting-related errors and the effects of cosmic variance.} 
\label{fig:errors}
 \end{center}
 \end{figure*}

\subsection{Schechter function fits}

The non-parametric Vmax estimator is fitted with a Schechter or double Schechter model, depending on the sample. Given that our sample is not large, especially  when it is divided into different morphological types, we preferentially use a single Schechter fit for $z>0.8$. Only in the lower redshift bins, where the SMFs reach lower stellar masses and an upturn is observed, do we adopt a double Schechter as done in previous works (Pozzetti et al. 2010, Ilbert et al. 2013, Muzzin et al. 2013). In all cases,  we only fit data points above the completeness limit to avoid biases related to the fact that the $1/V_{max}$ estimator tends to underestimate the number densities beyond the completeness limits. 

\section{Evolution of the morphological stellar mass functions}
\label{sec:evol_MF}

In this section, we discuss the evolution of the SMFs for different morphologies.  

\subsection{Full sample evolution}

Figure~\ref{fig:MFs} shows the stellar mass functions for each of the 4 morphological types defined previously; the different panels show results for 7 redshift bins.  Figure~\ref{fig:MFs_z} shows the same information in a different format:  each panel shows the evolution of the SMF for a fixed morphological type.  The redshift bin sizes were determined by a trade-off between number of objects and lookback time as seen in tables~\ref{tbl:fits_all} and~\ref{tbl:fits_SF}. The functions are only plotted above the mass completeness limit derived in section~\ref{sec:compl}. Best-fit parameters are shown in table~\ref{tbl:fits_all}. 

The global mass functions, i.e., not subdivided by morphology, (black region in figure~\ref{fig:MFs}) are also shown and compared with recent measurements in the UltraVista survey by Ilbert et al. (2013) and Muzzin et al. (2013).  There is good agreement despite the significantly smaller volume probed by the CANDELS fields.  This suggests that our completeness limits are well-estimated. Volume effects are mostly visible in the lowest redshift bin, where the CANDELS SMFs show a lack of very massive galaxies.  


Before we consider CANDELS in more detail, it is worth remarking on the cyan curve (same in each panel), which shows the SMF in the SDSS from \cite{2016arXiv160401036B}.  The large volume of the SDSS means Poisson errors are negligible, so the shaded region encompasses the systematic differences between different $M_*/L$ estimates.  At low $z$, the UltraVista measurements are in good agreement with the SDSS; moreover, they lie below it at higher redshifts, as one might expect.  In contrast, Figure~5 of Ilbert et al. (2013) shows that their $z\sim 0.5$ SMF lies {\em above} the $z\sim 0.1$ SDSS SMF, which does not make physical sense.  This is because their Figure~5 used the SDSS estimate of Moustakas et al. (2013).  Bernardi et al. (2016) discuss why their estimate is to be preferred; note that their work was not motivated by this problem, so the fact that evolution makes better physical sense when their SMF is used as the low redshift benchmark provides additional support for their analysis.  Very briefly, the main reason is that Moustakas et al. (2013) used SDSS Model magnitudes, which underestimate the total luminosity of bright galaxies (Bernardi et al. 2010, 2013;  D'Souza et al. 2015; see especially Figures 2 and 3 in Bernardi et al. 2016 and related discussion).  This accounts for about half the difference from Bernardi et al.; the remainder is due to M*/L.  Section 4.3 of Bernardi et al. (2016) discusses this in more detail (see, e.g., their Figures 14-16).  We refer the reader to Bernardi et al. (2016) for a more complete discussion. 

If the evolution is driven by star-formation (no mergers), then figure~\ref{fig:MFs_z} shows that the stellar mass of galaxies below $M^*$ increases by more than 1~dex in the redshift range  $0.5<z<3$. More massive galaxies increase their stellar mass by less than 0.5 dex.  Therefore, we confirm previous reports of a mass-dependent evolution for the global population.  In addition, Table~\ref{tbl:fits_all} shows that $Log(M^*/M_\odot)\sim10.85\pm 0.1$ is approximately independent of redshift.  

\begin{figure*}
\begin{center}
$\begin{array}{c c c}
\includegraphics[width=0.33\textwidth]{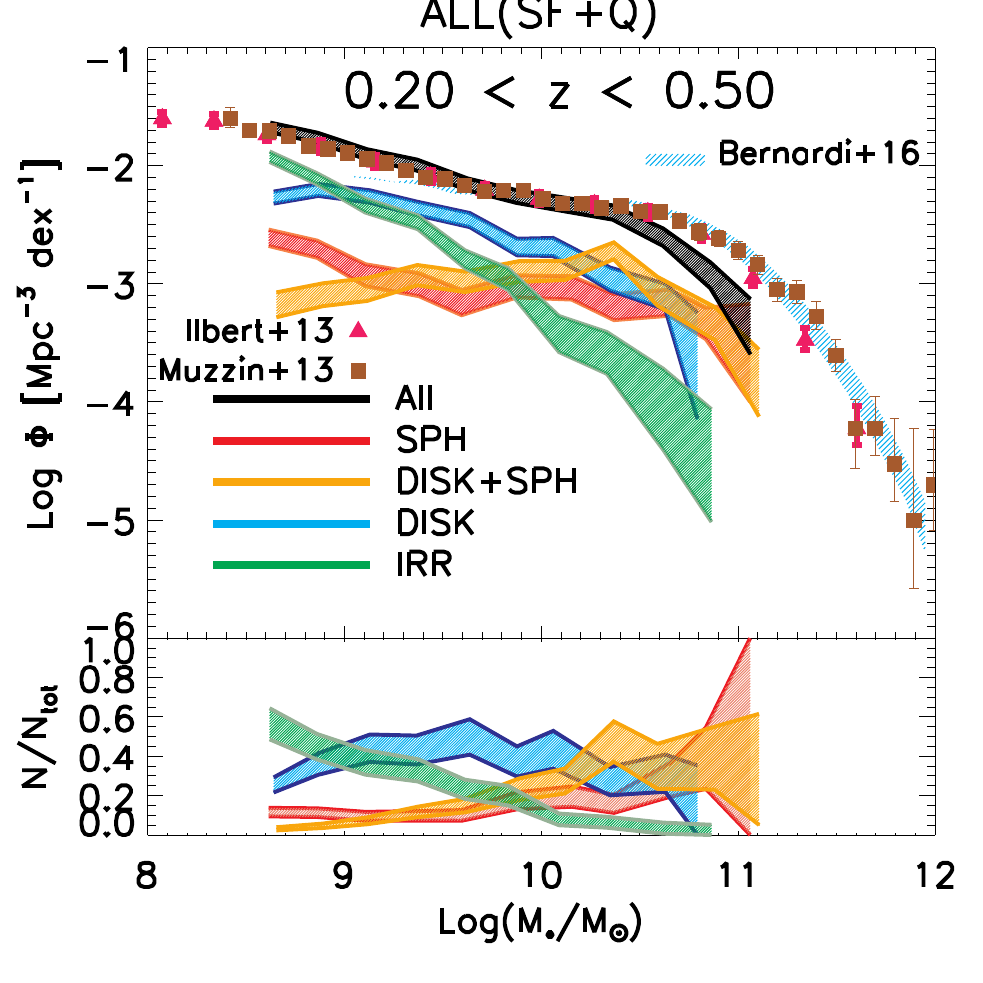} & \includegraphics[width=0.33\textwidth]{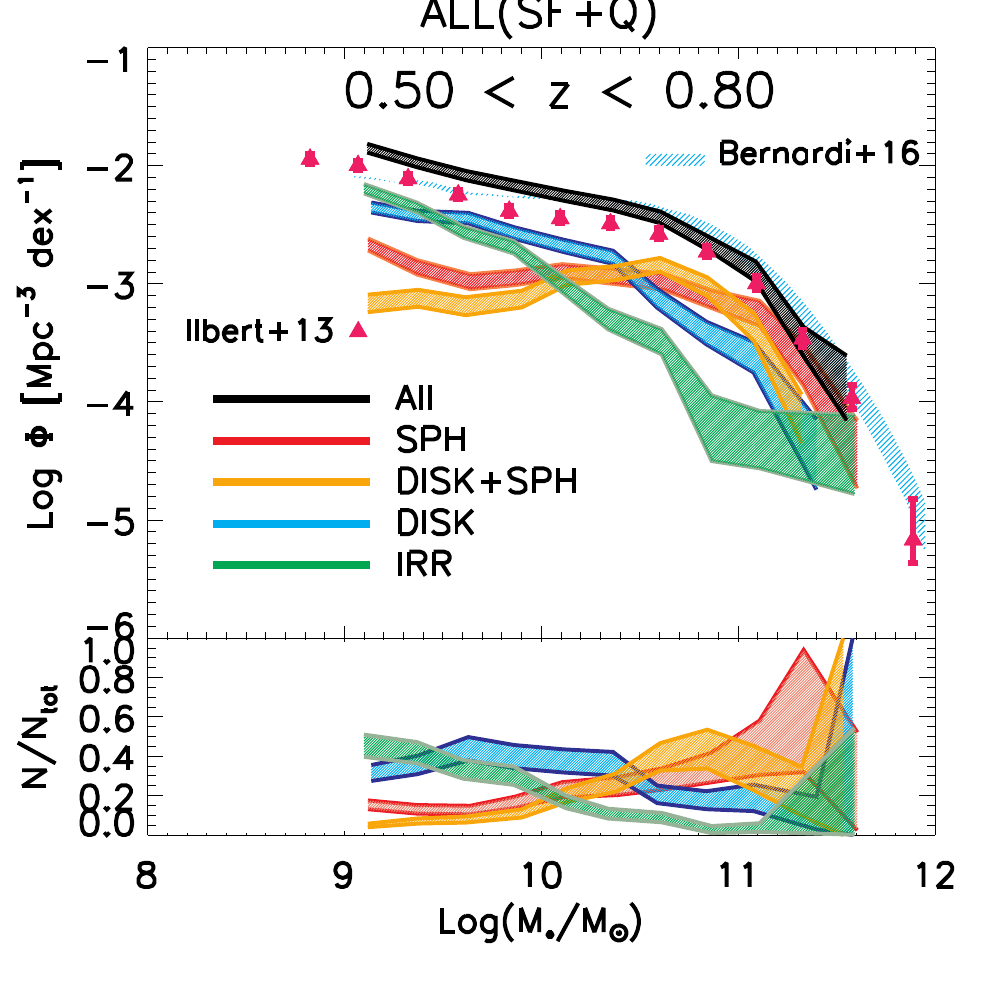} &  \includegraphics[width=0.33\textwidth]{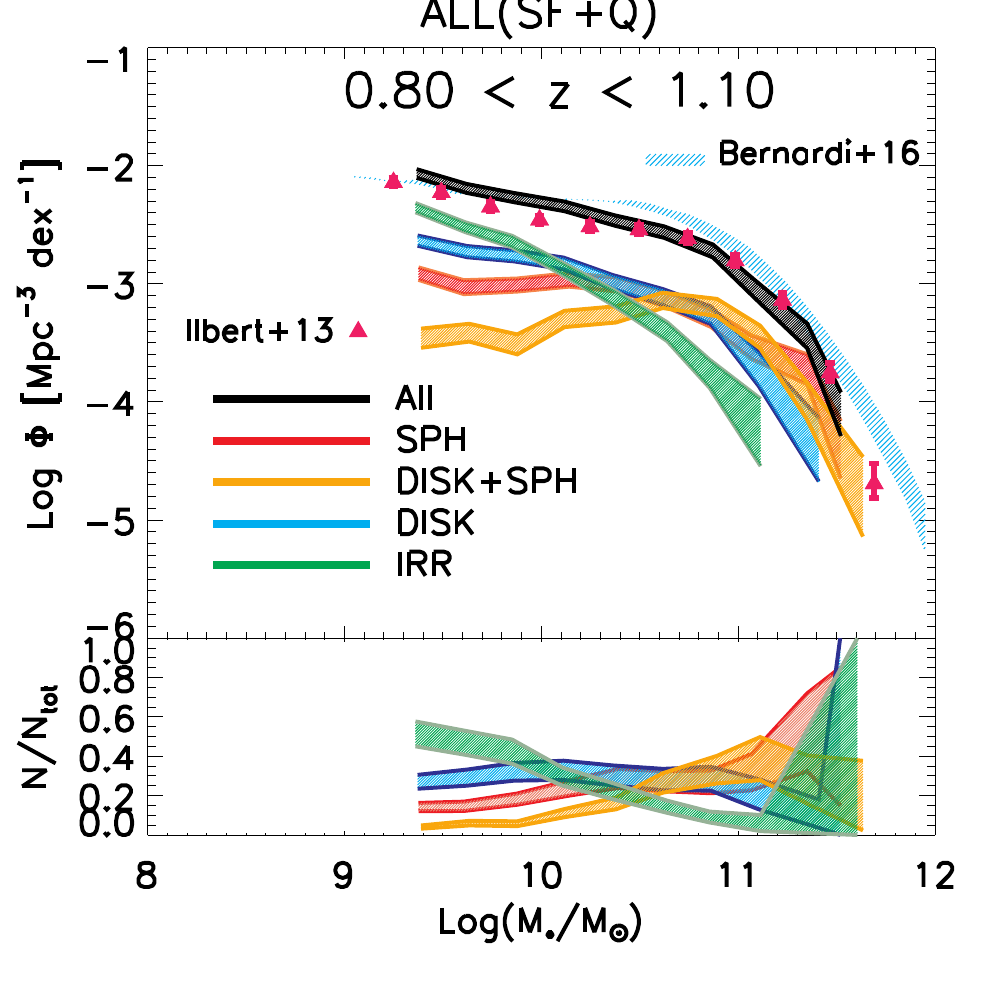} \\
\includegraphics[width=0.33\textwidth]{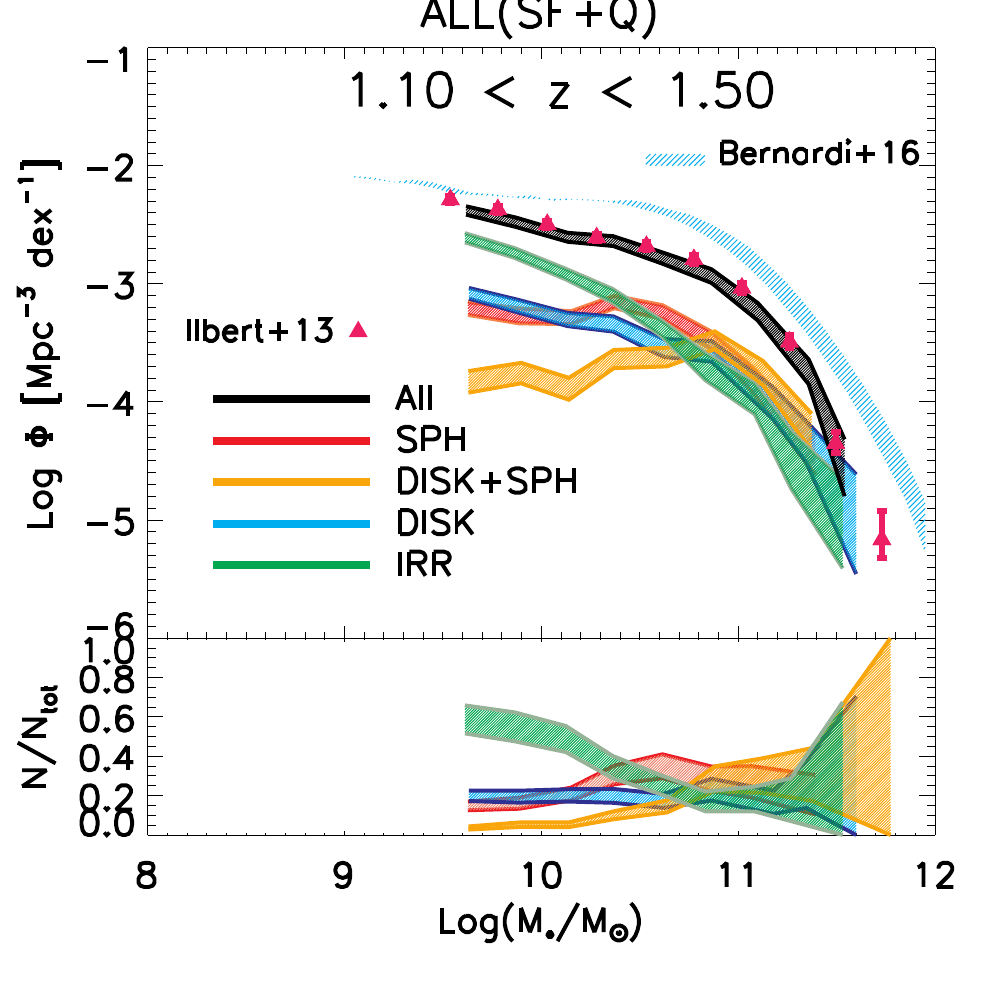} & \includegraphics[width=0.33\textwidth]{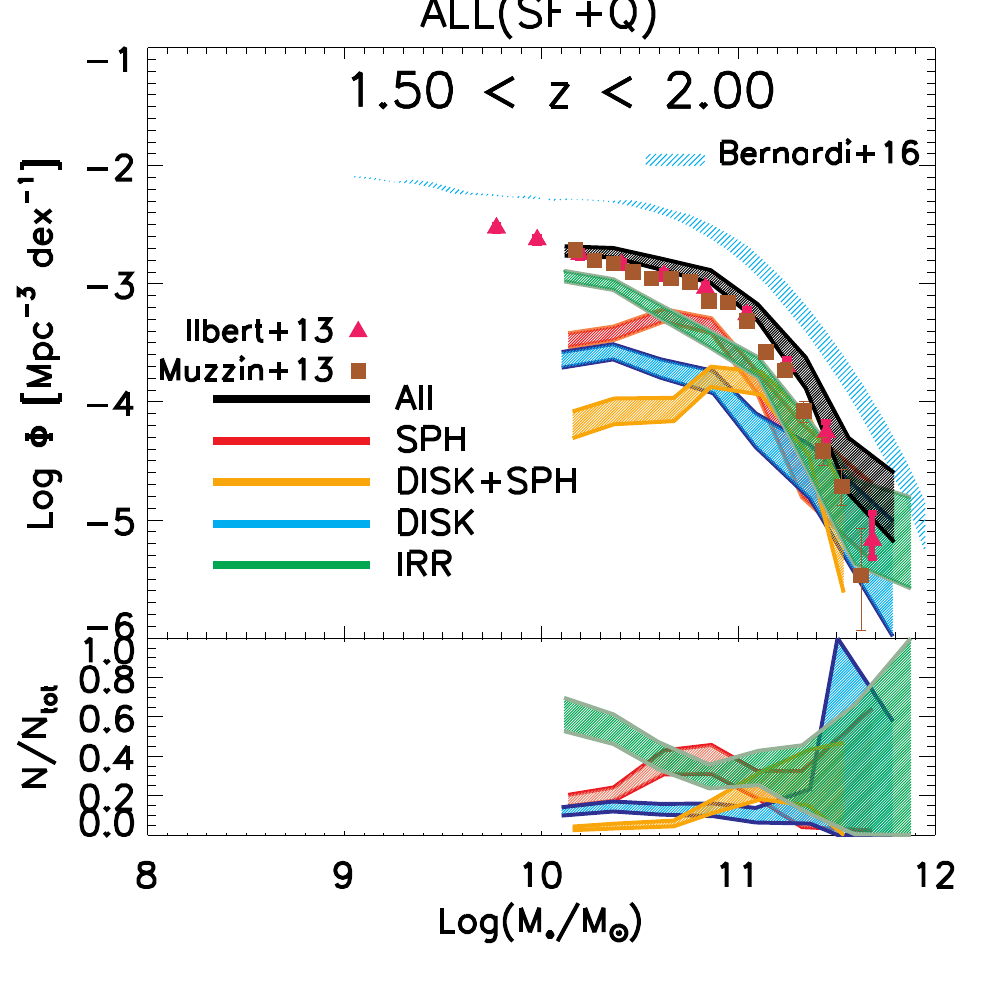} & \includegraphics[width=0.33\textwidth]{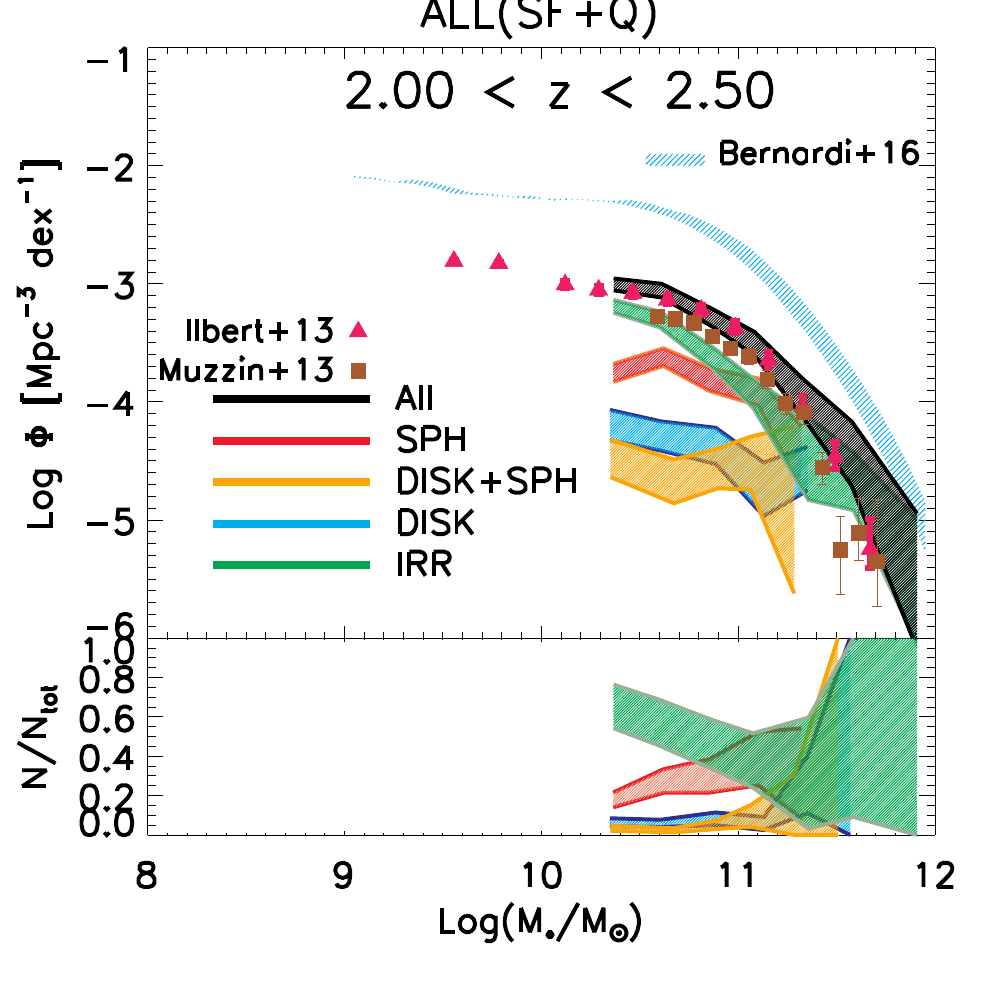}\\
\includegraphics[width=0.33\textwidth]{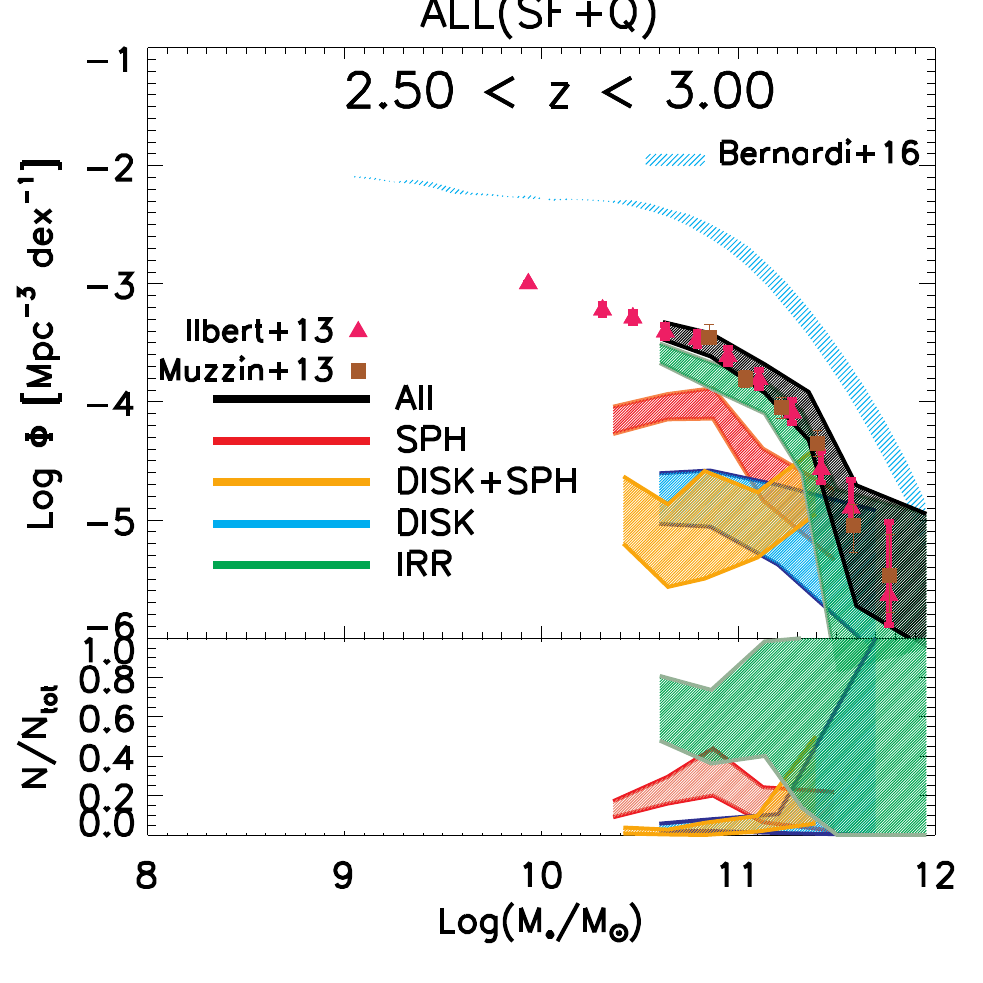}  & &\\
\end{array}$
\caption{Stellar mass functions for 4 morphological types in different redshift bins as labelled. Red, blue, orange and green shaded regions in the top panels show the number densities of spheroids, disks, disk+spheroids and irregular/clumpy systems respectively. The bottom panels show the fractions of each morphological type with the same color code. The black regions show the global stellar mass functions. The pink triangles and brown squares are the measurements by~\protect\cite{2013A&A...556A..55I} and~\protect\cite{2013ApJ...777...18M} respectively in the UltraVista survey. The~\protect\cite{2013ApJ...777...18M} points are only plotted when their redshift bins are the same than the ones used in this work. We also show for reference in all panels the SMF for all SDSS galaxies (cyan shaded region) from Bernardi et al. (2016). }
\label{fig:MFs}
\end{center}
\end{figure*}

\begin{figure*}
\begin{center}
$\begin{array}{c c}
\includegraphics[width=0.45\textwidth]{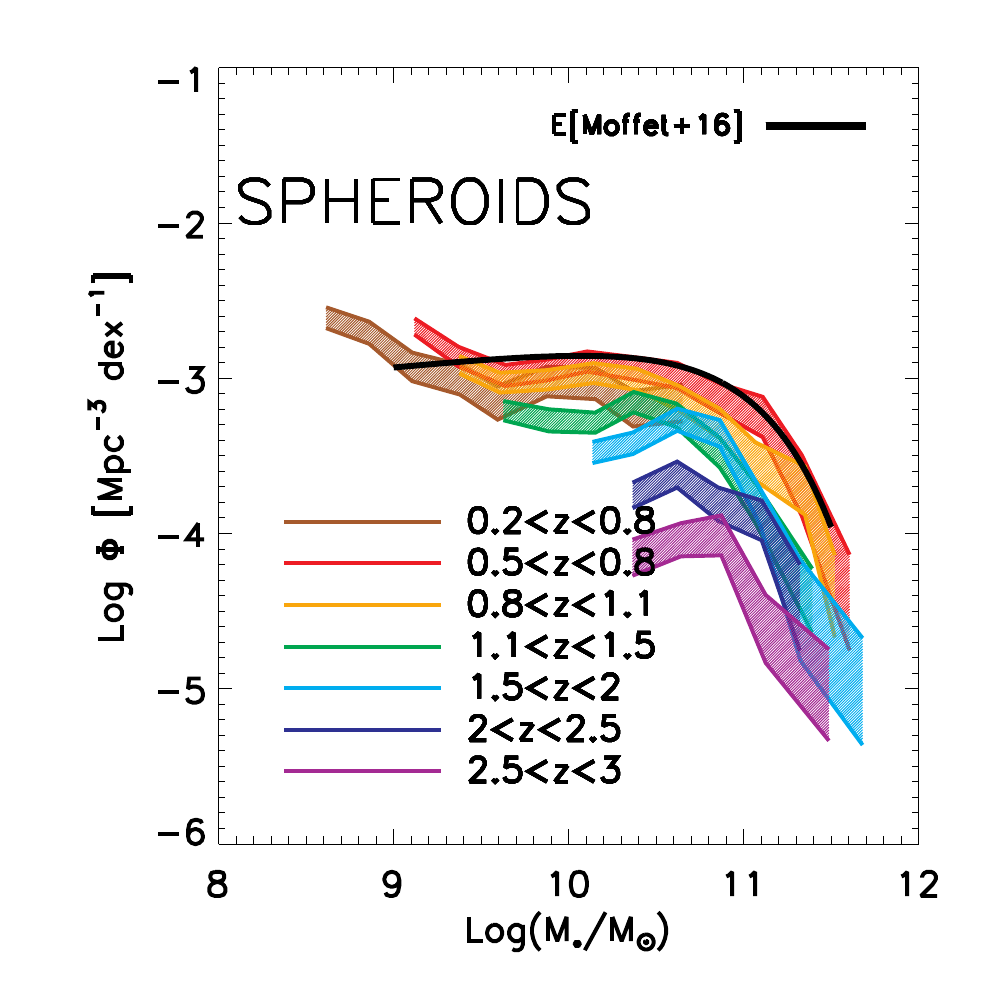} & \includegraphics[width=0.45\textwidth]{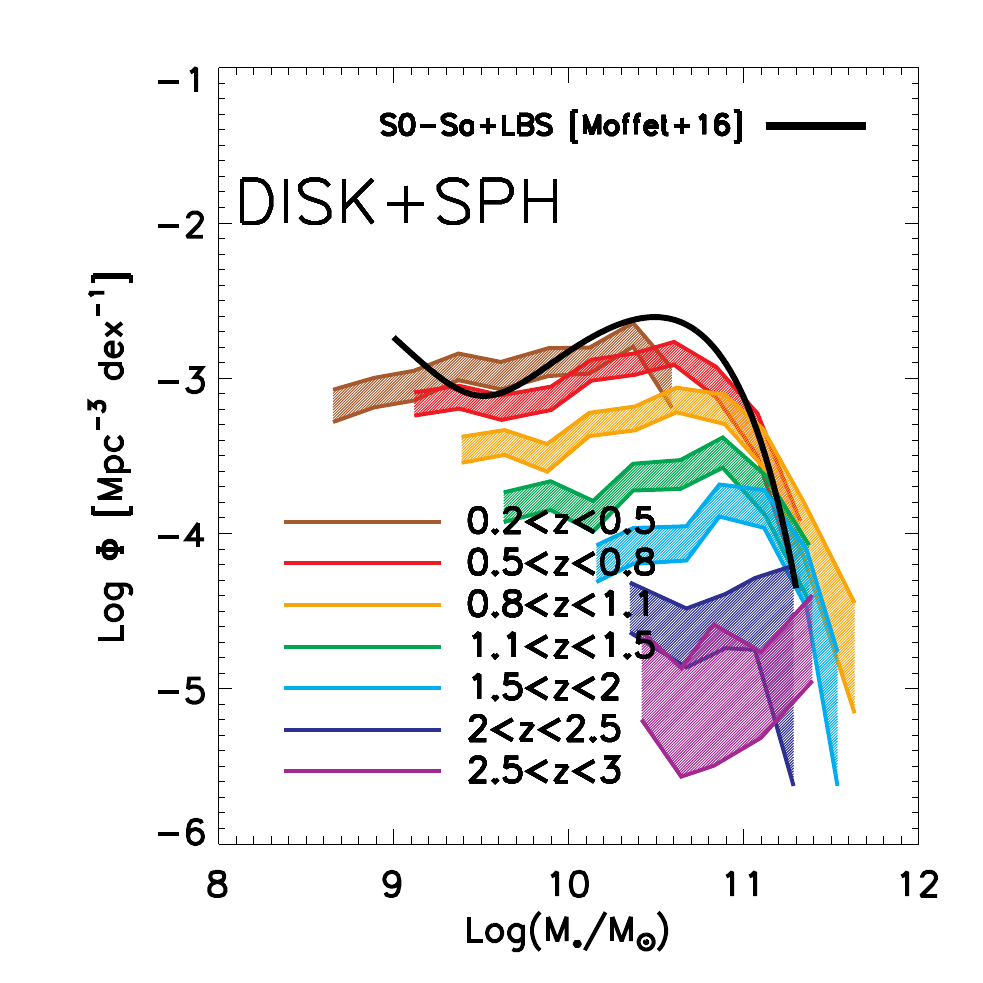}  \\
\includegraphics[width=0.45\textwidth]{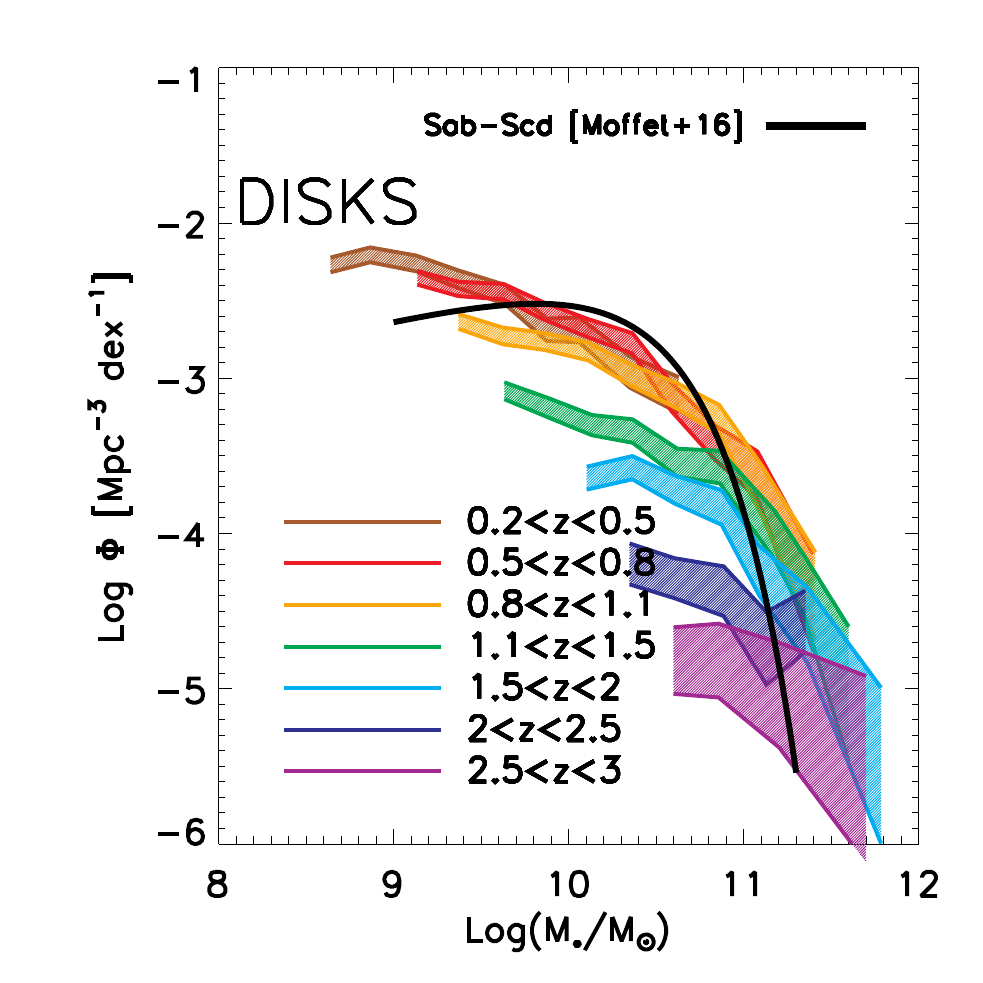} & \includegraphics[width=0.45\textwidth]{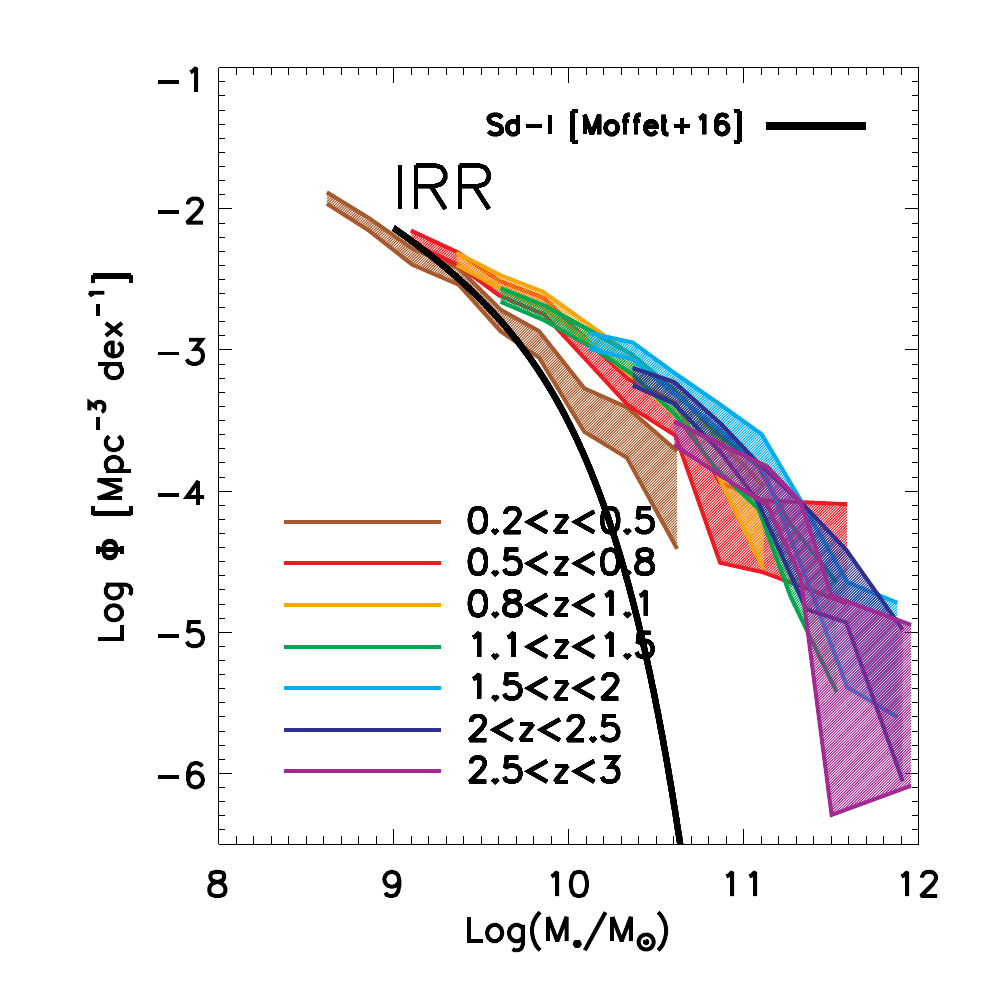} \\
\includegraphics[width=0.45\textwidth]{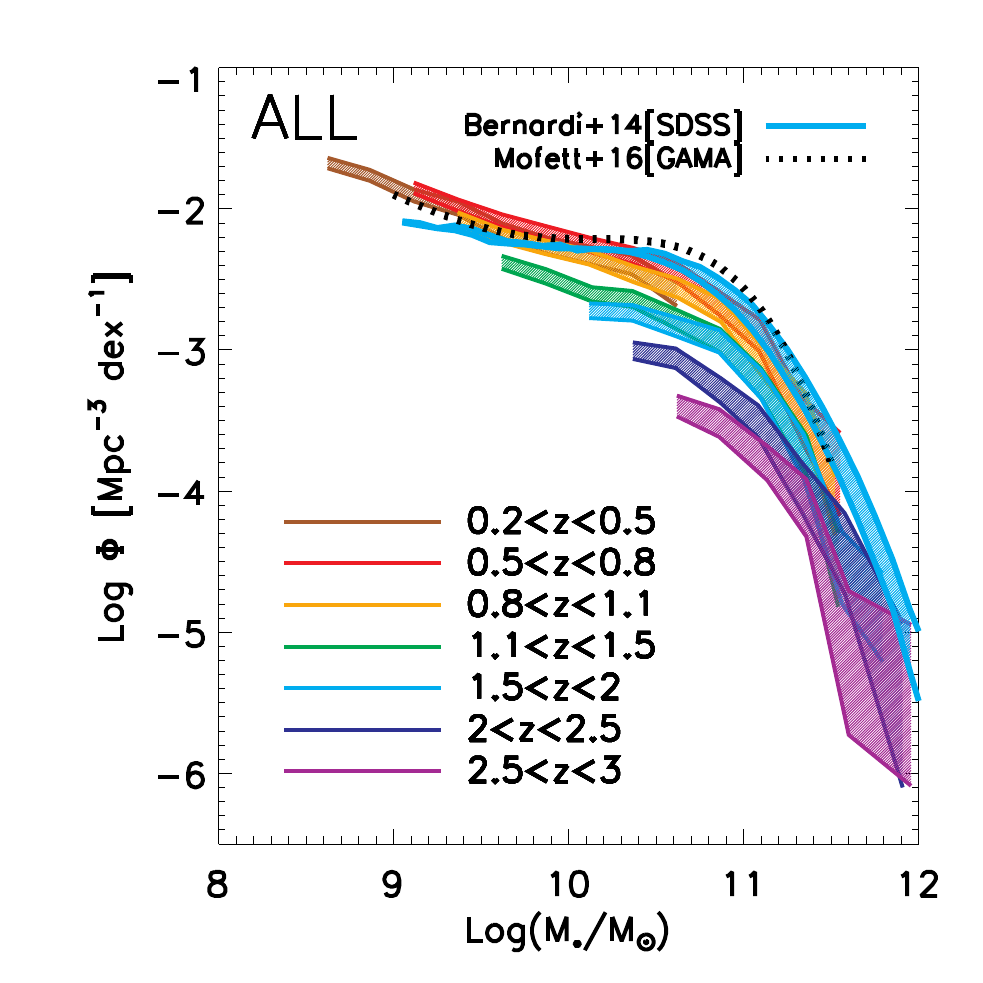}  & \\
\end{array}$
\caption{ Evolution of the stellar mass functions at fixed morphology. Same as figure~\protect\ref{fig:MFs} but binned by morphological type. Each color shows a different redshift bin. We also over-plot the local SMFs from Bernardi et al. (2016) for the total sample and the ones divided by morphology from Moffet et al. (2016) (best fit Schechter functions). }
\label{fig:MFs_z}
\end{center}
\end{figure*}

\begin{table*}
\begin{tabular}{cccccccccc}
\hline
\hline
 \multicolumn{1}{c}{Sample} &
 \multicolumn{1}{c}{redshift} &
 \multicolumn{1}{c}{$N_{gal}$} &
 \multicolumn{1}{c}{$log(M_{complete})$} &
 \multicolumn{1}{c}{$log(M^*$)} &
 \multicolumn{1}{c}{$\Phi_1^*$} &
 \multicolumn{1}{c}{$\alpha_1$} &
 \multicolumn{1}{c}{$\Phi_2^*$} &
 \multicolumn{1}{c}{$\alpha_2$} &
  \multicolumn{1}{c}{$log(\rho_*)$} \\
 
  \multicolumn{1}{c}{} &
 \multicolumn{1}{c}{} &
 \multicolumn{1}{c}{} &
 \multicolumn{1}{c}{\tiny{$M_\odot$}} & 
     \multicolumn{1}{c}{\tiny{$M_\odot$}} & 
    \multicolumn{1}{c}{\tiny{$10^{-3}$ Mpc$^{-3}$}} &
     \multicolumn{1}{c}{} &
      \multicolumn{1}{c}{\tiny{$10^{-3}$ Mpc$^{-3}$}} &
        \multicolumn{1}{c}{} &
         \multicolumn{1}{c}{\tiny{$M_\odot.Mpc^{-3}$}} \\

\hline
ALL & 0.2-0.5& 5233&  8.43& $10.86^{+0.10}_{-0.10}$ & $2.22^{+0.59}_{-0.59}$ & $-0.82^{+0.24}_{-0.24}$ &$ 0.45^{+ 0.15}_{- 0.15}$ &$ -1.60^{+ 0.00}_{- 0.00}$ &$ 8.34^{+0.02}_{-0.02}$\\ 
& 0.5-0.8& 8276&  8.94& $10.86^{+0.10}_{-0.10}$ & $2.22^{+0.59}_{-0.59}$ & $-0.82^{+0.24}_{-0.24}$ &$ 0.45^{+ 0.15}_{- 0.15}$ &$ -1.60^{+ 0.00}_{- 0.00}$ &$ 8.34^{+0.03}_{-0.03}$\\ 
& 0.8-1.1& 7602&  9.29& $11.03^{+0.05}_{-0.05}$ & $1.33^{+0.59}_{-0.59}$ & $-1.25^{+0.05}_{-0.05}$ &$-99.00^{+ 0.00}_{- 0.00}$ &$-99.00^{+ 0.00}_{- 0.00}$ &$ 8.24^{+0.03}_{-0.03}$\\ 
& 1.1-1.5& 7099&  9.61& $10.97^{+0.05}_{-0.05}$ & $0.94^{+0.59}_{-0.59}$ & $-1.20^{+0.06}_{-0.06}$ &$-99.00^{+ 0.00}_{- 0.00}$ &$-99.00^{+ 0.00}_{- 0.00}$ &$ 8.01^{+0.03}_{-0.03}$\\ 
& 1.5-2& 6802& 10.02& $10.89^{+0.06}_{-0.06}$ & $1.22^{+0.59}_{-0.59}$ & $-0.88^{+0.13}_{-0.13}$ &$-99.00^{+ 0.00}_{- 0.00}$ &$-99.00^{+ 0.00}_{- 0.00}$ &$ 7.95^{+0.03}_{-0.03}$\\ 
& 2-2.5& 4172& 10.21& $11.05^{+0.16}_{-0.16}$ & $0.41^{+0.59}_{-0.59}$ & $-1.19^{+0.26}_{-0.26}$ &$-99.00^{+ 0.00}_{- 0.00}$ &$-99.00^{+ 0.00}_{- 0.00}$ &$ 7.72^{+0.03}_{-0.03}$\\ 
& 2.5-3& 2147& 10.36& $10.90^{+0.18}_{-0.18}$ & $0.34^{+0.59}_{-0.59}$ & $-0.74^{+0.58}_{-0.58}$ &$-99.00^{+ 0.00}_{- 0.00}$ &$-99.00^{+ 0.00}_{- 0.00}$ &$ 7.39^{+0.05}_{-0.05}$\\ 
\hline
SPH  & 0.2-0.5& 0620&  8.43& $10.59^{+0.26}_{-0.26}$ & $0.75^{+0.24}_{-0.24}$ & $-0.25^{+0.45}_{-0.45}$ &$ 0.07^{+ 0.03}_{- 0.03}$ &$ -1.60^{+ 0.00}_{- 0.00}$ &$ 7.51^{+0.11}_{-0.11}$\\ 
& 0.5-0.8& 1385&  8.94& $10.89^{+0.10}_{-0.10}$ & $0.91^{+0.18}_{-0.18}$ & $-0.51^{+0.21}_{-0.21}$ &$ 0.06^{+ 0.02}_{- 0.02}$ &$ -1.60^{+ 0.00}_{- 0.00}$ &$ 7.85^{+0.05}_{-0.05}$\\ 
& 0.8-1.1& 1344&  9.29& $11.01^{+0.08}_{-0.08}$ & $0.48^{+0.09}_{-0.09}$ & $-1.00^{+0.07}_{-0.07}$ &$-99.00^{+ 0.00}_{- 0.00}$ &$-99.00^{+ 0.00}_{- 0.00}$ &$ 7.69^{+0.07}_{-0.07}$\\ 
& 1.1-1.5& 1260&  9.61& $10.80^{+0.08}_{-0.08}$ & $0.46^{+0.08}_{-0.08}$ & $-0.78^{+0.11}_{-0.11}$ &$-99.00^{+ 0.00}_{- 0.00}$ &$-99.00^{+ 0.00}_{- 0.00}$ &$ 7.42^{+0.05}_{-0.05}$\\ 
& 1.5-2& 1206& 10.02& $10.52^{+0.08}_{-0.08}$ & $0.59^{+0.04}_{-0.04}$ & $ 0.24^{+0.27}_{-0.27}$ &$-99.00^{+ 0.00}_{- 0.00}$ &$-99.00^{+ 0.00}_{- 0.00}$ &$ 7.34^{+0.05}_{-0.05}$\\ 
& 2-2.5& 0692& 10.21& $10.76^{+0.17}_{-0.17}$ & $0.23^{+0.05}_{-0.05}$ & $-0.26^{+0.46}_{-0.46}$ &$-99.00^{+ 0.00}_{- 0.00}$ &$-99.00^{+ 0.00}_{- 0.00}$ &$ 7.09^{+0.07}_{-0.07}$\\ 
& 2.5-3& 0291& 10.36& $10.45^{+0.19}_{-0.19}$ & $0.09^{+0.04}_{-0.04}$ & $ 0.74^{+0.93}_{-0.93}$ &$-99.00^{+ 0.00}_{- 0.00}$ &$-99.00^{+ 0.00}_{- 0.00}$ &$ 6.62^{+0.10}_{-0.10}$\\
\hline
DISK+SPH& 0.2-0.5& 0394&  8.35& $10.31^{+0.10}_{-0.10}$ & $0.93^{+0.77}_{-0.77}$ & $ 0.00^{+0.00}_{-0.00}$ &$ 0.64^{+ 0.53}_{- 0.53}$ &$ -0.78^{+ 0.20}_{- 0.20}$ &$ 7.49^{+0.08}_{-0.08}$\\ 
& 0.5-0.8& 0727&  8.86& $10.63^{+0.05}_{-0.05}$ & $1.27^{+0.15}_{-0.15}$ & $-0.49^{+0.07}_{-0.07}$ &$-99.00^{+ 0.00}_{- 0.00}$ &$-99.00^{+ 0.00}_{- 0.00}$ &$ 7.69^{+0.06}_{-0.06}$\\ 
& 0.8-1.1& 0568&  9.21& $10.73^{+0.06}_{-0.06}$ & $0.67^{+0.09}_{-0.09}$ & $-0.47^{+0.10}_{-0.10}$ &$-99.00^{+ 0.00}_{- 0.00}$ &$-99.00^{+ 0.00}_{- 0.00}$ &$ 7.50^{+0.07}_{-0.07}$\\ 
& 1.1-1.5& 0394&  9.53& $10.77^{+0.08}_{-0.08}$ & $0.29^{+0.04}_{-0.04}$ & $-0.35^{+0.14}_{-0.14}$ &$-99.00^{+ 0.00}_{- 0.00}$ &$-99.00^{+ 0.00}_{- 0.00}$ &$ 7.18^{+0.08}_{-0.08}$\\ 
& 1.5-2& 0271&  9.94& $10.77^{+0.12}_{-0.12}$ & $0.16^{+0.02}_{-0.02}$ & $ 0.12^{+0.34}_{-0.34}$ &$-99.00^{+ 0.00}_{- 0.00}$ &$-99.00^{+ 0.00}_{- 0.00}$ &$ 7.00^{+0.09}_{-0.09}$\\ 
& 2-2.5& 0092& 10.13& $10.73^{+0.26}_{-0.26}$ & $0.04^{+0.01}_{-0.01}$ & $ 0.16^{+0.91}_{-0.91}$ &$-99.00^{+ 0.00}_{- 0.00}$ &$-99.00^{+ 0.00}_{- 0.00}$ &$ 6.34^{+0.14}_{-0.14}$\\ 
& 2.5-3& 0035& 10.28& $99.00^{+0.00}_{-0.00}$ & $99.00^{+0.00}_{-0.00}$ & $99.00^{+0.00}_{0.00}$ &$-99.00^{+ 0.00}_{- 0.00}$ &$-99.00^{+ 0.00}_{- 0.00}$ &$ 99.00^{+0.00}_{-0.00}$\\ 
\hline
DISKS & 0.2-0.5& 1430&  8.35& $10.25^{+1.97}_{-1.97}$ & $0.09^{+6.56}_{-6.56}$ & $ 0.00^{+0.00}_{-0.00}$ &$ 1.48^{+ 1.03}_{- 1.03}$ &$ -1.12^{+ 0.05}_{- 0.05}$ &$ 7.47^{+0.05}_{-0.05}$\\ 
& 0.5-0.8& 2322&  8.86& $10.56^{+0.06}_{-0.06}$ & $1.04^{+0.17}_{-0.17}$ & $-1.15^{+0.05}_{-0.05}$ &$-99.00^{+ 0.00}_{- 0.00}$ &$-99.00^{+ 0.00}_{- 0.00}$ &$ 7.62^{+0.04}_{-0.04}$\\ 
& 0.8-1.1& 1826&  9.21& $10.80^{+0.08}_{-0.08}$ & $0.51^{+0.11}_{-0.11}$ & $-1.20^{+0.07}_{-0.07}$ &$-99.00^{+ 0.00}_{- 0.00}$ &$-99.00^{+ 0.00}_{- 0.00}$ &$ 7.56^{+0.05}_{-0.05}$\\ 
& 1.1-1.5& 1167&  9.53& $10.98^{+0.10}_{-0.10}$ & $0.15^{+0.04}_{-0.04}$ & $-1.27^{+0.09}_{-0.09}$ &$-99.00^{+ 0.00}_{- 0.00}$ &$-99.00^{+ 0.00}_{- 0.00}$ &$ 7.24^{+0.06}_{-0.06}$\\ 
& 1.5-2& 0780&  9.94& $10.85^{+0.16}_{-0.16}$ & $0.13^{+0.06}_{-0.06}$ & $-0.97^{+0.28}_{-0.28}$ &$-99.00^{+ 0.00}_{- 0.00}$ &$-99.00^{+ 0.00}_{- 0.00}$ &$ 6.96^{+0.07}_{-0.07}$\\ 
& 2-2.5& 0309& 10.13& $11.11^{+0.29}_{-0.29}$ & $0.03^{+0.02}_{-0.02}$ & $-0.97^{+0.55}_{-0.55}$ &$-99.00^{+ 0.00}_{- 0.00}$ &$-99.00^{+ 0.00}_{- 0.00}$ &$ 6.55^{+0.15}_{-0.15}$\\ 
& 2.5-3& 0104& 10.28& $11.42^{+0.85}_{-0.85}$ & $0.01^{+0.01}_{-0.01}$ & $-1.01^{+0.80}_{-0.80}$ &$-99.00^{+ 0.00}_{- 0.00}$ &$-99.00^{+ 0.00}_{- 0.00}$ &$ 6.20^{+0.05}_{-0.05}$\\ 
\hline
IRREGULARS & 0.2-0.5& 2673&  8.35& $ 9.87^{+0.43}_{-0.43}$ & $0.16^{+0.70}_{-0.70}$ & $ 0.00^{+0.00}_{-0.00}$ &$ 0.76^{+ 0.51}_{- 0.51}$ &$ -1.60^{+ 0.00}_{- 0.00}$ &$ 7.05^{+0.05}_{-0.05}$\\ 
& 0.5-0.8& 3593&  8.86& $10.45^{+0.83}_{-0.83}$ & $0.04^{+0.43}_{-0.43}$ & $ 0.00^{+0.00}_{-0.00}$ &$ 0.33^{+ 0.60}_{- 0.60}$ &$ -1.66^{+ 0.16}_{- 0.16}$ &$ 7.34^{+0.04}_{-0.04}$\\ 
& 0.8-1.1& 3613&  9.21& $10.72^{+0.11}_{-0.11}$ & $0.26^{+0.70}_{-0.70}$ & $-1.62^{+0.09}_{-0.09}$ &$-99.00^{+ 0.51}_{- 0.51}$ &$-99.00^{+ 0.00}_{- 0.00}$ &$ 7.45^{+0.04}_{-0.04}$\\ 
& 1.1-1.5& 3941&  9.53& $10.91^{+0.13}_{-0.13}$ & $0.17^{+0.70}_{-0.70}$ & $-1.60^{+0.09}_{-0.09}$ &$-99.00^{+ 0.51}_{- 0.51}$ &$-99.00^{+ 0.00}_{- 0.00}$ &$ 7.45^{+0.04}_{-0.04}$\\ 
& 1.5-2& 4210&  9.94& $10.91^{+0.11}_{-0.11}$ & $0.28^{+0.70}_{-0.70}$ & $-1.39^{+0.17}_{-0.17}$ &$-99.00^{+ 0.51}_{- 0.51}$ &$-99.00^{+ 0.00}_{- 0.00}$ &$ 7.52^{+0.03}_{-0.03}$\\ 
& 2-2.5& 2819& 10.13& $10.86^{+0.08}_{-0.08}$ & $0.23^{+0.70}_{-0.70}$ & $-1.39^{+0.00}_{-0.00}$ &$-99.00^{+ 0.51}_{- 0.51}$ &$-99.00^{+ 0.00}_{- 0.00}$ &$ 7.37^{+0.04}_{-0.04}$\\ 
& 2.5-3& 1560& 10.28& $11.09^{+0.32}_{-0.32}$ & $0.09^{+0.70}_{-0.70}$ & $-1.35^{+0.46}_{-0.46}$ &$-99.00^{+ 0.51}_{- 0.51}$ &$-99.00^{+ 0.00}_{- 0.00}$ &$ 7.18^{+0.05}_{-0.05}$\\ 

\end{tabular}

\caption{Best-fit parameters with single and double schechter functions to the stellar mass functions of the four morphological types defined in this work. The parameters of the double Schechter are set to -99 whenever a single Schechter was used. Values of -99 are also used when the fit did not converge. }
\label{tbl:fits_all}
\end{table*}

The key new ingredient of the present work is the evolution at fixed morphology. Morphological evolution, as well as the mass dependence of the dominant morphology, are both clearly observed. At $0.2<z<0.5$, the population of $\sim M^*$  galaxies ($10<log(M_*/M_\odot)<11$) is essentially uniformly distributed between disks with low bulge fractions, spheroids with large B/Ts and intermediate objects with 2 components meaning that here is no clear dominant morphology at this mass scale. Above $log(M_*/M_\odot)=11$), objects with a clear bulge component tend to dominate the population. Below $10^{10}M_\odot$, the population is basically dominated by objects with small bulges or without.  Irregular objects only start dominating the population at  $log(M_*/M_\odot) < 9$.  This morphological distribution remains globally unchanged from  $z\sim1$. 

Our low mass SMFs match the local SMFs recently derived in the GAMA survey (Moffet et al. 2016) quite well, as shown in figure~\ref{fig:MFs_z}.  We over-estimate the abundance of irregulars at $log(M_*/M_\odot)>10$ compared to them. This is probably a consequence of our definition of irregulars based on the asymmetry of the light profile. It has however little impact on the other morphologies since their abundance is still very low at the high mass end. The agreement with their work confirms the robustness of our automated classifications.

Above $z\sim1$ irregular objects start dominating even at higher-masses.  At $z>2$, the morphological mix changes radically:  there are basically only 2 types of galaxies at these redshifts -- irregulars account for 70\% of the objects, and bulge dominated galaxies (spheroids) for the remaining $30\%$ (based on the extrapolations of the Schechter fits). This has a number of interesting implications.  First, at these early epochs, the majority of disks are irregular (probably a signature of unstable disks as probed by recent IFU surveys, e.g.~\citealp{2015ApJ...799..209W}). Note that this is not a morphological k-correction effect, since we are probing the optical rest-frame band at this epoch. Second, symmetric disks and bulge+disks systems only begin to appear between $z\sim 2$ and $z\sim 1$; objects classified as DISK+SPH account for fewer than 5\% of the objects at $z>2$. This is also observed in the top right panel of figure~\ref{fig:MFs_z}. Disks and disk+spheroid mass functions experience the most dramatic evolution. One might worry that the apparent disappearance of $z\ge 2$ disks is due to surface brightness dimming. This is unlikely though for several reasons. Extensive simulations (e.g. Kartaltepe et al. 2015, van der Wel et al. 2014) have shown that disks should be detectable at the depth of the CANDELS survey for the magnitude selection used in this work.  Also, these are fairly massive galaxies so there is not much room for the presence of a massive disk. In fact, preliminary results of figure~\ref{fig:BT_visual} show a clear correlation between the morphological classification and the stellar mass bulge-to-disk ratio which would have been erased if surface brightness dimming was an issue.

These global trends are captured in the top left panel of figure~\ref{fig:mass_density} which summarizes the evolution of the stellar mass density (integrated over all galaxies with $log(M_*/M_\odot)\ge 8$). Since this lower limit lies below the completeness limit, the result relies on extrapolating the best Schechter fits to low masses. We first observe  the previously reported 2-speed growth of the mass density on the full sample (black line) in good agreement with previous measurements.  From $z\sim 4$ to 1, the total mass density increases by a factor of $\sim 6 $.  From $z\sim 1$ onwards the growth flattens: $\rho_*$ at $z=0$ is larger by only a factor of $\sim 2$.  As we discuss in the following sections, this is a consequence of both the decrease in the specific star-formation rate below $z\sim 1$  (e.g. \citealp{2016arXiv160104226S}) and of quenching at large stellar masses. 

Regarding the morphological evolution above $10^8$ solar masses (resulting from the extrapolation of the best Schechter fits), the key observed trends observed in figure~\ref{fig:mass_density} are
\begin{itemize}
\item At $z>2$, more than 70\% of the stellar mass density is in irregular galaxies (see also~\citealp{2005ApJ...620..564C}). The stellar mass density in irregulars decreases over time from $Log(\rho_*/M_\odot Mpc^{-3})\sim 7.7$ at $z\sim1.5$ to $\sim 7.1$ at $z\sim 0.3$. This is clear evidence of morphological transformations as we will discuss in the following sections.
\item At $z>2$, 30\% of the stellar mass density is in compact spheroids with large B/T. This suggests that bulge growth at this epoch destroys disks.
\item The emergence of \emph{regular disks} (S0a-Sbc) happens between $z\sim2$ and $z\sim1$. In this period, the stellar density in both pure disks and bulge+disk systems increases by a factor of $\sim 30$.
\item Below $z\sim 1$, the stellar mass density is equally distributed among disks, spheroids and mixed systems.
\end{itemize}


\begin{figure*}
\begin{center}
$\begin{array}{c c c}
\includegraphics[width=0.33\textwidth]{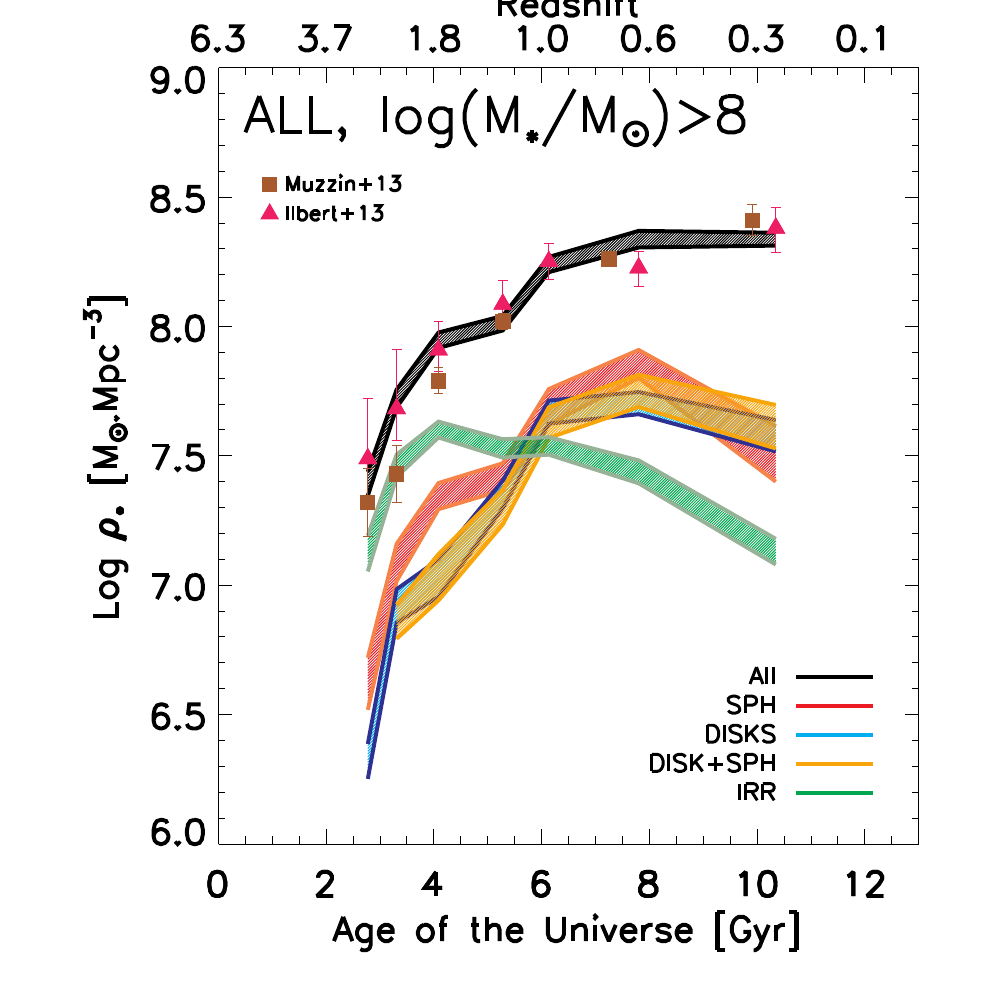}  & \includegraphics[width=0.33\textwidth]{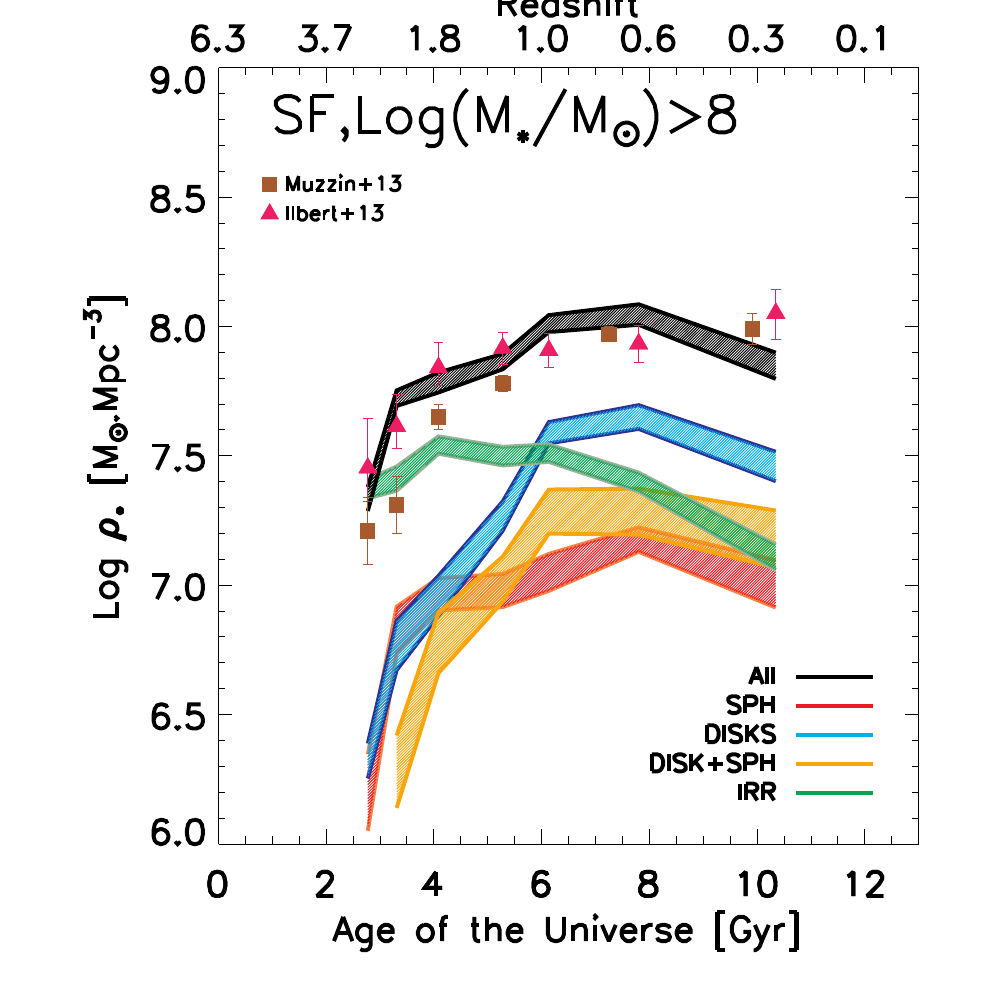} &  \includegraphics[width=0.33\textwidth]{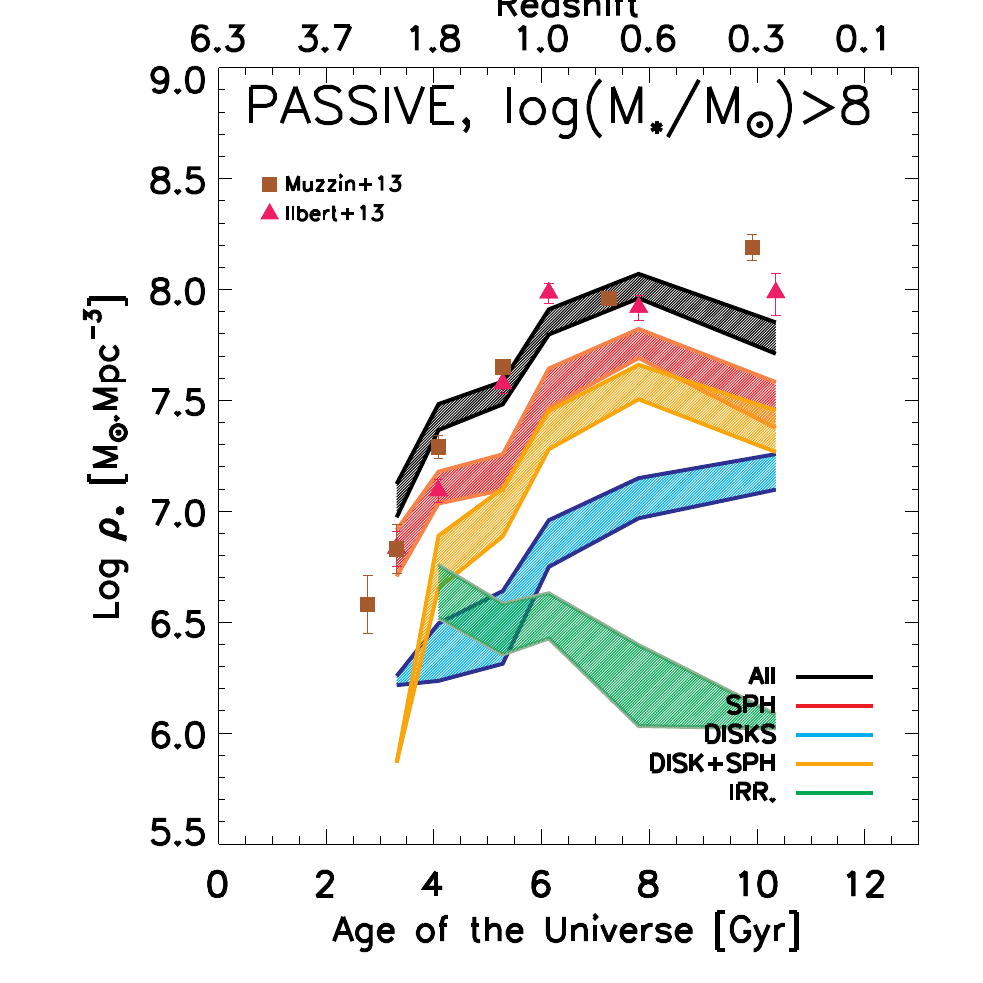}\\
\includegraphics[width=0.33\textwidth]{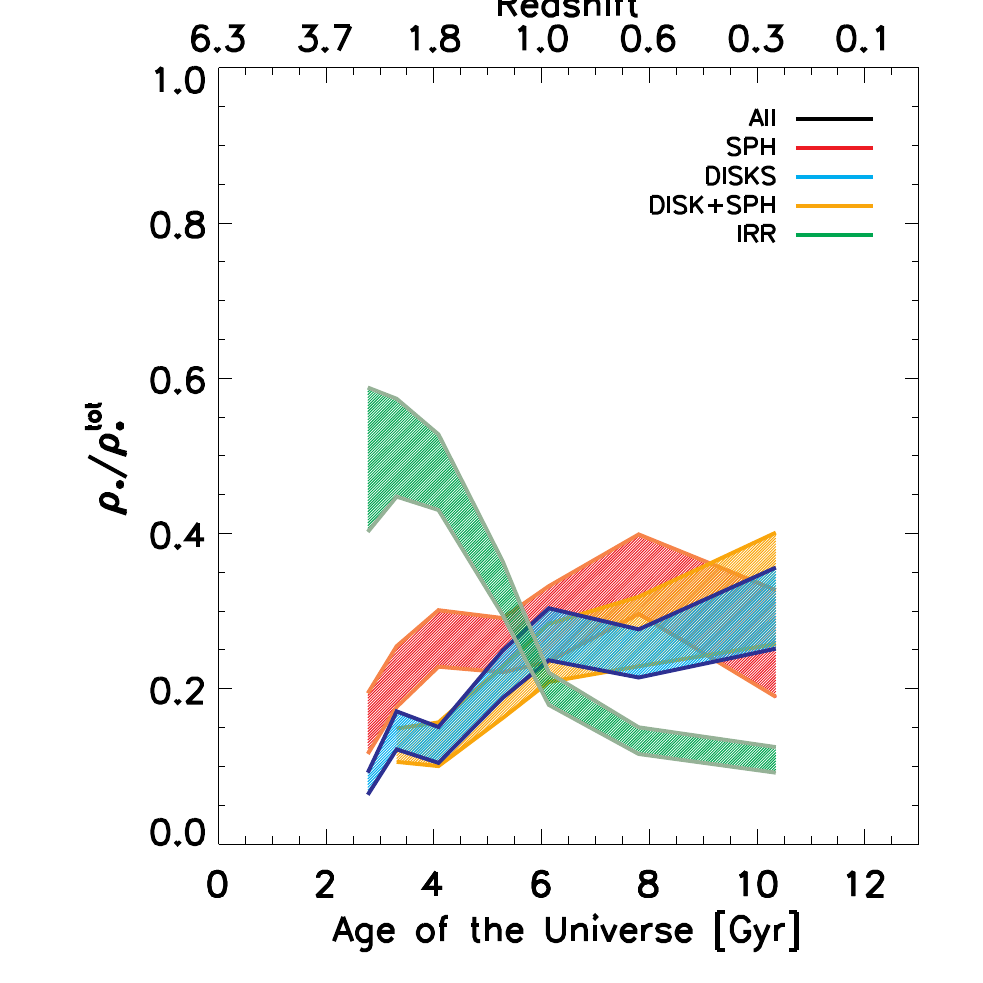}  & \includegraphics[width=0.33\textwidth]{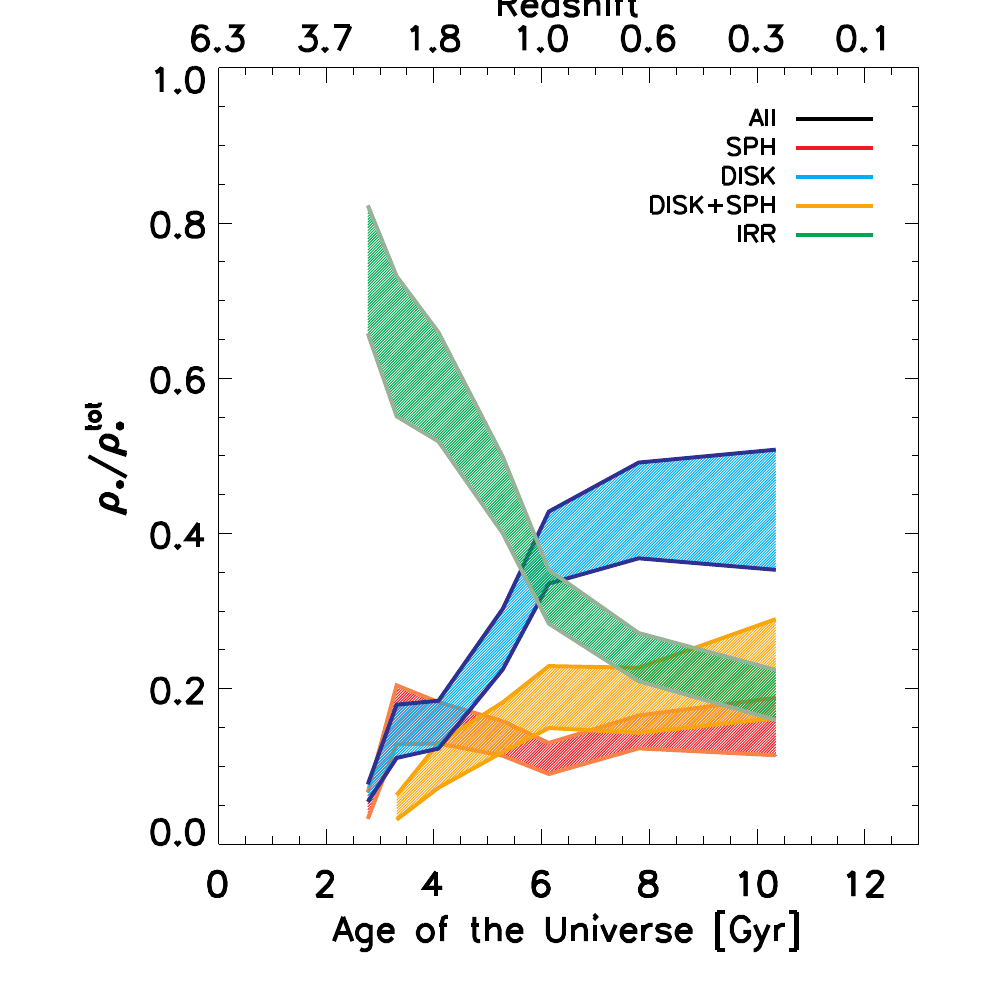} &  \includegraphics[width=0.33\textwidth]{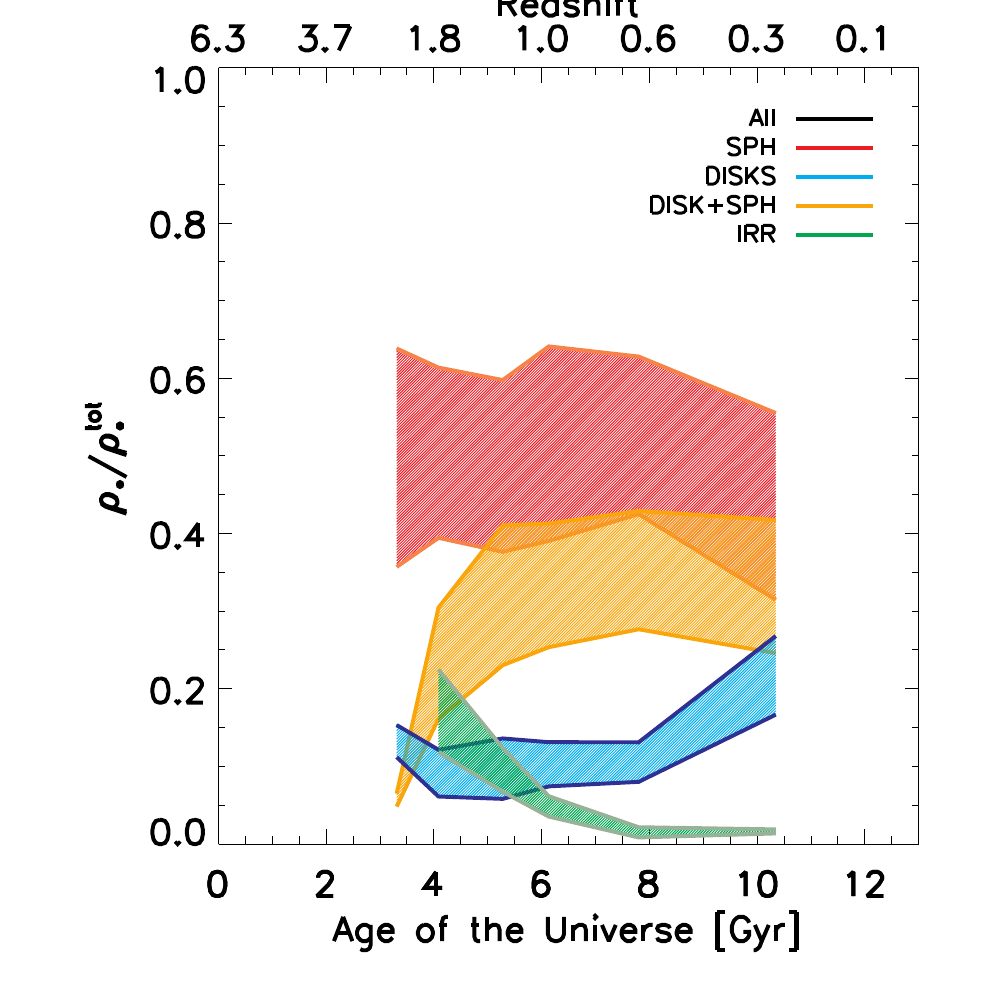}\\

\
\end{array}$
\caption{Evolution of the stellar mass density for galaxies with $log(M_*/M_\odot) > 10^8$. The left panels show the full sample. The middle and right panels show star-forming and quiescent galaxies respectively. The different colors correspond different morphologies as labelled. Bottom line are fractions. The pink triangles and brown squares are measurements from Ilbert et al. (2013), Muzzin et al. (2013). } 
\label{fig:mass_density}
\end{center}
\end{figure*}

\subsection{Evolution of the star-forming population}

Figure~\ref{fig:MFs_SF} shows the evolution of the stellar mass functions of star-forming galaxies as a function of morphological type.  To guide the eye, the cyan region (same in all panels) shows the $z\sim 0.1$ SDSS determination as a reference.  This curve was obtained by following the analysis of Bernardi et al. (2016), but selecting the subset of objects for which the log of the specific star formation rate determined by the MPA-JHU (e.g. Kauffmann et al. 2003) group is greater than $-11$~dex.

Table~\ref{tbl:fits_SF} summarizes the best fit Schechter function parameters for our CANDELS analysis.  In agreement with previous work, the SMF of all star-forming galaxies increases steeply at the low-mass end, and evolves very little at the high-mass end.  This is a consequence of quenching:  when star-forming galaxies exceed a critical mass, they quench and so are removed from the SF sample (e.g. Ilbert et al. 2013, Peng et al. 2010).  

\begin{figure*}
\begin{center}
$\begin{array}{c c c}
\includegraphics[width=0.33\textwidth]{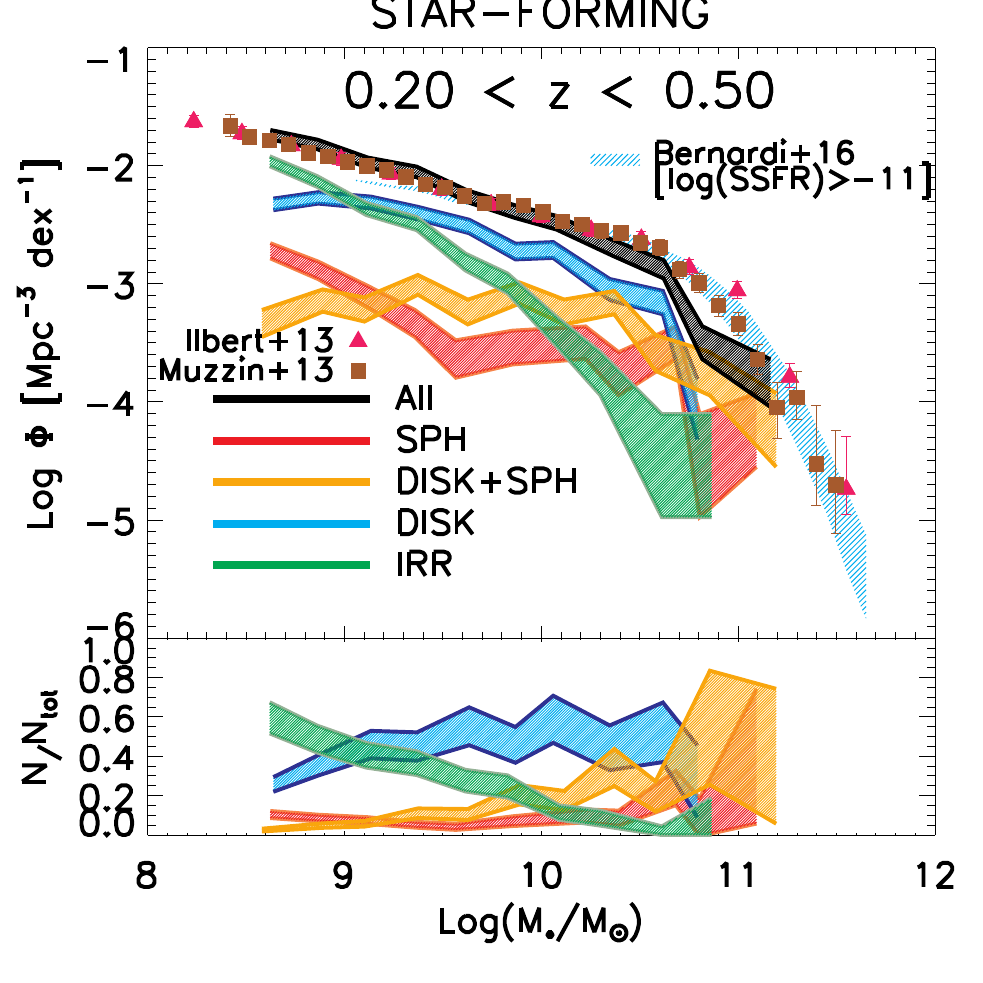} & \includegraphics[width=0.33\textwidth]{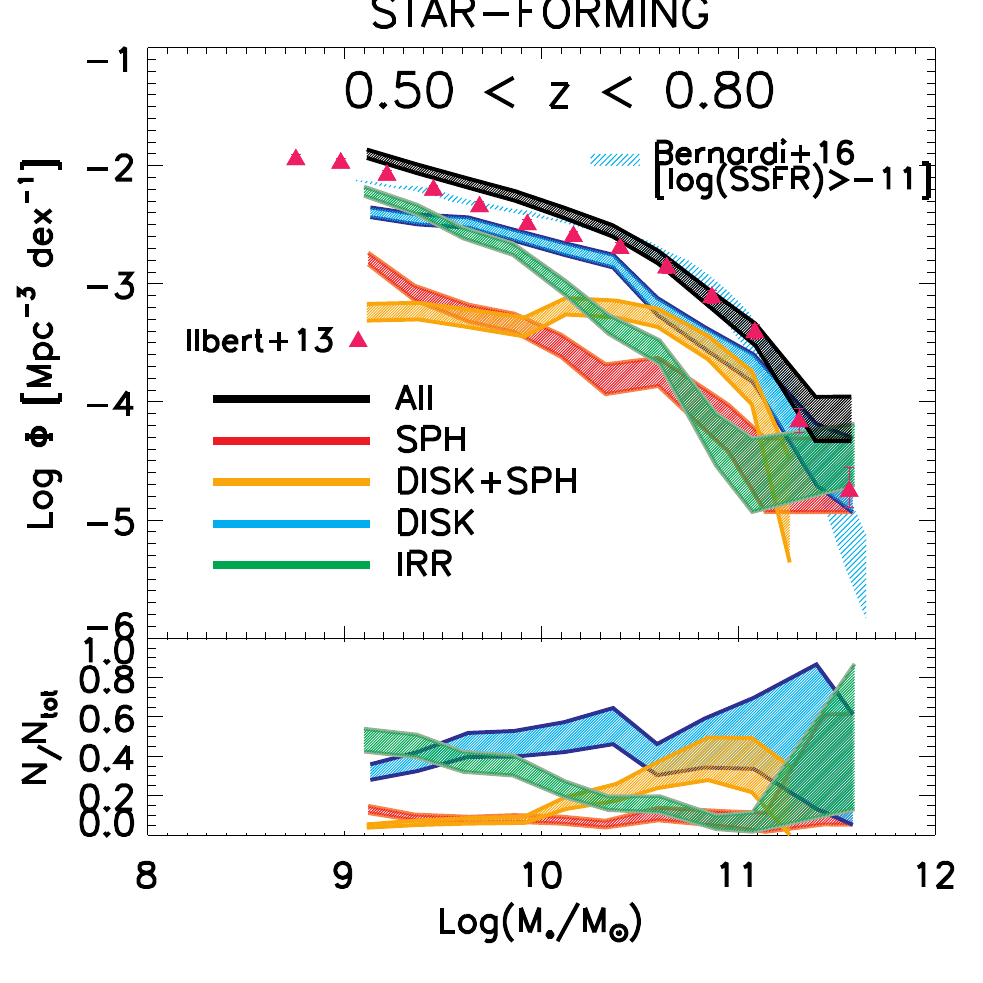} &  \includegraphics[width=0.33\textwidth]{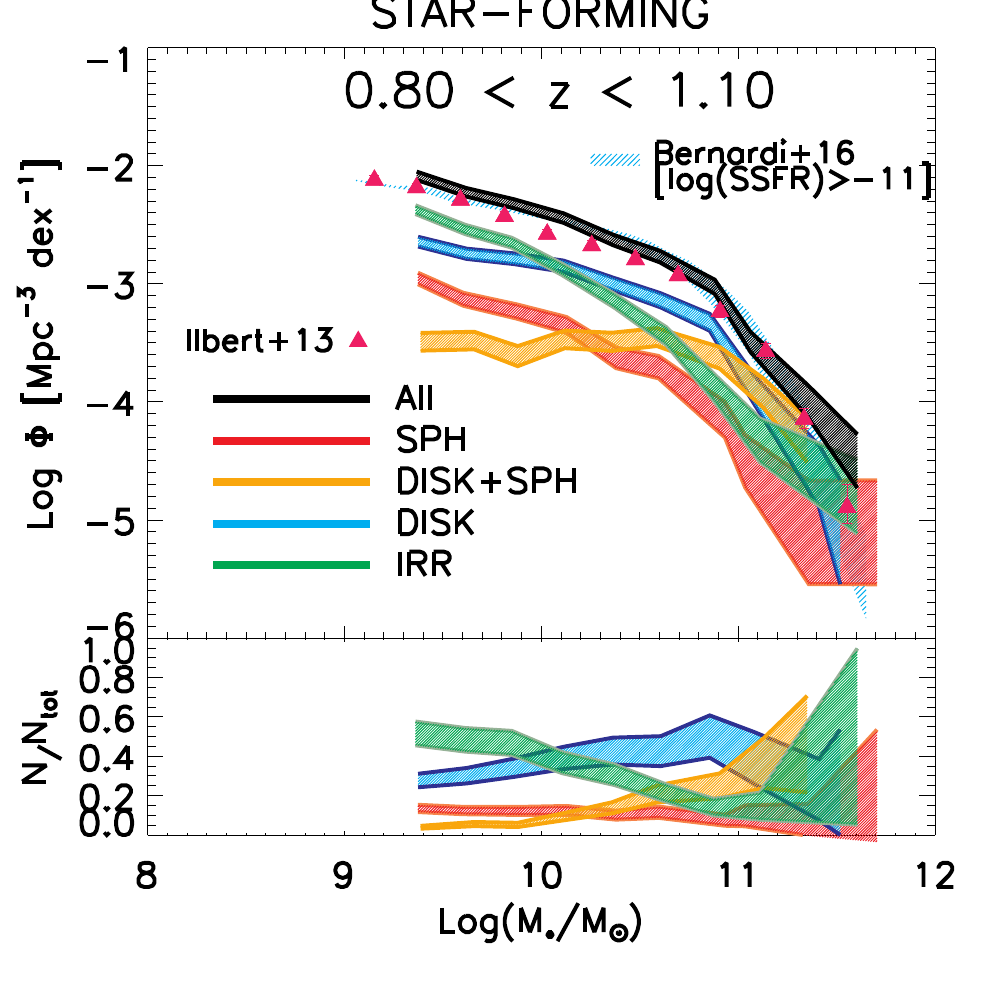} \\
\includegraphics[width=0.33\textwidth]{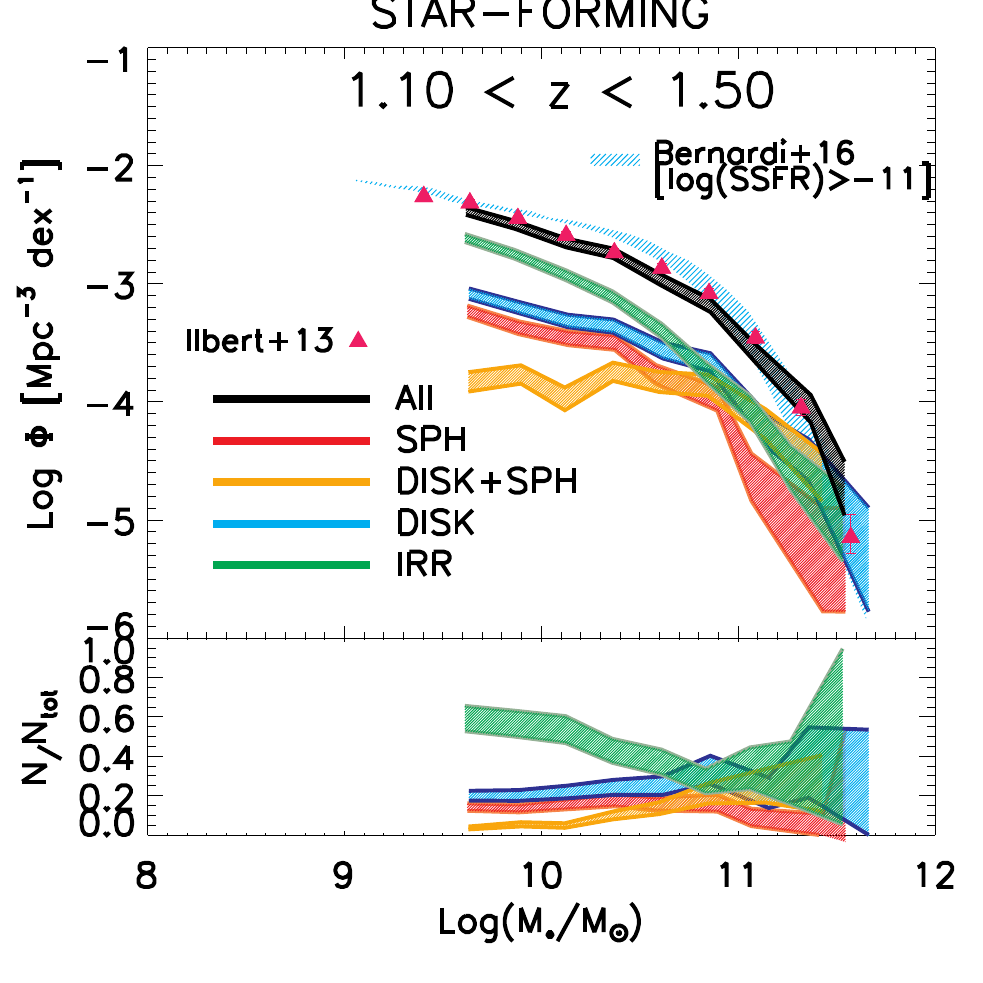} & \includegraphics[width=0.33\textwidth]{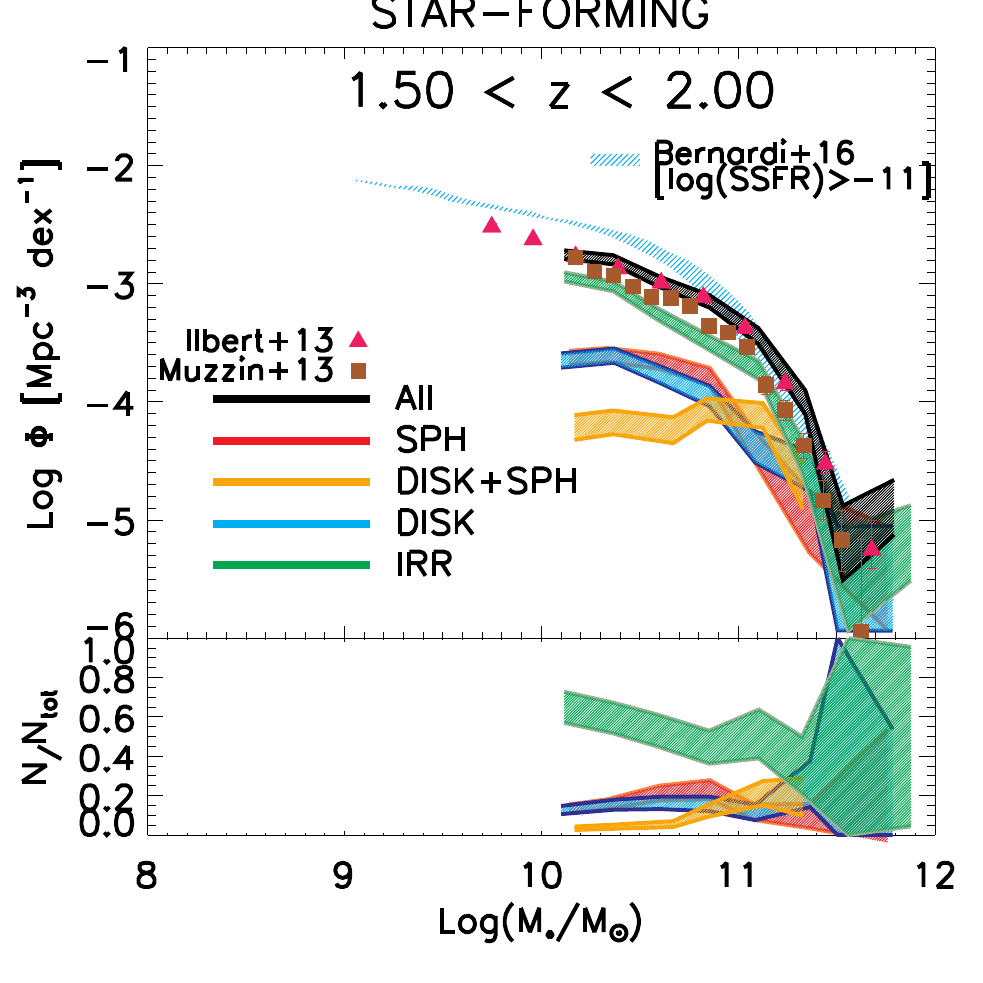} &  \includegraphics[width=0.33\textwidth]{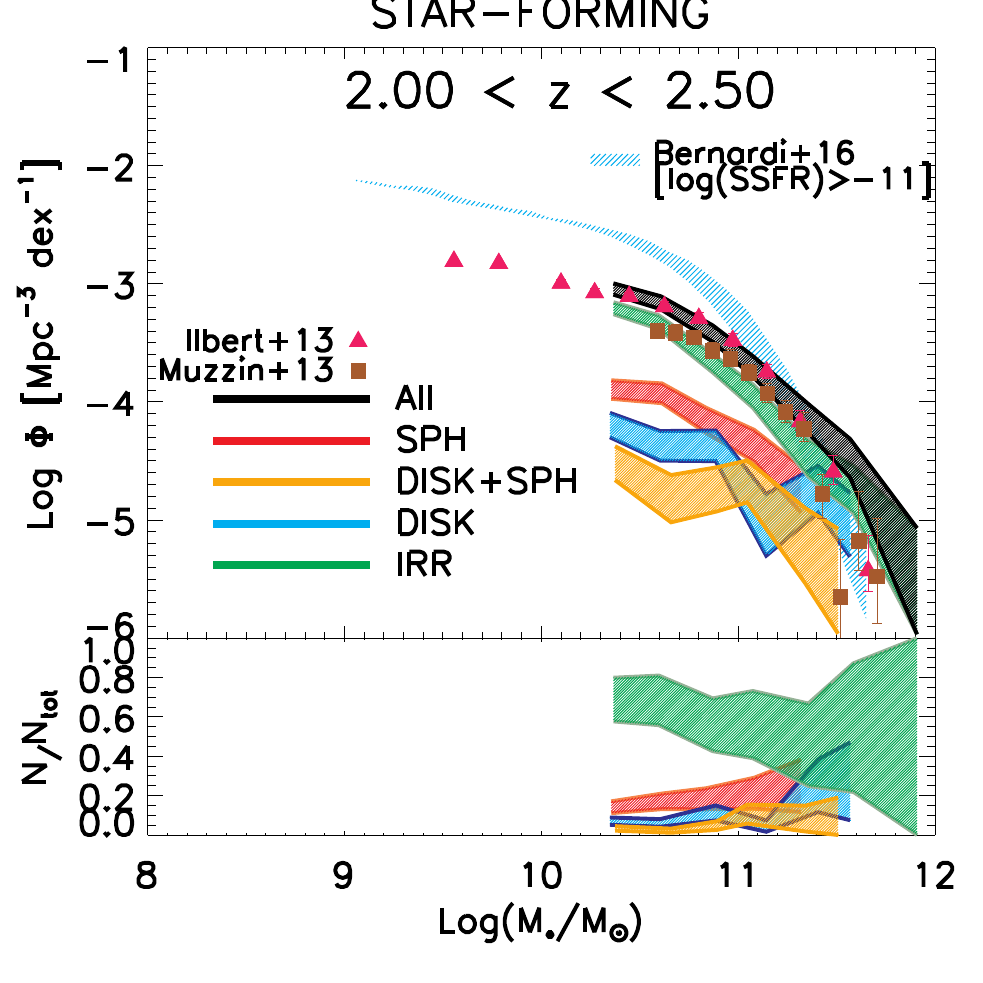} \\
\includegraphics[width=0.33\textwidth]{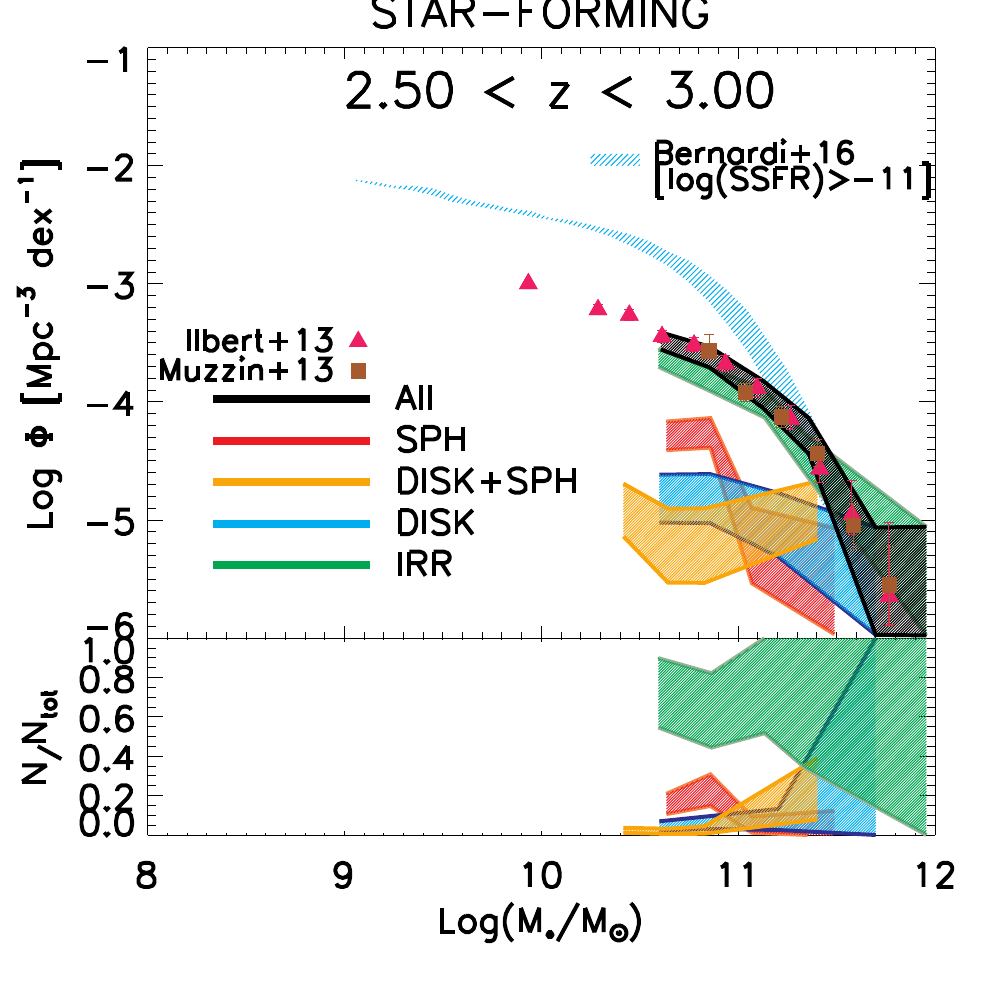}  & & \\
\end{array}$
\caption{Stellar-mass functions of star-forming galaxies divided in 4 morphological types and in different redshift bins as labelled. Red, blue, orange and green shaded regions in the top panels show the number densities of spheroids, disks, disk+spheroids and irregular/clumpy systems respectively. The black regions show the global mass functions. The pink triangles are the values measured by Ilbert et al. (2013) and the brown squares the values of Muzzin et al. (2013) in the same redshift bins. The bottom panels show the fractions of each morphological type with the same color code. The cyan shaded region shows the SMF for the SDSS star-forming galaxies from Bernardi et al. (2016) ($log(SSFR)>-11$).} 
\label{fig:MFs_SF}
\end{center}
\end{figure*}

Our new results show that the morphological mix of star-forming galaxies also experiences a pronounced evolution. At $0.2<z<0.5$, the typical morphology of a star-forming galaxy differs significantly from that in the full sample. Purely bulge dominated systems (spheroids) account for $\le 5\%$ of the objects at all stellar masses. Star-forming galaxies at low redshifts are therefore dominated by regular systems with no pronounced asymmetries and with low bulge fractions (i.e. disks) over 2 decades in stellar mass ($9<log(M_*/M_\odot)<11$).  Irregular disks start to dominate only at very low-masses ($log(M_*/M_\odot)\le 9$). Bulge+disk systems are also a minority, but account for $\sim 40\%$ of the population at stellar masses greater than $log(M_*/M_\odot)\sim 10.7$.   The presence of the bulge component is therefore tightly linked to the star-formation activity of the galaxy as widely documented in the recent literature (e.g.~\citealp{2012ApJ...753..114W}). As observed for the full sample, this morphological mix seems to have remained rather stable since $z\sim 1$. 

At higher redshifts, the relative abundance of irregulars and normal disks is inverted:  disturbed systems become the dominant morphological class of star-forming galaxies. The relative abundance steadily increases from $z\sim 1$ to $z\sim 2$.  At $z>2$, irregular systems are almost 100\% of the star-forming population. While we confirm a population of star-forming spheroids at $z>2$ (e.g. \citealp{2015ApJ...813...23V}), they account for only $\sim5-10\%$  of the star-forming population at these redshifts. This strongly suggests that bulge formation at these early epochs requires rapid consumption of gas and therefore the quenching of star-formation.

These trends are summarized in the middle panel of figure~\ref{fig:mass_density} which shows the evolution of the integrated stellar mass density of star-forming galaxies with $M_*/M_\odot>10^8$. The main observed features are:

\begin{itemize}

\item Stars are formed in systems with a disk. The abundance of star-forming spheroids is below $10\%$.

\item There is a transition of the galaxy morphology which hosts star-formation. At $z<1-1.5$ most of the stars in star-forming systems are in regular disks with low ($\sim50\%$) and intermediate ($\sim20\%$) B/T while at $z>1.5$ they are predominantly in irregular systems ($>80\%$). 

\item The stellar mass density in irregular galaxies decreases with redshift (by a factor of $\sim 3$); therefore, irregulars are being transformed into other morphologies.  
\end{itemize}

\begin{table*}
\resizebox{\textwidth}{!}{\begin{tabular}{ccccccccccccccccc}
\hline
\hline
 \multirow{3}{*}{S} & \multirow{3}{*}{z} &  \multicolumn{6}{c}{STAR-FORMING} & &  \multicolumn{6}{c}{QUIESCENT} \\
   \cmidrule{3-8}  \cmidrule{10-15}
   \multicolumn{2}{c}{} &
 \multicolumn{1}{c}{$N$} &
 \multicolumn{1}{c}{\small{$L(M_{c})$}} &
 \multicolumn{1}{c}{$L(M^*$)} &
 \multicolumn{1}{c}{$\Phi_1^*$} &
 \multicolumn{1}{c}{$\alpha_1$} &
  \multicolumn{1}{c}{L($\rho_*$)} &
    \multicolumn{1}{c}{} &
  \multicolumn{1}{c}{$N$} &
 \multicolumn{1}{c}{$L(M_{c})$} &
 \multicolumn{1}{c}{$L(M^*$)} &
 \multicolumn{1}{c}{$\Phi_1^*$} &
 \multicolumn{1}{c}{$\alpha_1$} &
\multicolumn{1}{c}{$L(\rho_*)$}  \\
 
  \multicolumn{1}{c}{} &
 \multicolumn{1}{c}{} &
 \multicolumn{1}{c}{} &
 \multicolumn{1}{c}{\tiny{$M_\odot$}} & 
     \multicolumn{1}{c}{\tiny{$M_\odot$}} & 
    \multicolumn{1}{c}{\tiny{$10^{-3}$ Mpc$^{-3}$}} &
     \multicolumn{1}{c}{} &
       \multicolumn{1}{c}{\tiny{$M_\odot.Mpc^{-3}$}} &
         \multicolumn{1}{c}{} &
      \multicolumn{1}{c}{} &
 \multicolumn{1}{c}{\tiny{$M_\odot$}} & 
     \multicolumn{1}{c}{\tiny{$M_\odot$}} & 
    \multicolumn{1}{c}{\tiny{$10^{-3}$ Mpc$^{-3}$}} &
     \multicolumn{1}{c}{}   &
     \multicolumn{1}{c}{\tiny{$M_\odot.Mpc^{-3}$}} \\

\hline
A & 0.2-0.5& 4465&  8.43& $10.68^{+0.09}_{-0.09}$ & $0.92^{+0.22}_{-0.22}$ & $-1.46^{+0.04}_{-0.04}$ &$ 7.85^{+0.05}_{-0.05}$& &0768&  8.60& $11.01^{+0.14}_{-0.14}$ & $0.56^{+0.14}_{-0.14}$ & $-1.09^{+0.06}_{-0.06}$ &$ 7.78^{+0.07}_{-0.07}$\\ 
& 0.5-0.8& 7032&  8.94& $10.92^{+0.07}_{-0.07}$ & $0.83^{+0.17}_{-0.17}$ & $-1.46^{+0.04}_{-0.04}$ &$ 8.05^{+0.04}_{-0.04}$& &1244&  9.04& $10.90^{+0.04}_{-0.04}$ & $1.47^{+0.16}_{-0.16}$ & $-0.68^{+0.05}_{-0.05}$ &$ 8.02^{+0.06}_{-0.06}$\\ 
& 0.8-1.1& 6743&  9.29& $10.93^{+0.06}_{-0.06}$ & $0.81^{+0.17}_{-0.17}$ & $-1.41^{+0.05}_{-0.05}$ &$ 8.01^{+0.03}_{-0.03}$& &0859&  9.48& $10.84^{+0.04}_{-0.04}$ & $1.17^{+0.11}_{-0.11}$ & $-0.40^{+0.08}_{-0.08}$ &$ 7.85^{+0.06}_{-0.06}$\\ 
& 1.1-1.5& 6534&  9.61& $10.98^{+0.06}_{-0.06}$ & $0.51^{+0.10}_{-0.10}$ & $-1.41^{+0.06}_{-0.06}$ &$ 7.86^{+0.03}_{-0.03}$& &0565&  9.79& $10.73^{+0.02}_{-0.02}$ & $0.64^{+0.04}_{-0.04}$ & $ 0.00^{+0.00}_{-0.00}$ &$ 7.53^{+0.05}_{-0.05}$\\ 
& 1.5-2& 6261& 10.02& $10.82^{+0.05}_{-0.05}$ & $0.93^{+0.15}_{-0.15}$ & $-1.00^{+0.12}_{-0.12}$ &$ 7.78^{+0.04}_{-0.04}$& &0541& 10.16& $10.78^{+0.03}_{-0.03}$ & $0.44^{+0.03}_{-0.03}$ & $ 0.00^{+0.00}_{-0.00}$ &$ 7.42^{+0.06}_{-0.06}$\\ 
& 2-2.5& 3961& 10.21& $11.29^{+0.16}_{-0.16}$ & $0.13^{+0.08}_{-0.08}$ & $-1.61^{+0.19}_{-0.19}$ &$ 7.72^{+0.03}_{-0.03}$& &0211& 10.51& $10.85^{+0.27}_{-0.27}$ & $0.18^{+0.08}_{-0.08}$ & $-0.56^{+0.86}_{-0.86}$ &$ 7.05^{+0.08}_{-0.08}$\\ 
& 2.5-3& 2057& 10.36& $11.02^{+0.17}_{-0.17}$ & $0.19^{+0.11}_{-0.11}$ & $-1.08^{+0.46}_{-0.46}$ &$ 7.33^{+0.05}_{-0.05}$& &0090& 11.05& $99.99^{+99.99}_{-99.99}$ & $99.99^{+99.99}_{-9.99}$ & $99.99^{+99.99}_{99.99}$ &$99.00^{+99.99}_{-99.99}$\\
\hline
S & 0.2-0.5& 0418&  8.43& $15.58^{+99}_{-99}$ & $0.00^{+0.24}_{-0.24}$ & $-1.71^{+0.10}_{-0.10}$ &$ 7.00^{+0.09}_{-0.09}$& &0202&  8.60& $11.04^{+0.22}_{-0.22}$ & $0.28^{+0.09}_{-0.09}$ & $-0.96^{+0.08}_{-0.08}$ &$ 7.48^{+0.10}_{-0.10}$\\ 
& 0.5-0.8& 0798&  8.94& $14.37^{+99}_{-99}$ & $0.00^{+0.02}_{-0.02}$ & $-1.77^{+0.08}_{-0.08}$ &$ 7.18^{+0.05}_{-0.05}$& &0586&  9.04& $10.88^{+0.05}_{-0.05}$ & $0.85^{+0.10}_{-0.10}$ & $-0.61^{+0.06}_{-0.06}$ &$ 7.76^{+0.07}_{-0.07}$\\ 
& 0.8-1.1& 0872&  9.29& $10.86^{+0.14}_{-0.14}$ & $0.10^{+0.04}_{-0.04}$ & $-1.46^{+0.10}_{-0.10}$ &$ 7.05^{+0.07}_{-0.07}$& &0470&  9.48& $10.81^{+0.06}_{-0.06}$ & $0.62^{+0.07}_{-0.07}$ & $-0.41^{+0.10}_{-0.10}$ &$ 7.56^{+0.09}_{-0.09}$\\ 
& 1.1-1.5& 0928&  9.61& $10.72^{+0.11}_{-0.11}$ & $0.16^{+0.05}_{-0.05}$ & $-1.20^{+0.13}_{-0.13}$ &$ 6.98^{+0.06}_{-0.06}$& &0331&  9.79& $10.52^{+0.08}_{-0.08}$ & $0.39^{+0.03}_{-0.03}$ & $ 0.29^{+0.22}_{-0.22}$ &$ 7.18^{+0.08}_{-0.08}$\\ 
& 1.5-2& 0889& 10.02& $10.58^{+0.10}_{-0.10}$ & $0.27^{+0.05}_{-0.05}$ & $-0.38^{+0.30}_{-0.30}$ &$ 6.97^{+0.06}_{-0.06}$& &0316& 10.16& $10.35^{+0.07}_{-0.07}$ & $0.14^{+0.06}_{-0.06}$ & $ 1.68^{+0.48}_{-0.48}$ &$ 7.11^{+0.07}_{-0.07}$\\ 
& 2-2.5& 0568& 10.21& $11.00^{+0.29}_{-0.29}$ & $0.06^{+0.05}_{-0.05}$ & $-1.08^{+0.46}_{-0.46}$ &$ 6.83^{+0.09}_{-0.09}$& &0120& 10.51& $10.72^{+0.23}_{-0.23}$ & $0.13^{+0.02}_{-0.02}$ & $-0.14^{+1.02}_{-1.02}$ &$ 6.81^{+0.10}_{-0.10}$\\ 
& 2.5-3& 0248& 10.36& $15.58^{+99}_{-99}$ & $0.00^{+0.24}_{-0.24}$ & $-1.71^{+0.10}_{-0.10}$ &$ 6.20^{+0.15}_{-0.15}$& &0039& 11.05& $99.99^{+99.99}_{-99.99}$ & $99.99^{+99.99}_{-9.99}$ & $99.99^{+99.99}_{99.99}$ &$99.00^{+99.99}_{-99.99}$\\
\hline
DS  & 0.2-0.5& 0233&  8.43& $10.53^{+0.12}_{-0.12}$ & $0.48^{+0.11}_{-0.11}$ & $-0.85^{+0.09}_{-0.09}$ &$ 7.18^{+0.11}_{-0.11}$& &0161&  8.60& $10.42^{+0.07}_{-0.07}$ & $0.94^{+0.13}_{-0.13}$ & $-0.17^{+0.13}_{-0.13}$ &$ 7.36^{+0.09}_{-0.09}$\\ 
& 0.5-0.8& 0408&  8.94& $10.66^{+0.06}_{-0.06}$ & $0.46^{+0.08}_{-0.08}$ & $-0.78^{+0.08}_{-0.08}$ &$ 7.28^{+0.09}_{-0.09}$& &0319&  9.04& $10.63^{+0.05}_{-0.05}$ & $0.89^{+0.08}_{-0.08}$ & $-0.01^{+0.11}_{-0.11}$ &$ 7.58^{+0.08}_{-0.08}$\\ 
& 0.8-1.1& 0360&  9.29& $10.94^{+0.10}_{-0.10}$ & $0.24^{+0.05}_{-0.05}$ & $-0.82^{+0.10}_{-0.10}$ &$ 7.28^{+0.09}_{-0.09}$& &0208&  9.48& $10.67^{+0.06}_{-0.06}$ & $0.43^{+0.04}_{-0.04}$ & $ 0.29^{+0.18}_{-0.18}$ &$ 7.37^{+0.09}_{-0.09}$\\ 
& 1.1-1.5& 0281&  9.61& $11.02^{+0.12}_{-0.12}$ & $0.11^{+0.03}_{-0.03}$ & $-0.82^{+0.13}_{-0.13}$ &$ 7.02^{+0.09}_{-0.09}$& &0113&  9.79& $10.57^{+0.08}_{-0.08}$ & $0.10^{+0.02}_{-0.02}$ & $ 1.29^{+0.36}_{-0.36}$ &$ 7.00^{+0.11}_{-0.11}$\\ 
& 1.5-2& 0196& 10.02& $10.91^{+0.16}_{-0.16}$ & $0.08^{+0.02}_{-0.02}$ & $-0.38^{+0.34}_{-0.34}$ &$ 6.78^{+0.12}_{-0.12}$& &0074& 10.16& $10.65^{+0.09}_{-0.09}$ & $0.03^{+0.02}_{-0.02}$ & $ 1.64^{+0.62}_{-0.62}$ &$ 6.77^{+0.12}_{-0.12}$\\ 
& 2-2.5& 0081& 10.21& $11.21^{+0.48}_{-0.48}$ & $0.01^{+0.01}_{-0.01}$ & $-1.06^{+0.69}_{-0.69}$ &$ 6.28^{+0.14}_{-0.14}$& &0010& 10.51& $10.56^{+0.72}_{-0.72}$ & $0.00^{+0.02}_{-0.02}$ & $ 1.94^{+4.85}_{-4.85}$ &$ 5.87^{+0.00}_{-0.00}$\\ 
& 2.5-3& 0028& 10.36& $10.53^{+0.12}_{-0.12}$ & $0.48^{+0.11}_{-0.11}$ & $-0.85^{+0.09}_{-0.09}$ &$ 6.66^{+0.06}_{-0.06}$& &0007& 11.05& $99.99^{+99.99}_{-99.99}$ & $99.99^{+99.99}_{-9.99}$ & $99.99^{+99.99}_{99.99}$ &$99.00^{+99.99}_{-99.99}$\\ 
\hline
D  & 0.2-0.5& 1263&  8.43& $10.33^{+0.07}_{-0.07}$ & $1.21^{+0.23}_{-0.23}$ & $-1.17^{+0.05}_{-0.05}$ &$ 7.46^{+0.06}_{-0.06}$& &0202&  8.60& $16.02^{+99}_{-99}$ & $0.00^{+3.45}_{-3.45}$ & $-1.55^{+0.11}_{-0.11}$ &$ 7.18^{+0.08}_{-0.08}$\\ 
& 0.5-0.8& 2162&  8.94& $10.66^{+0.08}_{-0.08}$ & $0.80^{+0.18}_{-0.18}$ & $-1.24^{+0.06}_{-0.06}$ &$ 7.65^{+0.05}_{-0.05}$& &0179&  9.04& $15.81^{+99}_{-99}$ & $0.00^{+1.22}_{-1.22}$ & $-1.54^{+0.14}_{-0.14}$ &$ 7.06^{+0.09}_{-0.09}$\\ 
& 0.8-1.1& 1752&  9.29& $10.80^{+0.07}_{-0.07}$ & $0.54^{+0.11}_{-0.11}$ & $-1.18^{+0.07}_{-0.07}$ &$ 7.59^{+0.04}_{-0.04}$& &0084&  9.48& $11.52^{+0.58}_{-0.58}$ & $0.02^{+0.02}_{-0.02}$ & $-1.17^{+0.22}_{-0.22}$ &$ 6.85^{+0.11}_{-0.11}$\\ 
& 1.1-1.5& 1126&  9.61& $11.01^{+0.10}_{-0.10}$ & $0.14^{+0.04}_{-0.04}$ & $-1.30^{+0.09}_{-0.09}$ &$ 7.27^{+0.06}_{-0.06}$& &0044&  9.79& $11.15^{+0.26}_{-0.26}$ & $0.02^{+0.01}_{-0.01}$ & $-0.60^{+0.38}_{-0.38}$ &$ 6.48^{+0.16}_{-0.16}$\\ 
& 1.5-2& 0748& 10.02& $10.84^{+0.12}_{-0.12}$ & $0.13^{+0.04}_{-0.04}$ & $-0.96^{+0.21}_{-0.21}$ &$ 6.96^{+0.08}_{-0.08}$& &0039& 10.16& $10.79^{+0.43}_{-0.43}$ & $0.04^{+0.01}_{-0.01}$ & $-0.11^{+1.01}_{-1.01}$ &$ 6.36^{+0.13}_{-0.13}$\\ 
& 2-2.5& 0299& 10.21& $15.81^{+99}_{-99}$ & $0.00^{+0.05}_{-0.05}$ & $-1.76^{+0.43}_{-0.43}$ &$ 6.77^{+0.09}_{-0.09}$& &0014& 10.51& $11.25^{+0.46}_{-0.46}$ & $0.01^{+0.00}_{-0.00}$ & $ 0.07^{+1.28}_{-1.28}$ &$ 6.24^{+0.02}_{-0.02}$\\ 
& 2.5-3& 0103& 10.36& $10.33^{+0.07}_{-0.07}$ & $1.21^{+0.23}_{-0.23}$ & $-1.17^{+0.05}_{-0.05}$ &$ 6.32^{+0.07}_{-0.07}$& &0002& 11.05& $99.99^{+99.99}_{-99.99}$ & $99.99^{+99.99}_{-9.99}$ & $99.99^{+99.99}_{99.99}$ &$99.00^{+99.99}_{-99.99}$\\
\hline
I & 0.2-0.5& 2472&  8.43& $10.15^{+0.15}_{-0.15}$ & $0.40^{+0.19}_{-0.19}$ & $-1.71^{+0.09}_{-0.09}$ &$ 7.11^{+0.05}_{-0.05}$& &0192&  8.60& $14.48^{+99.9}_{-99.9}$ & $0.00^{+0.00}_{-0.00}$ & $-2.07^{+0.19}_{-0.19}$ &$ 6.05^{+0.03}_{-0.03}$\\ 
& 0.5-0.8& 3460&  8.94& $10.56^{+0.11}_{-0.11}$ & $0.31^{+0.11}_{-0.11}$ & $-1.66^{+0.07}_{-0.07}$ &$ 7.40^{+0.04}_{-0.04}$& &0130&  9.04& $16.17^{+99.9}_{-99.9}$ & $0.00^{+0.00}_{-0.00}$ & $-1.92^{+0.16}_{-0.16}$ &$ 6.21^{+0.18}_{-0.18}$\\ 
& 0.8-1.1& 3532&  9.29& $10.72^{+0.10}_{-0.10}$ & $0.33^{+0.11}_{-0.11}$ & $-1.57^{+0.08}_{-0.08}$ &$ 7.51^{+0.03}_{-0.03}$& &0081&  9.48& $18.87^{+99.9}_{-99.9}$ & $0.00^{+0.00}_{-0.00}$ & $-1.57^{+0.21}_{-0.21}$ &$ 6.53^{+0.10}_{-0.10}$\\ 
& 1.1-1.5& 3887&  9.61& $10.86^{+0.09}_{-0.09}$ & $0.23^{+0.07}_{-0.07}$ & $-1.57^{+0.08}_{-0.08}$ &$ 7.50^{+0.04}_{-0.04}$& &0048&  9.79& $20.15^{+99.9}_{-99.9}$ & $0.00^{+0.00}_{-0.00}$ & $-1.45^{+0.19}_{-0.19}$ &$ 6.47^{+0.11}_{-0.11}$\\ 
& 1.5-2& 4132& 10.02& $10.88^{+0.08}_{-0.08}$ & $0.37^{+0.10}_{-0.10}$ & $-1.28^{+0.14}_{-0.14}$ &$ 7.54^{+0.03}_{-0.03}$& &0065& 10.16& $19.21^{+99.9}_{-99.9}$ & $0.00^{+0.00}_{-0.00}$ & $-1.59^{+0.19}_{-0.19}$ &$ 6.64^{+0.12}_{-0.12}$\\ 
& 2-2.5& 2776& 10.21& $10.93^{+0.06}_{-0.06}$ & $0.24^{+0.04}_{-0.04}$ & $-1.28^{+0.00}_{-0.00}$ &$ 7.41^{+0.05}_{-0.05}$& &0038& 10.51& $17.85^{+99.9}_{-99.9}$ & $0.00^{+0.00}_{-0.00}$ & $-1.93^{+0.39}_{-0.39}$ &$ 6.61^{+0.04}_{-0.04}$\\ 
& 2.5-3& 1530& 10.36& $11.53^{+0.37}_{-0.37}$ & $0.03^{+0.04}_{-0.04}$ & $-1.67^{+0.35}_{-0.35}$ &$ 7.37^{+0.04}_{-0.04}$& &0023& 11.05& $99.99^{+99.99}_{-99.99}$ & $99.99^{+99.99}_{-9.99}$ & $99.99^{+99.99}_{99.99}$ &$99.00^{+99.99}_{-99.99}$\\
\end{tabular}}

\caption{Best-fit parameters for a single Schechter function to the star-forming and quiescent SMFs of the four morphological types defined in this work. A=all. S=spheroids, D=disks, DS=disks+spheroids and I=irregulars. Quiescent galaxies at $z>2.5$ are not fitted because there are too few values above completeness. Quiescent irregulars are also very few. Although the fit works, the mass function is not always well described by a Schechter function. To emphasize this we have set the error on $M^*$ to 99.9. }
\label{tbl:fits_SF}
\end{table*}

\subsection{Evolution of the quiescent population}
\label{sec:passive}

Figure~\ref{fig:MFs_Q} shows the evolution of the SMFs of quiescent galaxies as a function of morphological type. We also show in all panels the SMF of quiescent galaxies ($log(SSFR)<-11$) in the SDSS based on the recent measurements of Bernardi et al. (2016). The quiescent SMF, summed over all morphological types, agrees with the one measured in Ilbert et al. (2013). There are some discrepancies, especially in the  $0.5<z<0.8$ bin, which can be a consequence of both cosmic variance and of a difference in the colors used to select passive galaxies. In any case, the broad evolution trends remain the same. Quiescent galaxies first appear at the high-mass end. The quiescent SMF increases rapidly at the high-mass end up to $z\sim 1$. From $z\sim 1$ to the present, the low mass population of passive galaxies starts to emerge. 

\begin{figure*}
\begin{center}
$\begin{array}{c c c}
\includegraphics[width=0.33\textwidth]{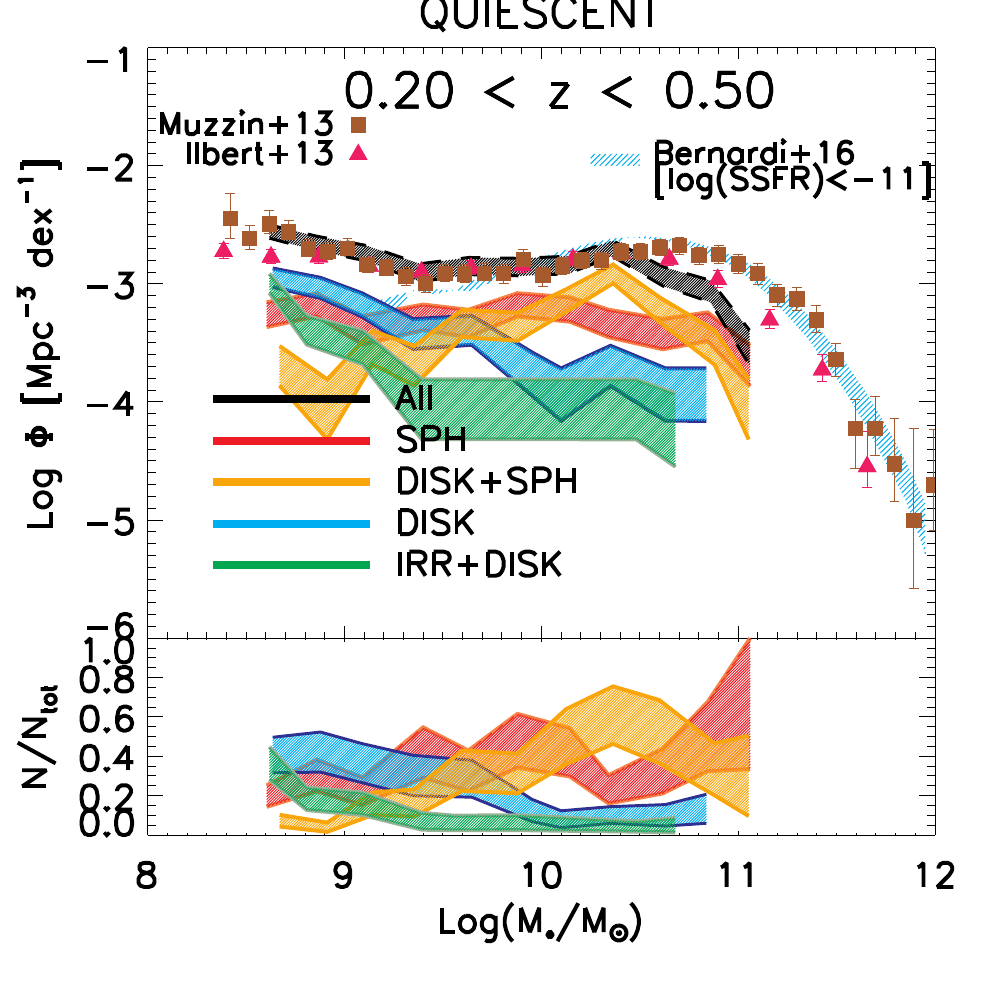} & \includegraphics[width=0.33\textwidth]{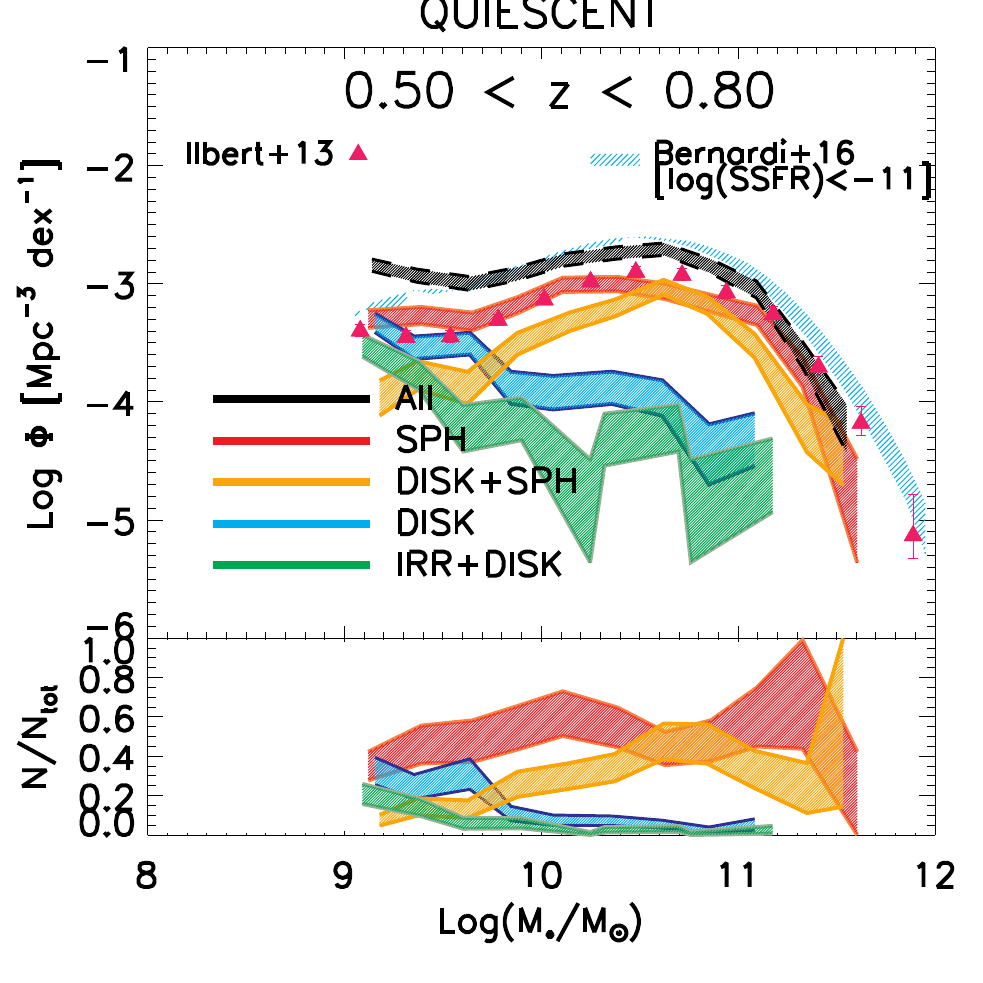} &  \includegraphics[width=0.33\textwidth]{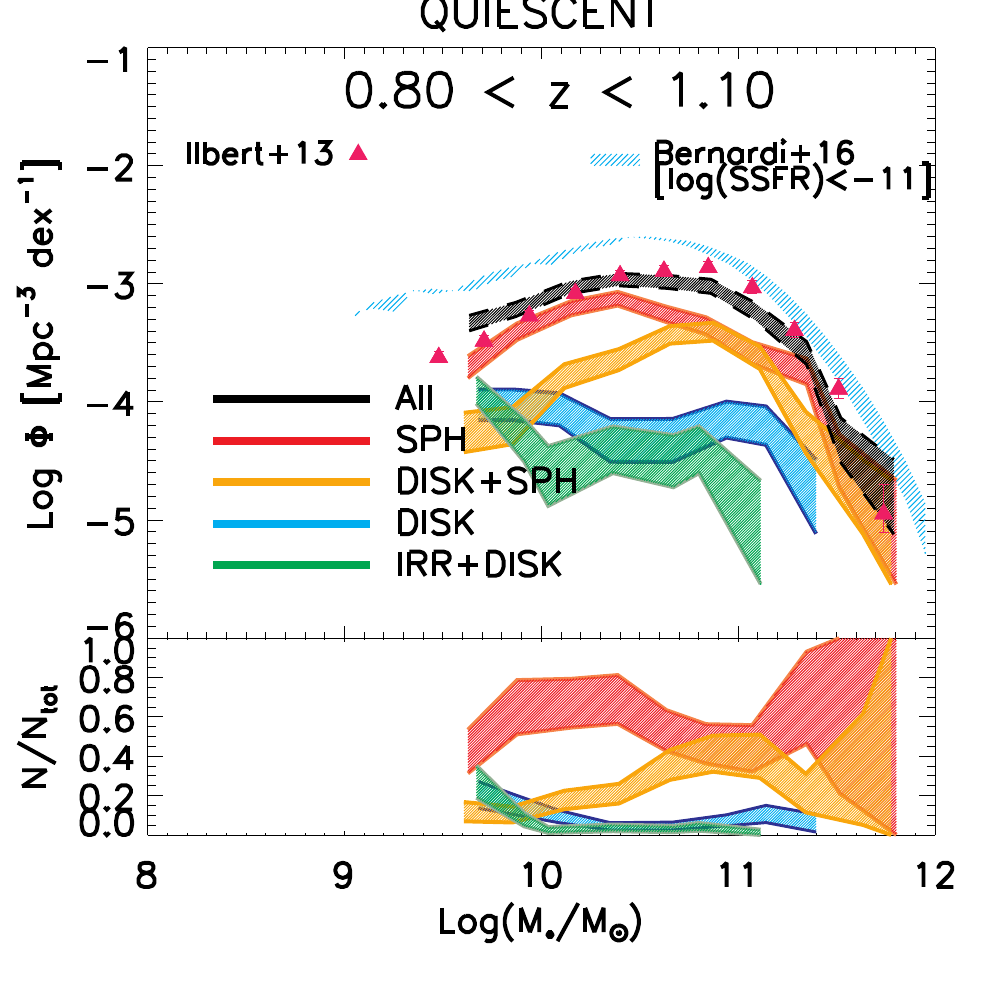} \\
\includegraphics[width=0.33\textwidth]{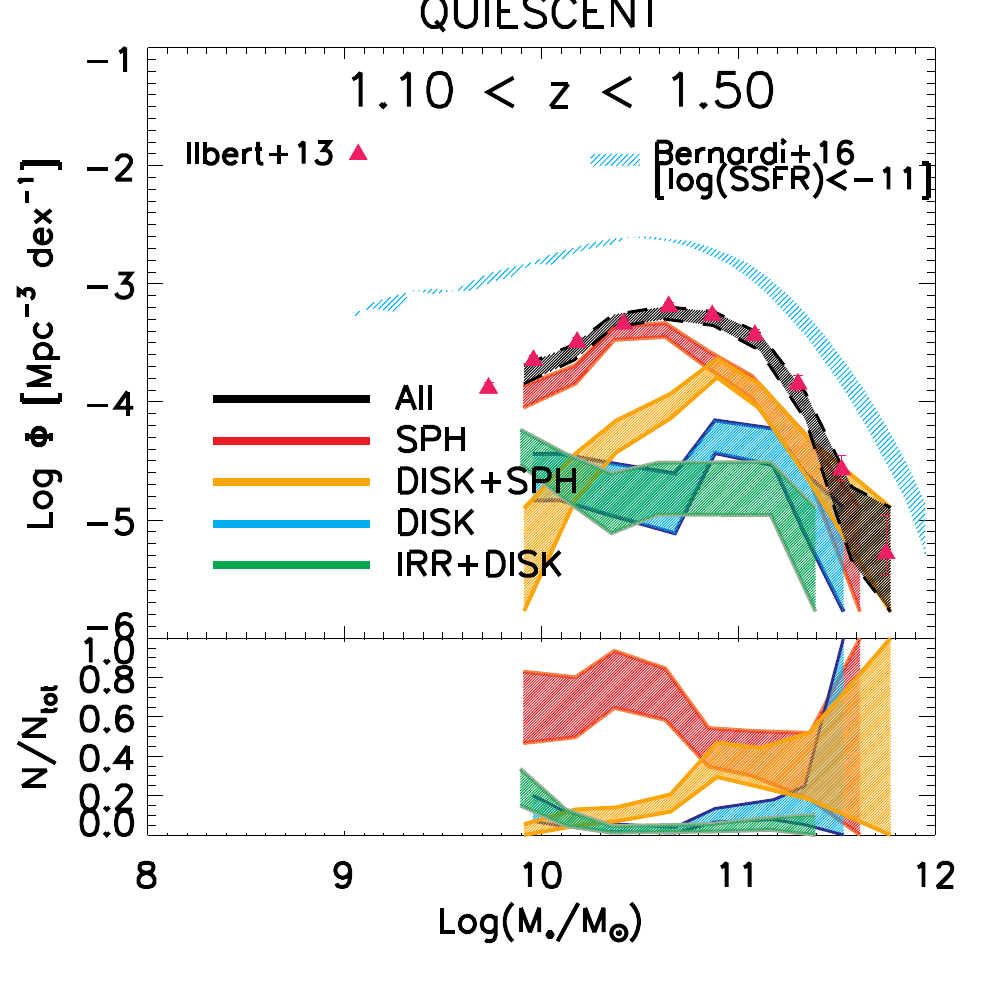} & \includegraphics[width=0.33\textwidth]{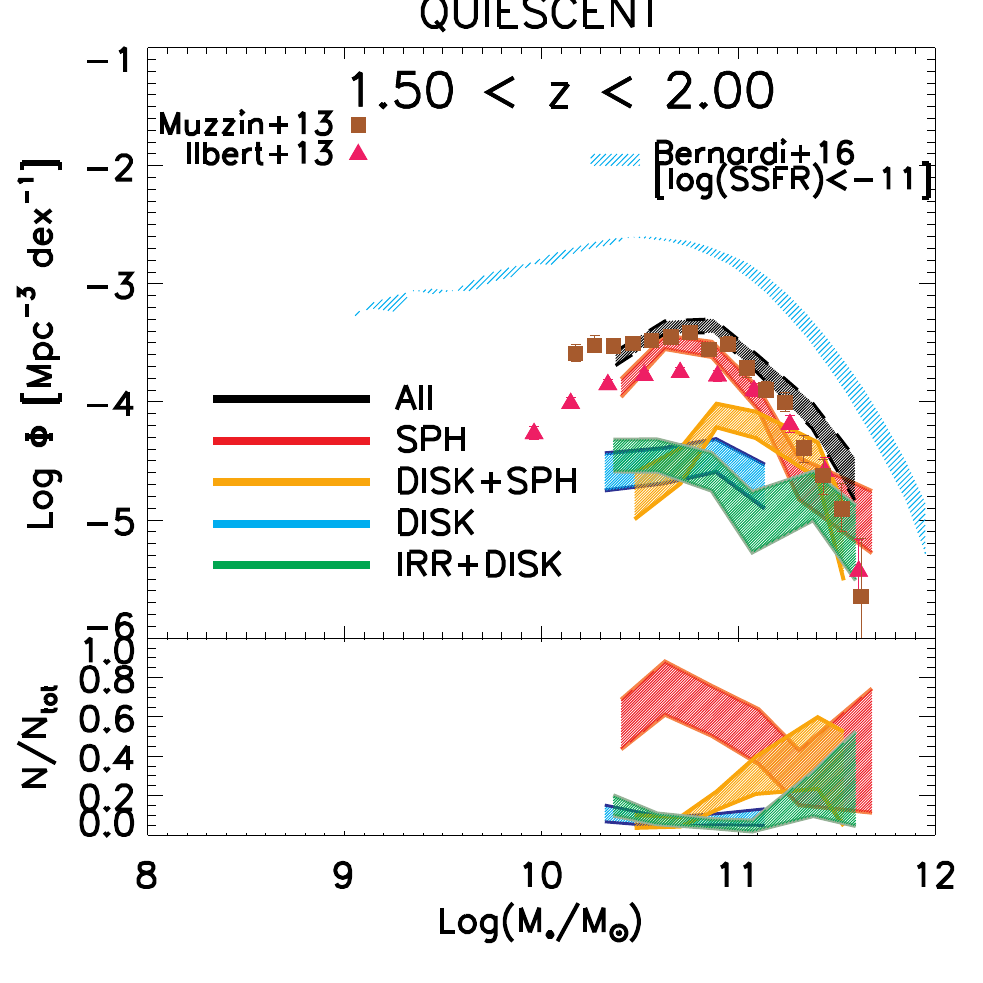} &  \includegraphics[width=0.33\textwidth]{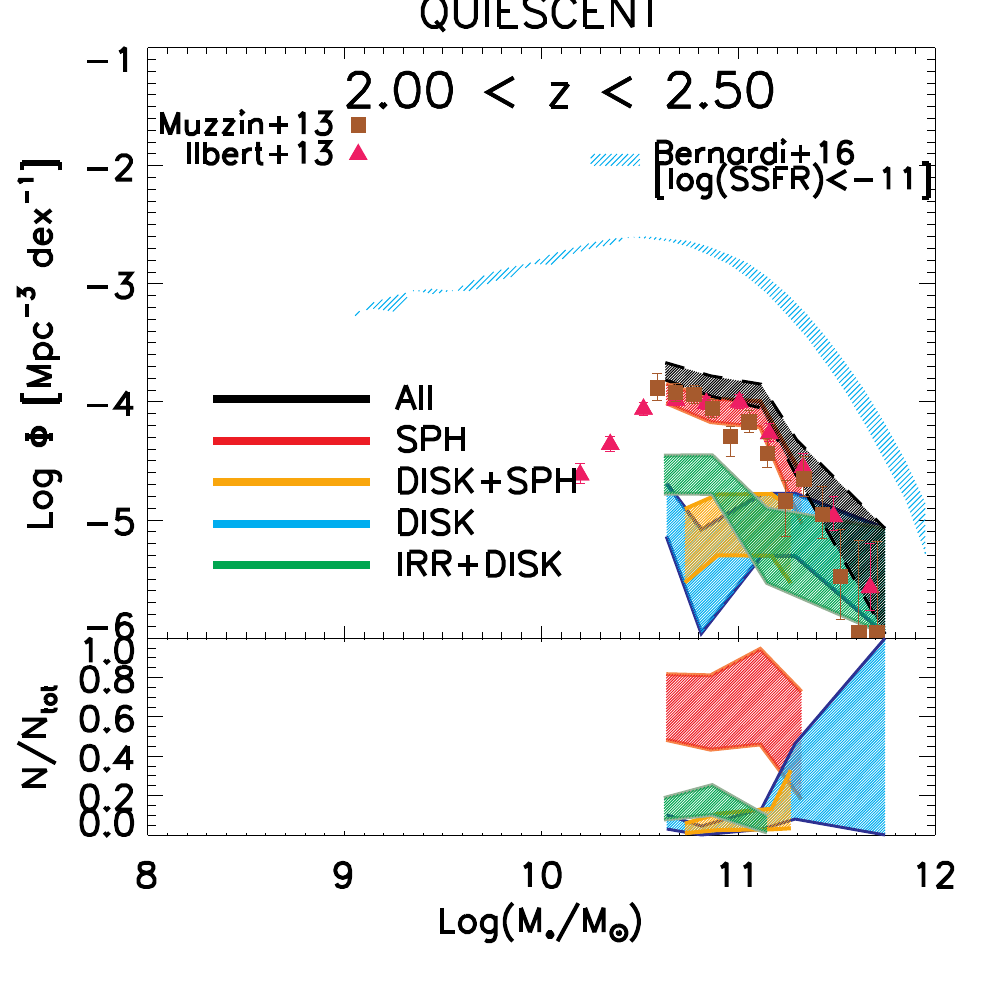} \\
\includegraphics[width=0.33\textwidth]{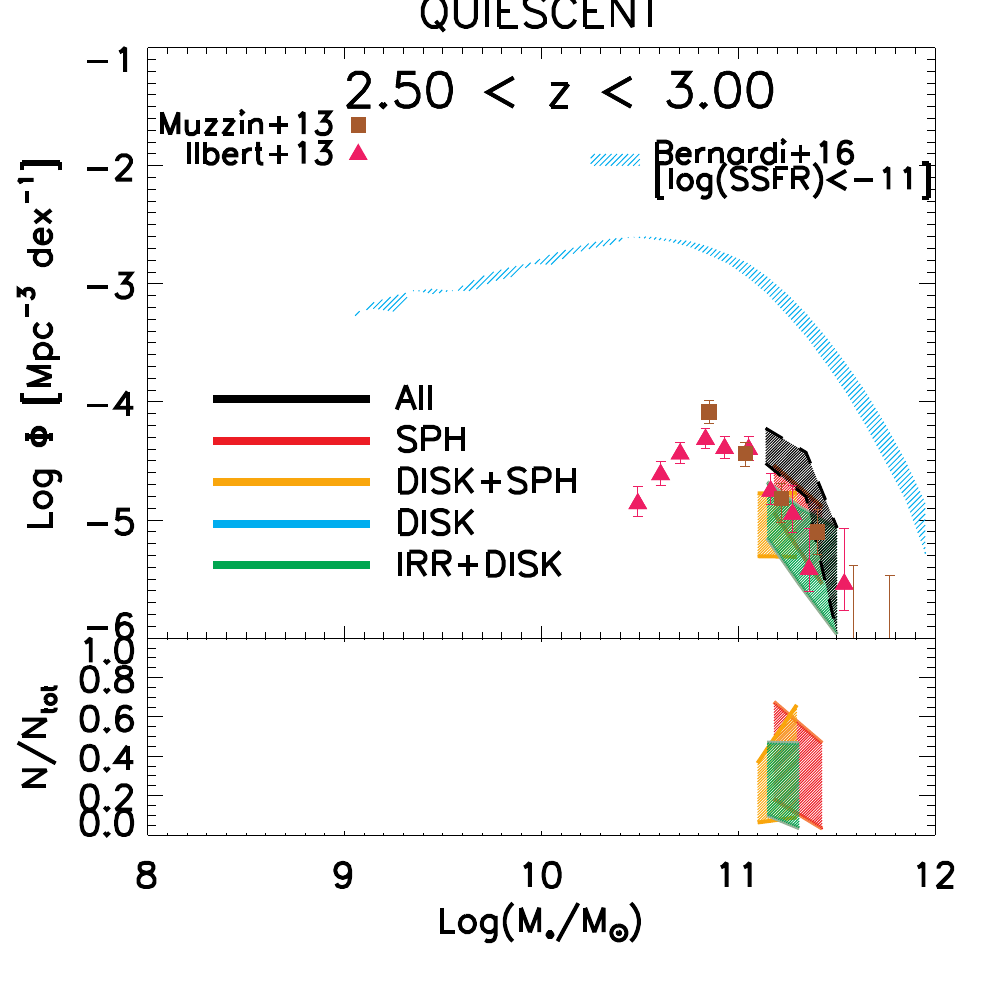}  & & \\
\end{array}$
\caption{Stellar mass functions of quiescent galaxies divided in four morphological types and in different redshift bins as labelled. Red, blue, orange and green shaded regions in the top panels show the number densities of spheroids, disks, disk+spheroids and irregular/clumpy systems respectively. The black regions show the global mass function. Pink filled triangles and brown squares show the recent SMFs by Ilbert et al. (2013) and Muzzin et al. (2013) in the UltraVista survey respectively. The bottom panels show the fractions of each morphological type with the same color code. The cyan shaded region shows the SMF for the SDSS quiescent galaxies from Bernardi et al. (2016) ($log(SSFR)<-11$).} 
\label{fig:MFs_Q}
\end{center}
\end{figure*}

The morphological dissection which our analysis allows provides additional information on how quenching mechanisms affect the morphologies of galaxies. At low redshift, the morphologies of passive galaxies are dominated by two types, pure spheroids and disk+bulge systems. The fraction of quenched late-type spirals is almost negligible ($~\sim5\%$), in agreement with measurements in the local universe \citep{2010MNRAS.405..783M}. Only below $10^9M_\odot$ do red disk systems seem to be more abundant. The quenching mechanisms above $10^{10}M_\odot$ are therefore linked to the presence of a bulge. 

At $z>2$, the population of quiescent galaxies is dominated by pure spheroids while the abundance of passive disk+bulge systems (intermediate B/T) is less than $5\%$ at $z\sim2$. The fraction of disk+bulge systems grows at lower $z$.  This suggests that most of the newly quenched galaxies between $z\sim 2$ and $z\sim 0$ have a disk component. 

These trends are also captured in figure~\ref{fig:mass_density} (which is integrated down to $10^8M_\odot$). To summarize, the main observed trends are:
\begin{itemize}
 \item $\sim80-90\%$ of the stellar mass density of quiescent galaxies is in galaxies with bulges (spheroids and bulge+disk systems)  from z>2. Stars in dead galaxies are therefore almost exclusively in systems with a bulge component.
\item The relative distribution between the two types changes with time. At $z>2$, almost all stars are in spheroids while at $z<1$ stars are equally distributed in disky passive galaxies and spheroids. The stellar mass density in disky passive galaxies increases therefore much faster than in spheroids (a factor of $\sim40$ compared to a factor $4$ for spheroids). The majority of newly quenched galaxies between $z\sim 2$ and $z\sim 0.5$ preserve a disk component. This constrains the dominant quenching mechanisms as discussed in section~\ref{sec:quenching}. 

\end{itemize}

\section{Discussion}
\label{sec:discussion}

The evolution of stellar mass functions can be used to indirectly infer the star formation histories of the different morphologies. The evolution also allows an estimate of when different morphologies emerge. We now discuss the implications of our results for morphological transformations and quenching processes at different stellar mass scales from $z\sim 3$ to the present.

\subsection{Inferred star-formation histories at fixed morphology}
\label{sec:SFHs}

As is well known, the stellar mass density is the integral of the star-formation rate density corrected for the amount of mass loss:
\begin{equation}
 \rho_*(t)=\int_{0}^{t}SFRD(t^{'})(1-0.05ln(1+(t-t^{'})/0.3))dt^{'}
\label{eq:SFRD}
\end{equation}
SFRD stands here for star-formation rate density. The previous equation assumes a parametrization of the return fraction provided by \cite{2009ApJ...696..620C} with a Chabrier IMF. 

Several works have already done this exercise and compared the inferred SFRD evolution with the one obtained from direct measurements, finding different results. \cite{2008MNRAS.385..687W} first reported  a discrepancy of $\sim0.6$ dex between inferred  and direct measurement of the SFHs. Ilbert et al. (2013) revisited this issue with more recent measurements. They found a reasonable agreement with direct measurements from the data compilation of \cite{2013ApJ...777L..10B}, especially at $z<2$, reducing the previous tensions.  We repeat those efforts here, but add morphological information. This enables the first estimates of the formation of stars in different morphologies. 

We first assume that the SFRD evolution can be parametrized with a \emph{Lilly-Madau} law as done by \cite{2013ApJ...777L..10B}:
\begin{equation}
 SFRD(z)=\frac{C}{10^{A(z-z_0)}+10^{B(z-z_0)}}
\label{eq:SFRD_fit}
\end{equation}
Then we fit, for each morphological type, an SFRD following the parametrization of equation~\ref{eq:SFRD_fit} using equation~\ref{eq:SFRD} and the measured of stellar mass densities reported in figure~\ref{fig:mass_density}. The results are shown in figure~\ref{fig:SFH}. The global inferred SFRD evolution agrees reasonably well with the one derived by Ilbert et al. (2013) using the same methodology but on a completely different dataset and with different assumptions on stellar masses. This suggests that the method is robust. With $A$ fixed to $-1$, we find best fit parameters of: $C=0.11\pm0.02$, $B=0.21\pm0.04$ and $z_0=0.98\pm0.13$. This confirms a peak of star formation activity at $z\sim 2$. Our measurements also agree reasonably well with the most recent compilation of different direct measurements performed by \cite{2014ARA&A..52..415M}, especially at low redshifts. At $z>2$, direct measurements estimate a star-formation rate density that is $\sim1.25$ times larger than our inferred values.

The interpretation of the SFHs at fixed morphologies  is more complex since galaxies can transform their morphologies over time. Hence the SFRD we infer cannot be directly interpreted as the star-formation activity of a single morphological type.  Rather, it captures the combined effect of stars formed in-situ in a given morphology and of new stars which were formed in another morphological type and then merged or transformed. For morphologies with a very low quiescent fraction, it is reasonable to assume that the SFRD will be dominated by in-situ star-formation. However, for morphologies which are mostly quiescent, the inferred history is most probably driven by morphological transformations and mergers. To help in the interpretation of the evolution of the SFRD, figure~\ref{fig:Q_frac} shows the evolution of the quiescent fractions of the different morphological types. As we noted before, it appears that irregulars and disks with low B/T fractions are predominantly star-forming. The quiescent fraction in this population is below $10\%$.  In contrast, the quiescent fraction in (massive) spheroids exceeds $\sim 70\%$ at all redshifts.  Of course, this is based on the assumption that the SFHs of individual morphologies can indeed be properly parametrized by equation~\ref{eq:SFRD_fit}.


Nevertheless, the analysis of the SFRDs reveals some interesting first-order trends. There is a clear transition in the dominant morphology hosting star-formation. At $z>2$, star-formation mostly takes place in irregular systems. The fact that the best fit exceeds the global SFRD is clearly an effect of the over simplification of our model. The SFH of irregulars peaks indeed at $z\sim2.5-3$ and sharply decreases thereafter.  But this does not mean that they stop forming stars:  their quiescent fraction is $\le 5\%$ at all redshifts.  Rather, they must transform into other morphologies.  At $z\sim 1.5$, stars are indeed formed in normal symmetric disks both with and without bulge. 

Also interesting is that, at high redshift, the star formation rate density in spheroids is significantly larger than in disks. Again, this does not mean that spheroids are forming more stars than disks, since stars which formed in another morphology and then transformed into spheroids would be credited to spheroids by our parametrization. Since most of spheroids are quiescent, it is likely that the SFH in spheroids is actually dominated by the contribution from transformations. The SFH suggests that the formation of spheroids was most efficient at $z>2$.

\begin{figure}
\begin{center}
\includegraphics[width=0.45\textwidth]{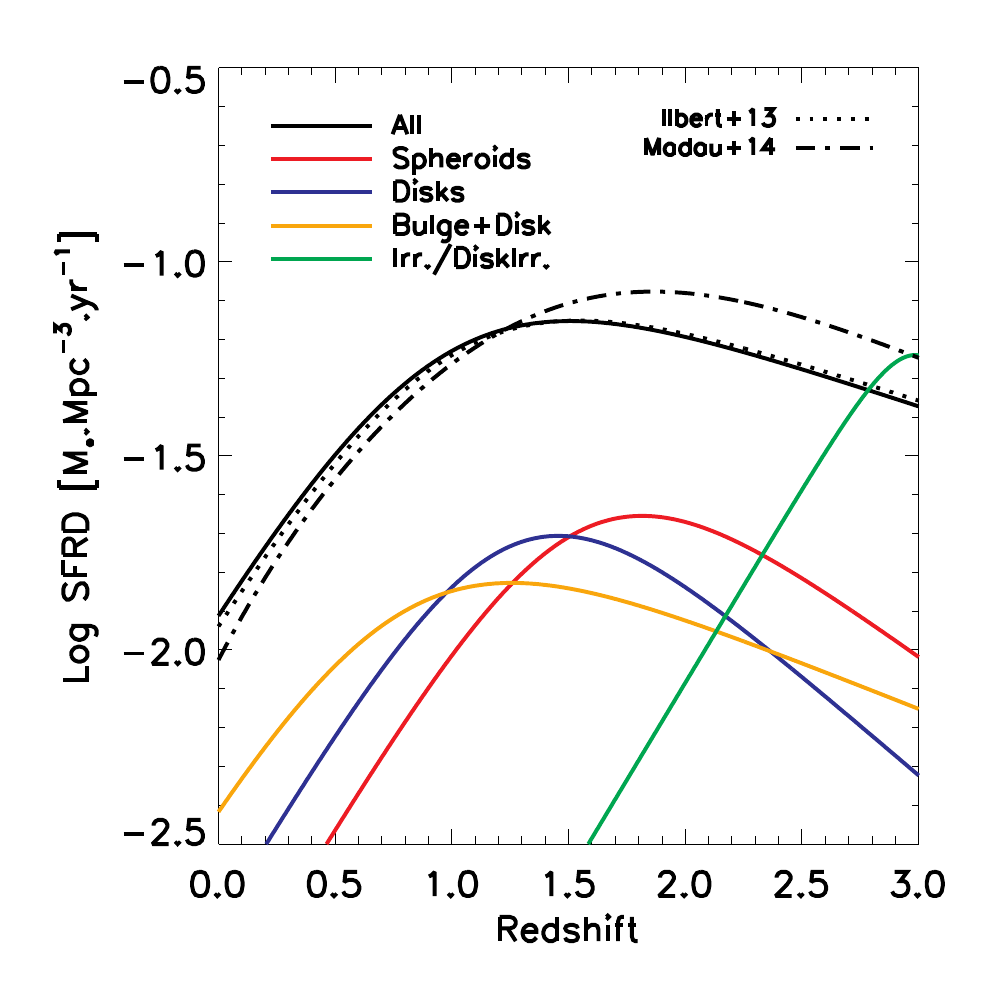}
\caption{Inferred star-formation histories for different morphological types. The black solid line shows the global sample and the different colors show the star-formation histories of different morphologies. The dotted line is the SFRD inferred by~\protect\cite{2013A&A...556A..55I} on UltraVista following the same methodology. The black dashed-dotted line is the most recent compilation of direct measurements by~\protect\cite{2014ARA&A..52..415M}.} 
\label{fig:SFH}
\end{center}
\end{figure}

\subsection{Constraints on morphological transformations of star-forming galaxies: rejuvenation or continuous star-formation?}
\label{sec:SF_trans}
The SFHs suggest a transition in the morphology where most stars are forming. A more complete understanding requires accounting for the impact of mergers and morphological transformations.  Figure~\ref{fig:MF_SF_z} shows the redshift evolution of the SMF of the two morphologies which dominate the star-forming population, i.e. disks (left) and irregulars (right).  In the panel on the left, the abundances increase monotically with time; the opposite is true in the panel on the right.  Since the vast majority of irregulars are star-forming, the decrease in the panel on the right indicates that massive irregulars disappear from the irregular class, so they must begin contributing to another morphological class.  


What do these objects become?  If they continue forming stars, then it is plausible that they transform into symmetric disks.  This would be consistent with the low quiescent fractions ($<10\%$) of both (the irregular and the disk) populations.  It would also be qualitively consistent with the obvious increase with time of the number of normal star-forming disks (left hand panel of figure~\ref{fig:MF_SF_z}).
%
%
I.e., the measured evolution of the disk SMF must result from the combined effects of morphological transformations and in-situ star-formation. At $z>1.5$, where irregulars still dominate the population of star-forming galaxies, the evolution in the panel on the right is probably dominated by transformations.  Below $z\sim 1$, the reservoir of massive irregular galaxies is exhausted, so from this point on, the evolution becomes dominated by genuine star-formation within disks, rather than transformations from irregulars.

This does not actually constrain the individual detailed channels. It does not mean that all star-forming galaxies move straight from an irregular to a pure disk. Only general statistical trends are reflected in the SMF. The individual paths followed by galaxies are necessarily more complex and diverse. As a matter of fact, star-formation in a galaxy is not necessarily continuous and galaxies can experience several morphological transformations during their lives.  A galaxy can easily destroy a disk, quench and then rejuvenate by rebuilding a disk and going back to the disk SF population (e.g. \citealp{2009A&A...507.1313H,2009ApJ...691.1168H}). Also a galaxy might appear as an irregular if seen in a merger phase and then go-back to the disk population. Although this last possibility does not seem to be very common since the evolution of the high mass end of the spheroid mass function does not significantly evolve.} \cite{2016arXiv160107907T} recently looked at the individual paths followed by massive galaxies in the EAGLE simulation. They found however that the fraction of rejuvenated disks represents less than $2\%$ of the star-forming population, suggesting that most of the SF galaxies should keep star-forming and experience a gradual morphological transformation as suggested by the global evolution of the SMFs. Separating rejuvenated disks from disks with continuous star-formation in our sample requires an accurate age determination of our disk population.  While this is not currently possible with the available data (broad or medium band photometry), we can try to place some constraints. 

As shown in \cite{2016arXiv160107907T} most of the rejuvenated galaxies go through a compact quenched phase which lasts $\sim 4$ Gyrs, after which they rebuild a disk. Since rejuvenated objects also build a bulge, once they accrete the disk, an important fraction of them should end up as a star-forming 2-component system in our classification system. Thus, to first-order, star-forming bulge+disks systems are potential good candidates for being rejuvenated objects. The fraction of these systems among the star-forming population is $\sim 5\%$ which suggests rejuvenation is not common. Only above $\sim 10^{11}M_\odot$, and at low redshifts, does the fraction increase to $30-40\%$. Major mergers should play an important role at these mass scales \citep{2010ApJ...721..193P, 2016arXiv160204267C}. So these massive systems might result from mergers followed by the accretion of a disk as suggested in numerical simulations \citep{2009ApJ...691.1168H}. Another possible explanation for this population of star-forming bulge+disks systems is that they are in fact transiting in the other direction. i.e. they are in the process of inside-out quenching (e.g \citealp{2015Sci...348..314T}). The fact that they are a small fraction of the population would then suggest that the transitioning phase is short, in line with expectations (e.g \citealp{2016arXiv160107907T}). We discuss this in the following section.

At the the low-mass end, the irregular SMFs show little evidence of evolution.  This suggests an equilibrium between the arrival of new systems and galaxy growth followed by morphological transformations. This is expected in a CDM scenario where the halo mass function also evolves little at low masses (e.g. \citealp{2016MNRAS.456.2361B}). The growth of haloes is indeed compensated by the emergence of new smaller systems.  The assumption that these new small haloes will be populated by irregular galaxies initially is in qualitative agreement with the observed mild evolution of the abundance of low mass irregulars.

\begin{figure*}
\begin{center}
$\begin{array}{c c}
\includegraphics[width=0.45\textwidth]{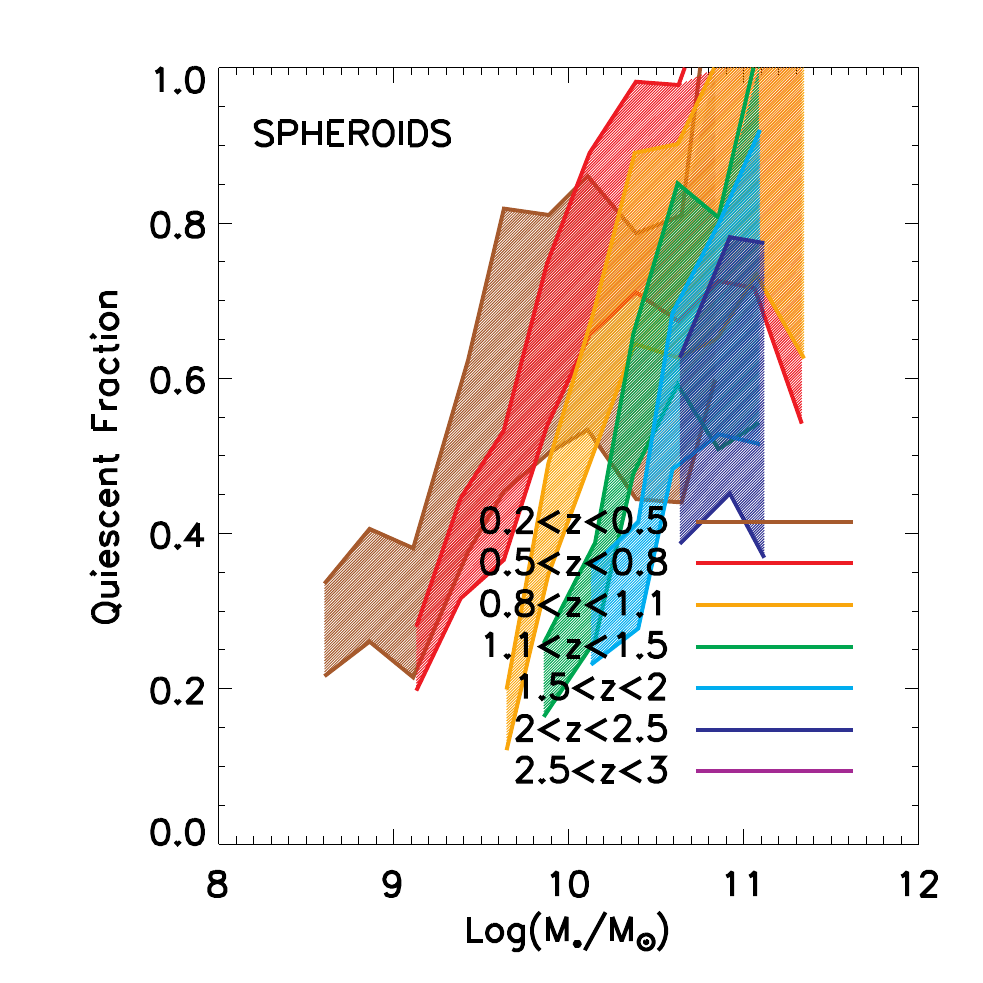} & \includegraphics[width=0.45\textwidth]{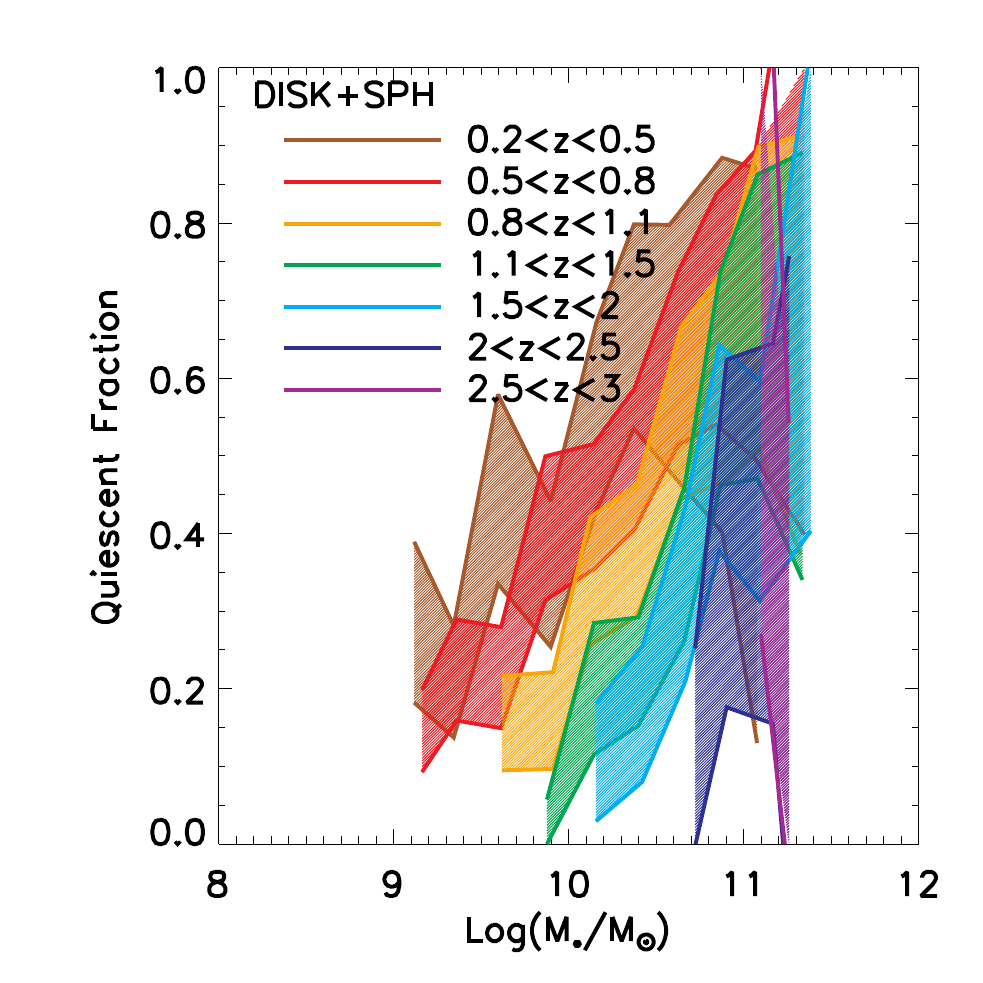} \\
\includegraphics[width=0.45\textwidth]{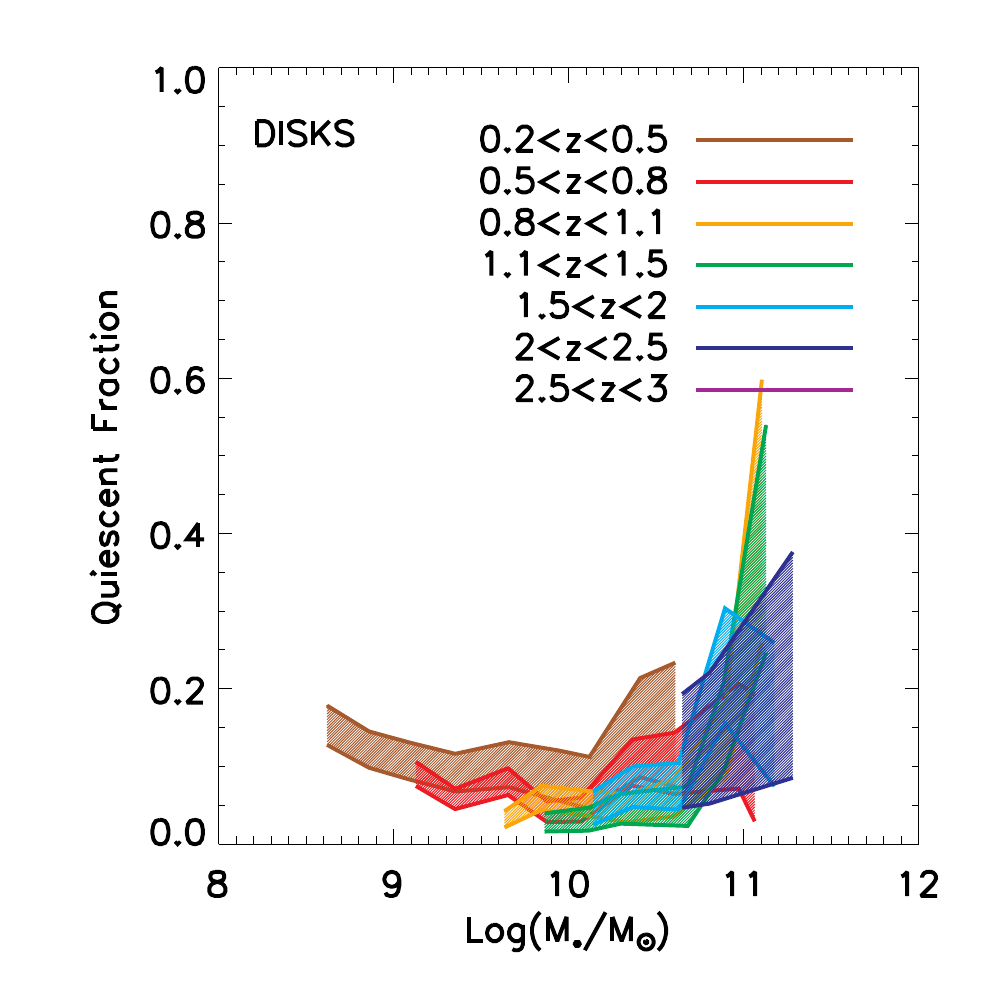} & \includegraphics[width=0.45\textwidth]{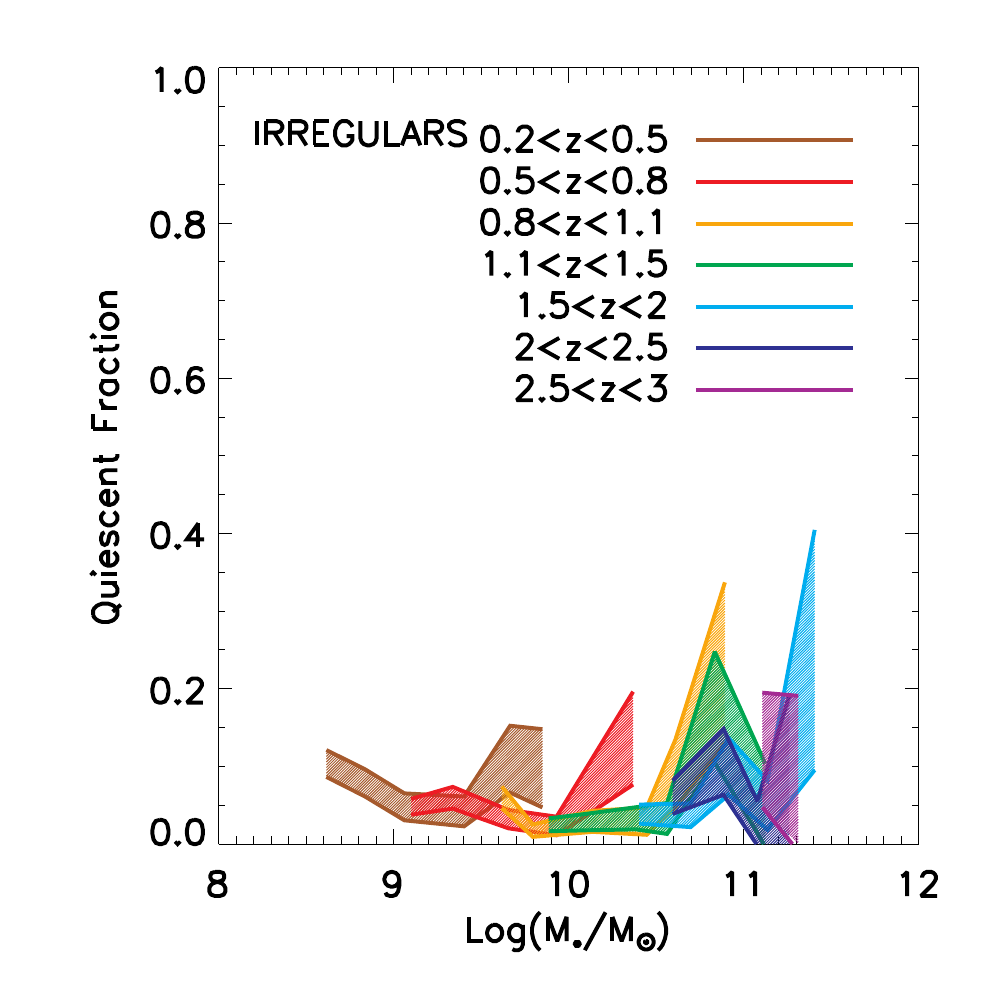} \\
\includegraphics[width=0.45\textwidth]{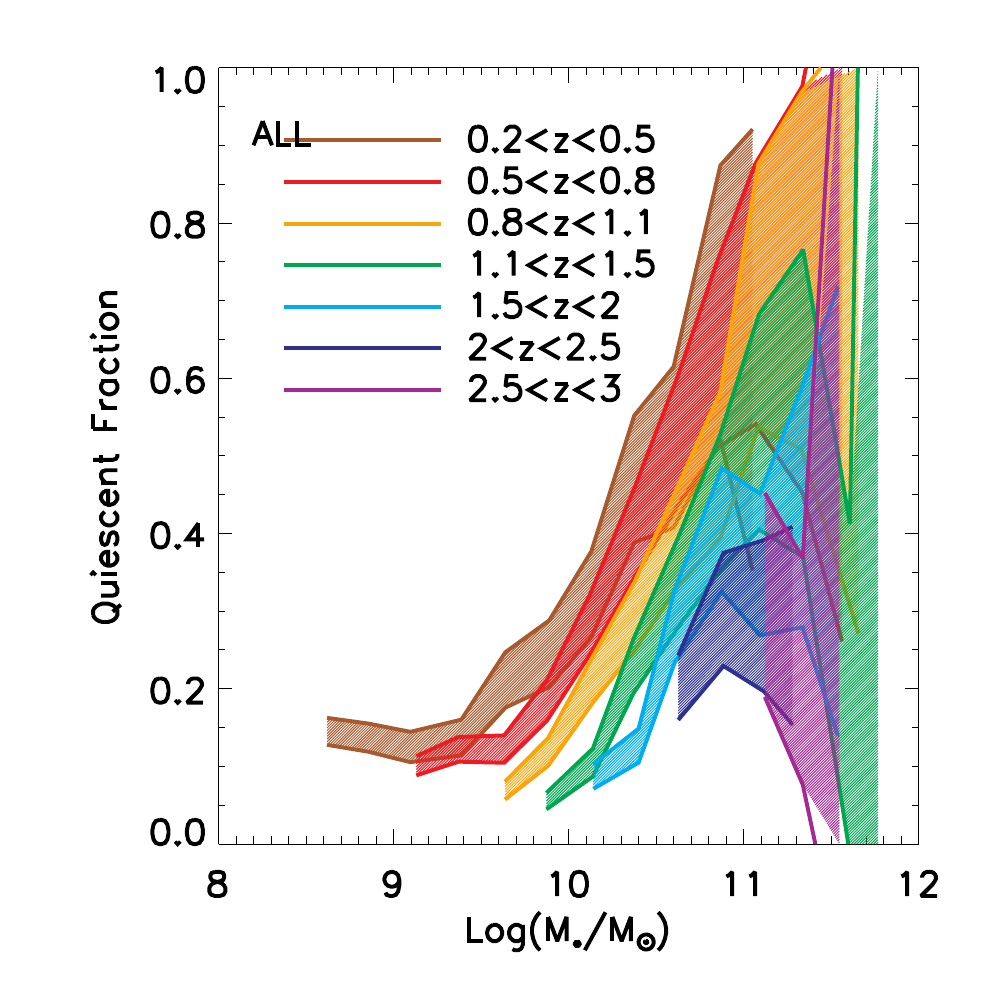}
\end{array}$
\caption{Evolution of the quiescent fraction at fixed morphology. Different colors show different redshift bins as labeled.} 
\label{fig:Q_frac}
\end{center}
\end{figure*}


\begin{figure*}
\begin{center}
$\begin{array}{c c}
\includegraphics[width=0.45\textwidth]{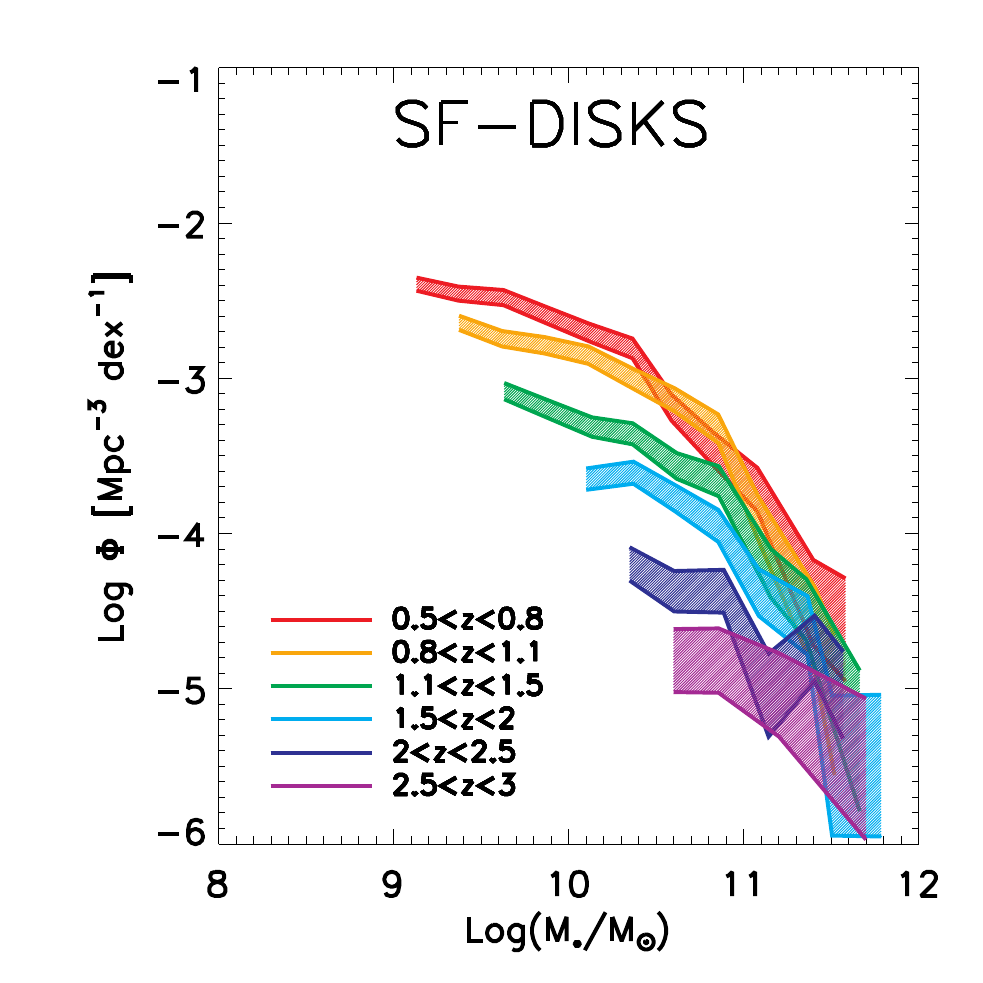} & \includegraphics[width=0.45\textwidth]{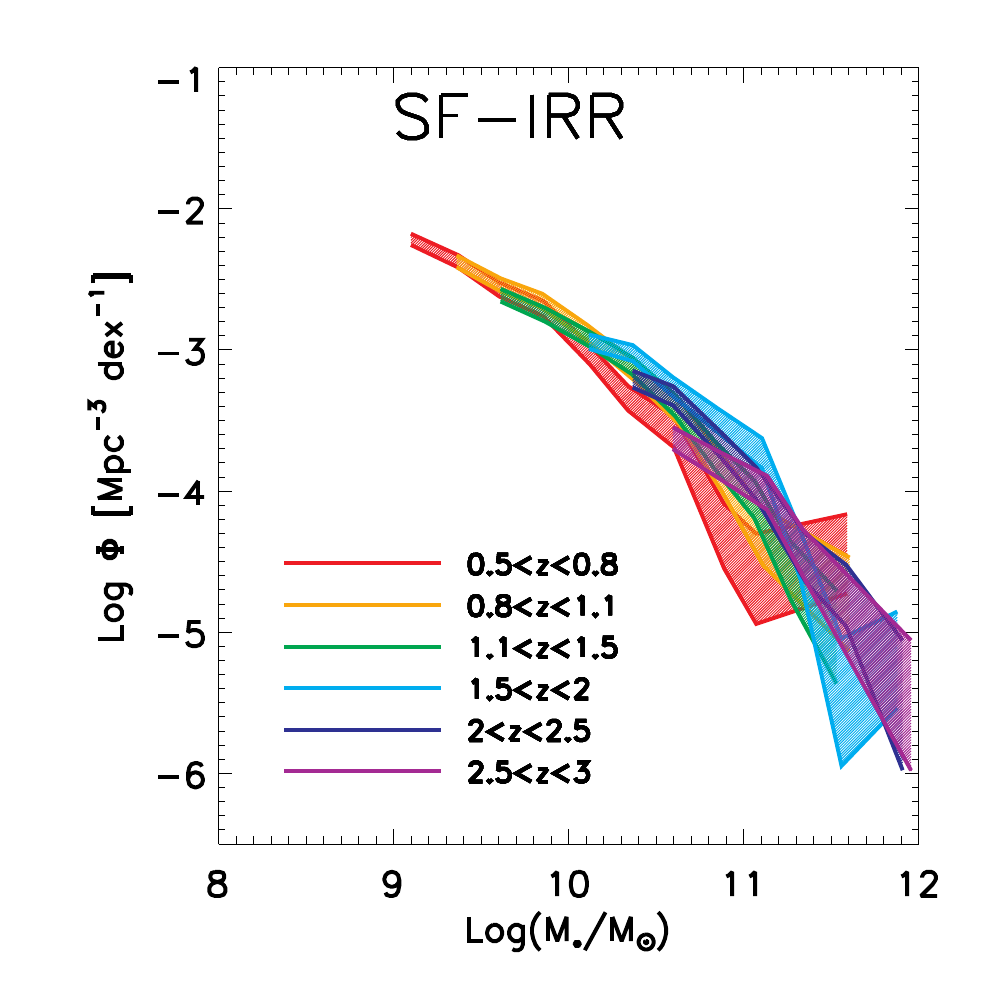}\\
\end{array}$
\caption{Evolution of the stellar mass function of star-forming galaxies at fixed morphological type. The two dominant morphologies of star-forming galaxies, disks (left) and irregulars (right), are shown.} 
\label{fig:MF_SF_z}
\end{center}
\end{figure*}

\subsection {Constraints on quenching processes}
\label{sec:quenching}

The morphological evolution of quiescent and star-forming galaxies contrains the quenching mechanisms which operate at different epochs.  

\subsubsection{Quenching at $M^*$}
The analysis of the evolution of the global SMFs suggests that galaxies reaching masses close to $M^*$ ($\sim10^{10.8}M\odot$) tend to quench and populate the quiescent stellar mass function (e.g. Peng et al. 2010, Ilbert et al. 2013). This is now known as mass quenching (i.e. Peng et al. 2010). However, the physical cause of this decrease in star-formation is still unclear. It may be a consequence of halo heating, which prevents the inflow of gas that is required to feed further star formation.  On the other hand, quenching may be due to more violent processes such as AGN feedback, or some kind of violent disk instabilities (VDI), which are expected to have a stronger impact on the morphology. 

At low redshift, by studying the metallicity of local quiescent galaxies Peng et al. (2015) suggested that strangulation might be a dominant process. At high redshift, more violent mechanisms such as VDI might be needed to \emph{compactify} objects (e.g Barro et al. 2013, 2015). In section~\ref{sec:passive}, we showed that the quiescent population at $z>2$ is dominated by spheroids, and that a population of disky passive galaxies emerges at later epochs. In figure~\ref{fig:MF_Q_z}, we show the evolution of the dominant morphologies of quiescent galaxies, i.e. spheroids and disk+spheroids. The plot shows that the SMFs evolve differently:  Especially around $10.5<log(M_*)<11$, where mass quenching is expected to be the dominant process, the spheroidal SMF seems to increase with time by a smaller amount than that of disk+spheroids. We quantify this effect in figure~\ref{fig:ND_sph}: in this mass range, the number density of pure spheroids evolves little from $z\sim 2$ to the present, whereas that of quiescent galaxies with disks increases $\sim 100\times$.  As discussed in section~\ref{sec:passive}, this is unlikely an observational bias due to cosmic dimming.

\begin{figure*}
\begin{center}
$\begin{array}{c c}
\includegraphics[width=0.45\textwidth]{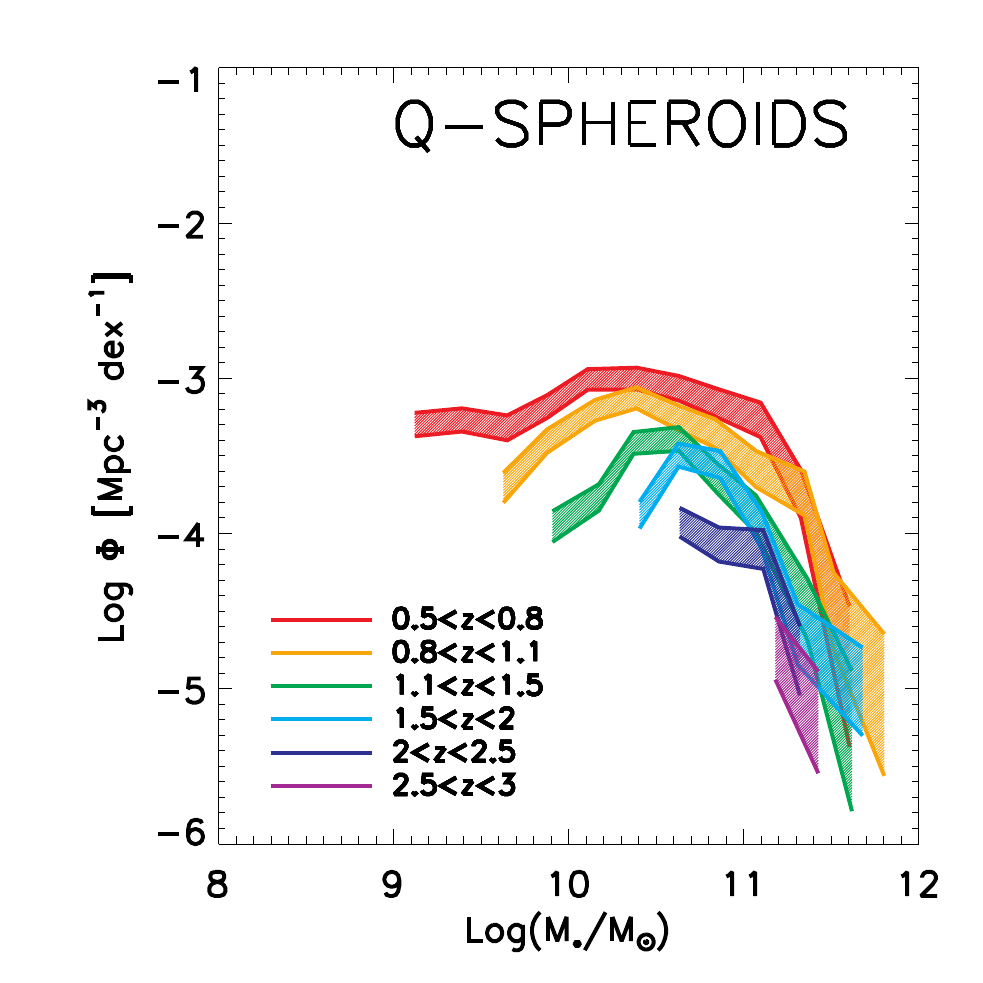} & \includegraphics[width=0.45\textwidth]{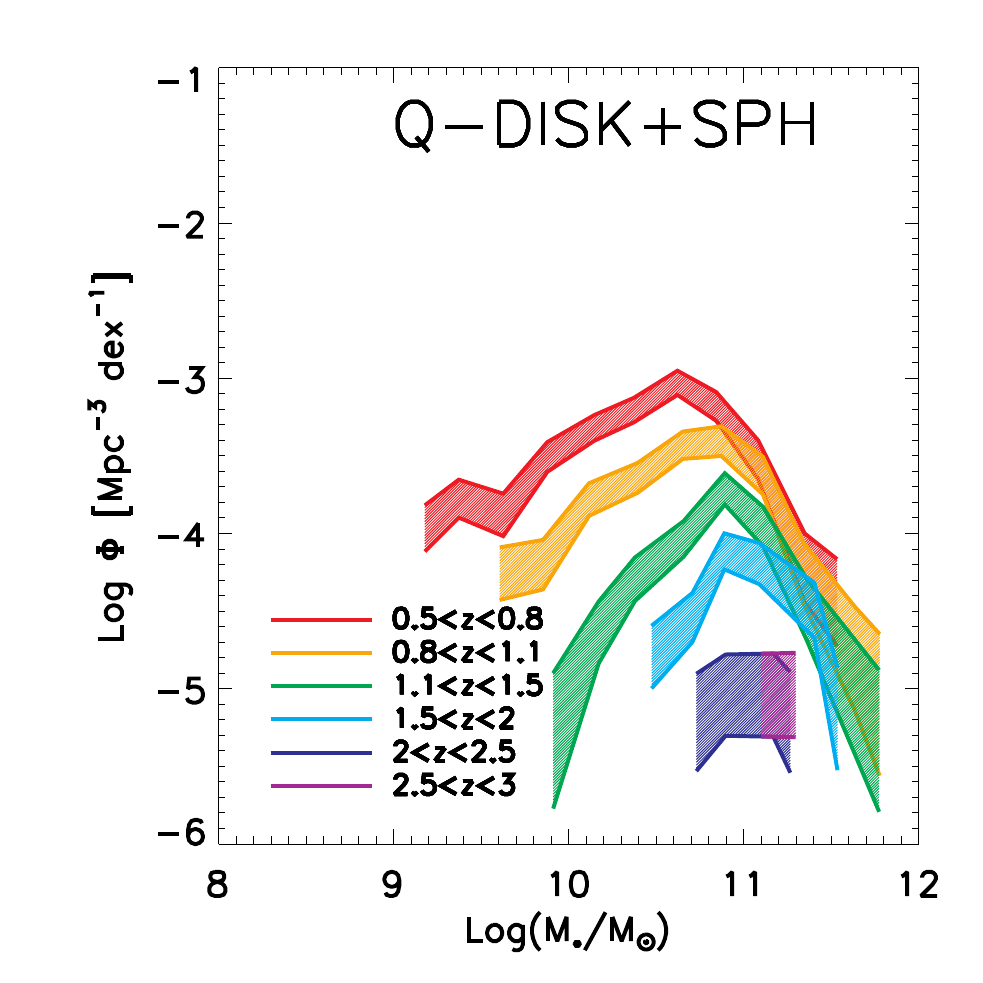}\\
\end{array}$
\caption{Evolution of the stellar mass function of quiescent spheroids (left) and quiescent disk+spheroids (right).  } 
\label{fig:MF_Q_z}
\end{center}
\end{figure*}

One can take a step forward in the interpretation by making the simplistic assumption that pure spheroids are created after some kind of violent dissipative process that destroys the disk  and rapidly quenches star-formation. Pure SPHs are indeed compact, round and dense (Huertas-Company et al. 2015a). Gas-rich major mergers or violent disk instabilities followed by some AGN feedback (e.g. Ceverino et al., Bongiorno et al. 2016) are possible processes. In contrast, disky passive galaxies can be assumed to be predominantly a consequence of a more gradual mechanism related to the lack of available fresh gas (e.g. strangulation, Peng et al. 2015) or morphological quenching~\citep{2009ApJ...707..250M}. Then, the evolution observed in figure~\ref{fig:ND_sph} can be interpreted as a signature of a transition in the dominant quenching mechanism. At $2<z<3$ violent processes such as mergers and VDIs seem to be rather common channel for quenching since the number of spheroids increases in this period by a factor of $\sim10$. At lower redshift though, VDIs appear to be less common in light of the weak evolution of the abundance of spheroids and the passive evolution of their stellar populations. At $z<2$, the majority of newly quenched $\sim M^*$ galaxies preserves a passive disk component. Therefore the most common mass quenching path could be more related to some kind of strangulation that provokes an aging of the stellar populations without significantly altering the morphology. This agrees with Peng et al. (2015). It also means that the population of $\sim M^*$ star-forming bulge+disk systems are probably in the process of quenching from the inside-out (e.g.~\citealp{2015Sci...348..314T}):  Although they have already built a bulge, the star-formation has not yet ceased. Since this phase is expected to last at most $\sim 2$~Gyrs this would explain why these objects are so uncommon.

The previous discussion starts from an assumption that the two populations of passive galaxies formed in different processes. This does not need to be true. The fact that galaxies which quenched at later epochs appear larger and less dense can also simply be a consequence of the fact that their star-forming progenitors are also larger given the observed size increase in star-forming galaxies (e.g. \citealp{2014ApJ...788...28V, 2015ApJS..219...15S}). If so, this would not imply a change in the dominant quenching mechanism. The recent simulations of \cite{2016MNRAS.458L..14F} suggest that the formation of massive quiescent galaxies at very high redshift is also predominantly a consequence of a low amount of gas accretion.  Size measurements of star-forming galaxies at $z>3$ \citep{2016arXiv160201840R, 2015ApJS..219...15S} also show that the typical effective radii of star-forming galaxies at these redshifts are $1-2kpc$, consistent with the sizes of spheroids at $z<2-3$. This would imply that the amount of required \emph{compaction} might be less.

\begin{figure}
\begin{center}
\includegraphics[width=0.45\textwidth]{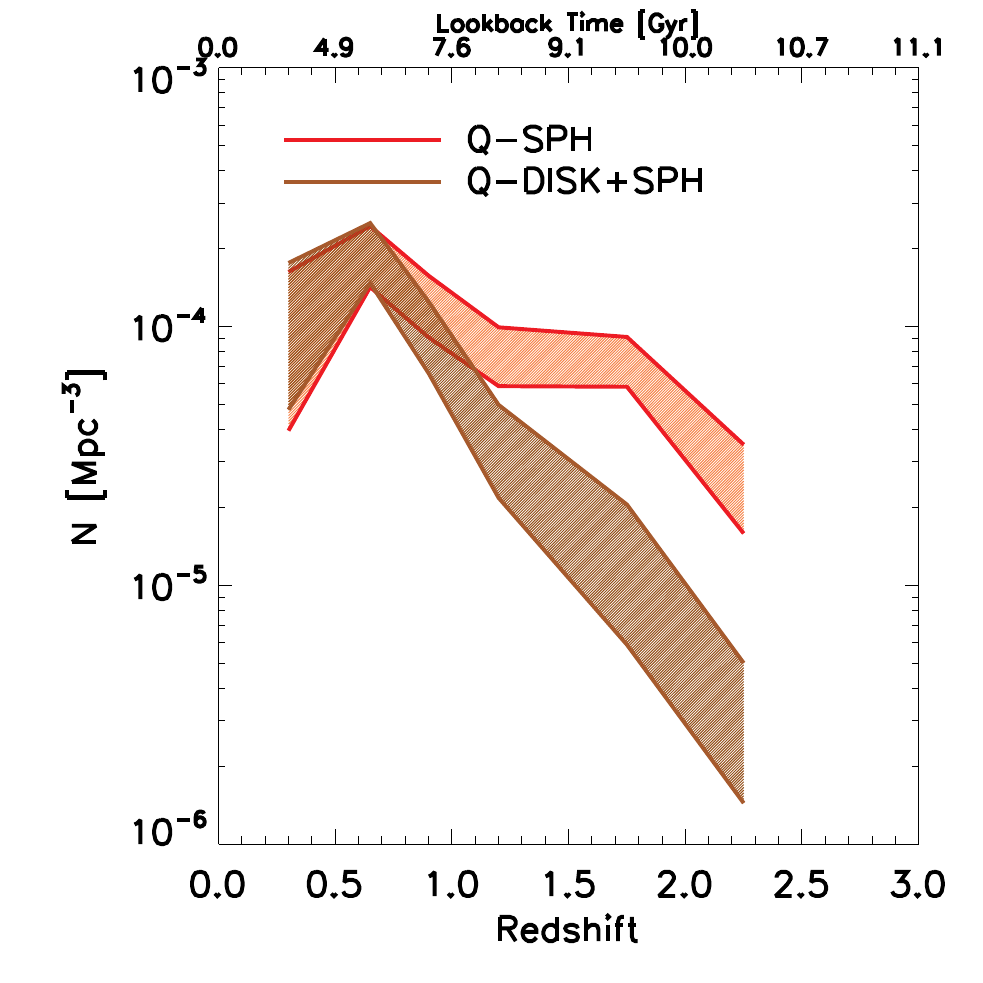}
\caption{Number density evolution of quiescent galaxies with $10.5<log(M_*/M_\odot)<11$ divided by morphology. Red regions show pure spheroids and brown are early-type disks. The last redshift bin is not shown due to incompleteness.} 
\label{fig:ND_sph}
\end{center}
\end{figure}

\subsubsection {Quenching in sub-$M^*$ galaxies}

At $z\sim1.5$, the SMF of quiescent galaxies with stellar masses close to $M_*$ is mostly in place as seen in the bottom panel of figure~\ref{fig:MF_SF_z} and also in previous works analyzing the global mass functions. This means that most of the newly quenched galaxies at $z<1$ have lower stellar masses, i.e. $log(M_*/M_\odot)<10$. They are  expected to be}predominantly satellite galaxies and the quenching process for these systems is generally known as environmental quenching. The effect of environmental quenching is clearly observed in the low mass end of the quiescent mass function which begins to turn upward at $z<1$.

The analysis of the morphological SMFs derived in this work also reveals interesting properties of the mechanisms of environmental quenching. In fact, at these mass scales, the shape of the SMF of passive bulge+disks systems and spheroids is significantly different. While the abundance of disky systems clearly decreases at the low mass end (Fig.~\ref{fig:MF_Q_z}) the one of spheroids tends to increase and mimic the upturn of the global quiescent SMF. Therefore, the low mass end of the quiescent SMF is in fact predominantly populated by spheroids which are significantly more abundant than disk+bulge systems. This means that the environmental quenching process happening at these mass scales will in general destroy the disk and keep only the central component so that the morphology appears like a roundish bulge dominated system. Mechanisms like ram-pressure stripping could indeed remove the disk as a satellite galaxy enters a massive halo. Also relevant is that the abundance of red spirals, i.e. passive disk galaxies with no bulge also increases at $z<1$ (see figures~\ref{fig:mass_density} and~\ref{fig:MFs_Q}). One possible formation mechanism is strangulation as they enter a massive halo as satellites. These two mechanisms seem to coexist at the low mass-end.

We emphasize that this is only a first-order interpretation. A proper quantification of the environments of these low-mass galaxies needs to be undertaken in order to conclude on the effect of environment on quenching. This is beyond the scope of this work and will be explored in the near future.

\section{Summary and conclusions}
\label{sec:summ}

\begin{figure*}
\begin{center}
\includegraphics[width=0.99\textwidth]{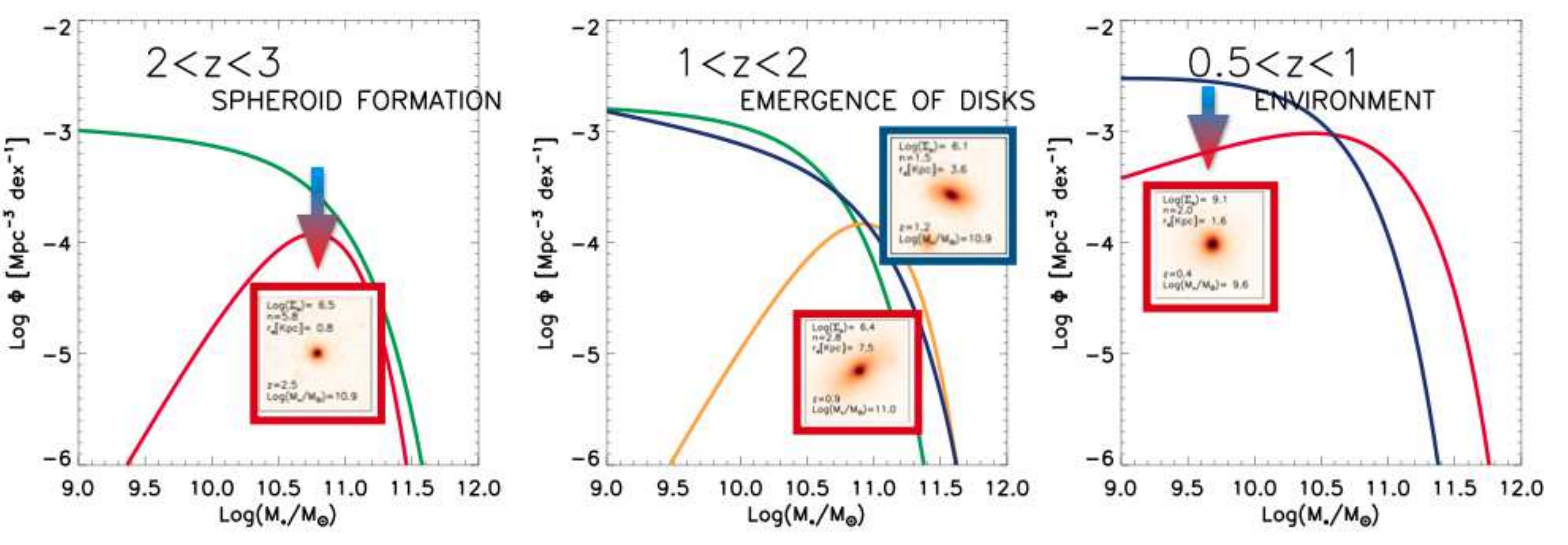}
\caption{Cartoon summarizing the main statistical trends reported in this work. The green, blue, orange and red curves are the real measured SMFs of irregulars, disks, early-type disks and spheroids respectively. The arrows indicate quenching. Stamps are typical morphologies emerging at a given epoch. Blue frames for star-forming and red for quiescent. } 
\label{fig:cartoon}
\end{center}
\end{figure*}

We have studied the evolution of the stellar mass functions of quiescent and star-forming galaxies in the redshift range $0.2<z<3$ at fixed morphological type covering an area of $\sim880$ $arcmin^2$. Our sample consists of $\sim 50,000$ galaxies with $H<24.5$. The stellar mass completeness goes from $log(M_*/M_\odot)\sim 8$ at $z\sim0.2$ to $log(M_*/M_\odot)\sim 10.3$ at $z\sim 3$. Galaxies are divided into four main morphological classes based on a deep-learning classification, i.e pure bulge dominated spheroids, pure disks, intermediate 2-component systems and irregular or disturbed objects.  Each morphology has clearly differentiated structural properties. Our main conclusions are summarized below:

\begin{itemize}

\item Our global SMFs agree with recent measurements from large NIR ground-based surveys. Volume effects are only seen in the lowest redshift bins. We find mass-dependent evolution of the global and star-forming stellar mass function:  the low mass end evolves faster than the high mass end in agreement with previous work. This is a consequence of \emph{mass-quenching} being efficient for galaxies which reach a typical stellar mass of $log(M_*/M_\odot)\sim 10.8$.  The stellar mass density of quiescent galaxies with $log(M_*/M_\odot)>8$ increases by a factor of 5 between $z\sim 3$ and $z\sim 1$. At $z<1$ the passive SMF flattens at the low mass-end; this is usually interpreted a signature of \emph{environment quenching} on satellite galaxies. 

\item  The inclusion of statistical morphological information brings additional insight. See also figure~\ref{fig:cartoon}.

\begin{itemize}
\item At $z>2$, the morphological distribution of massive galaxies is bimodal: spheroids and irregulars. All star-forming galaxies are irregulars. Taking into account recent dynamical studies of star-forming objects at $z>1$, this might be a signature of unstable and turbulent disks. The quiescent galaxies are pure compact spheroids with no clear evidence of a disk component. At these redshifts, the high mass end of the passive population is building-up rapidly. The morphological distribution suggests therefore a violent quenching mechanism as main channel to quench galaxies at $z>2$. Strong dissipative processes such as very-gas rich mergers or violent disk instabilities are known to rapidly bring a large amount of gas into the central parts of the galaxy, leading to a massive, compact and dense remnant as observed. Alternatively, they might be the result of quenching of small star-forming systems at higher redshifts. 


\item Between $1<z<2$ the majority of \emph{normal} disks observed in the local universe emerge. The SMF of normal star-forming spiral disks evolves rapidly during this time. The evolution is a combination of in-situ star-formation and morphological transformations from irregular disks.
At $z\sim1.5$, star-formation occurs primarily in normal spiral disks.  To lowest order, this morphological transition does not seem to interrupt star-formation. Rejuvenation does not play an important role, although this has to be confirmed with a careful age analysis of the stellar population of late-type disks.  The morphological mix of quiescent galaxies also evolves significantly between $1<z<2$. Most of the newly massive quenched galaxies in this redshift range have a disk component but with a larger bulge than the star-forming ones. The number density increases $100\times$ while that of quiescent spheroidals stays roughly constant. The efficiency with which spheroids form decreases and the dominant quenching process does not destroy the disk. This suggests a transition in the main quenching mechanism. Strangulation and/or morphological quenching are possible explanations.

\item At $z<1$, there is little evolution of the morphological mix above $10^{10.8}M_\odot$.  At the highest masses, the abundance of bulgeless systems decreases; nearly $100\%$ of the population has a significant bulge; star-forming objects with a large bulge represent $\sim 40\%$ of the population.  Galaxies with masses $\sim 10^{10.8}M_\odot$ are equally likely to be spheroids as symmetric late and early-type spirals.  Most ($95\%$) passive galaxies are spheroids or early-type spiral/S0 galaxies while most (90\%) star-forming galaxies are late-type spirals.  Below $\sim10^{10}M_\odot$, irregular objects dominate the star-forming population. Quenching mostly happens at this low mass-end: it creates a population of low-mass bulge dominated systems and leads to an increase in the fraction of red spirals.  This suggests both ram-pressure and strangulation as the main quenching mechanisms.

\end{itemize}

\end{itemize}

\section*{Acknowledgements}
Thanks to R. Sheth for comments on an early draft. Thanks to the anonymous referee for a constructive and quick report.

\appendix

\section{Morphologies}
\label{app:morph_stamps}

\begin{figure*}
\begin{center}

\includegraphics[width=\textwidth]{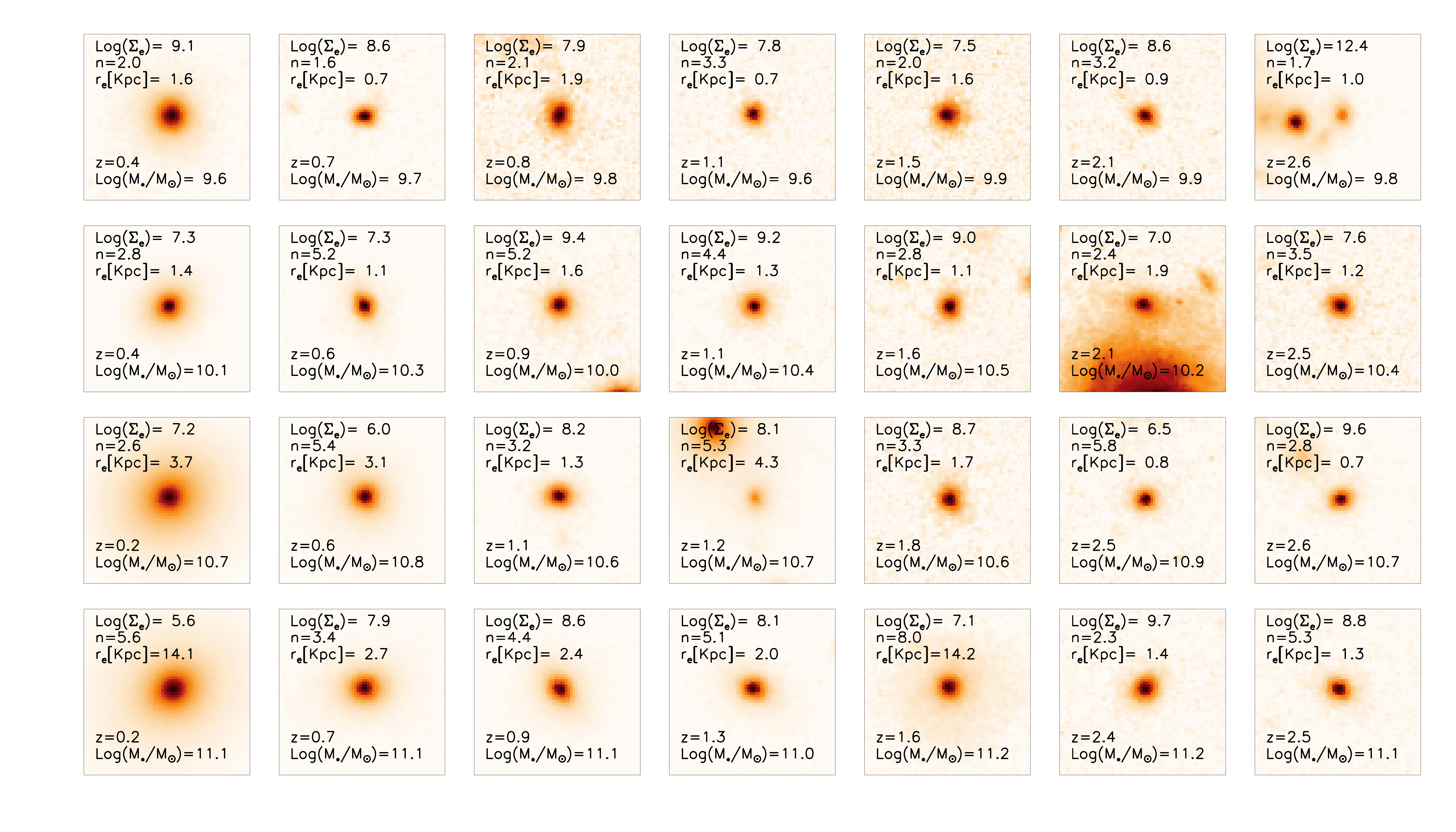}

\caption{Postage stamps of galaxies classified as spheroids ($\sim Es$) sorted by increasing stellar mass (vertical direction) and redshift (horizontal direction). } 
\label{fig:sph_stamps}
\end{center}
\end{figure*}

\begin{figure*}
\begin{center}

\includegraphics[width=\textwidth]{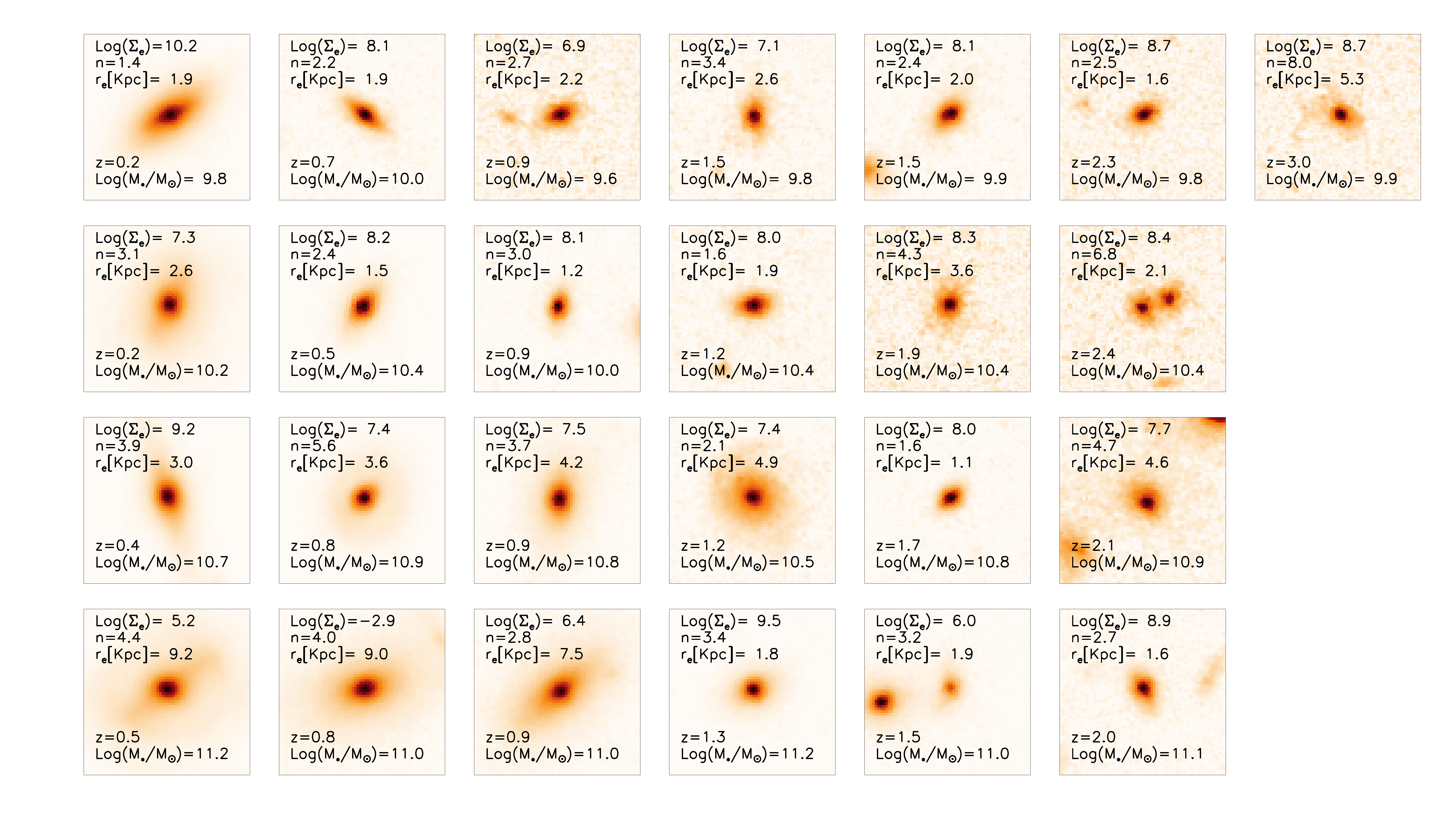}

\caption{Postage stamps of galaxies classified as disk+spheroids ($\sim$S0s and early-type spirals) sorted by increasing stellar mass (vertical direction) and redshift (horizontal direction). } 
\label{fig:disksph_stamps}
\end{center}
\end{figure*}

\begin{figure*}
\begin{center}

\includegraphics[width=\textwidth]{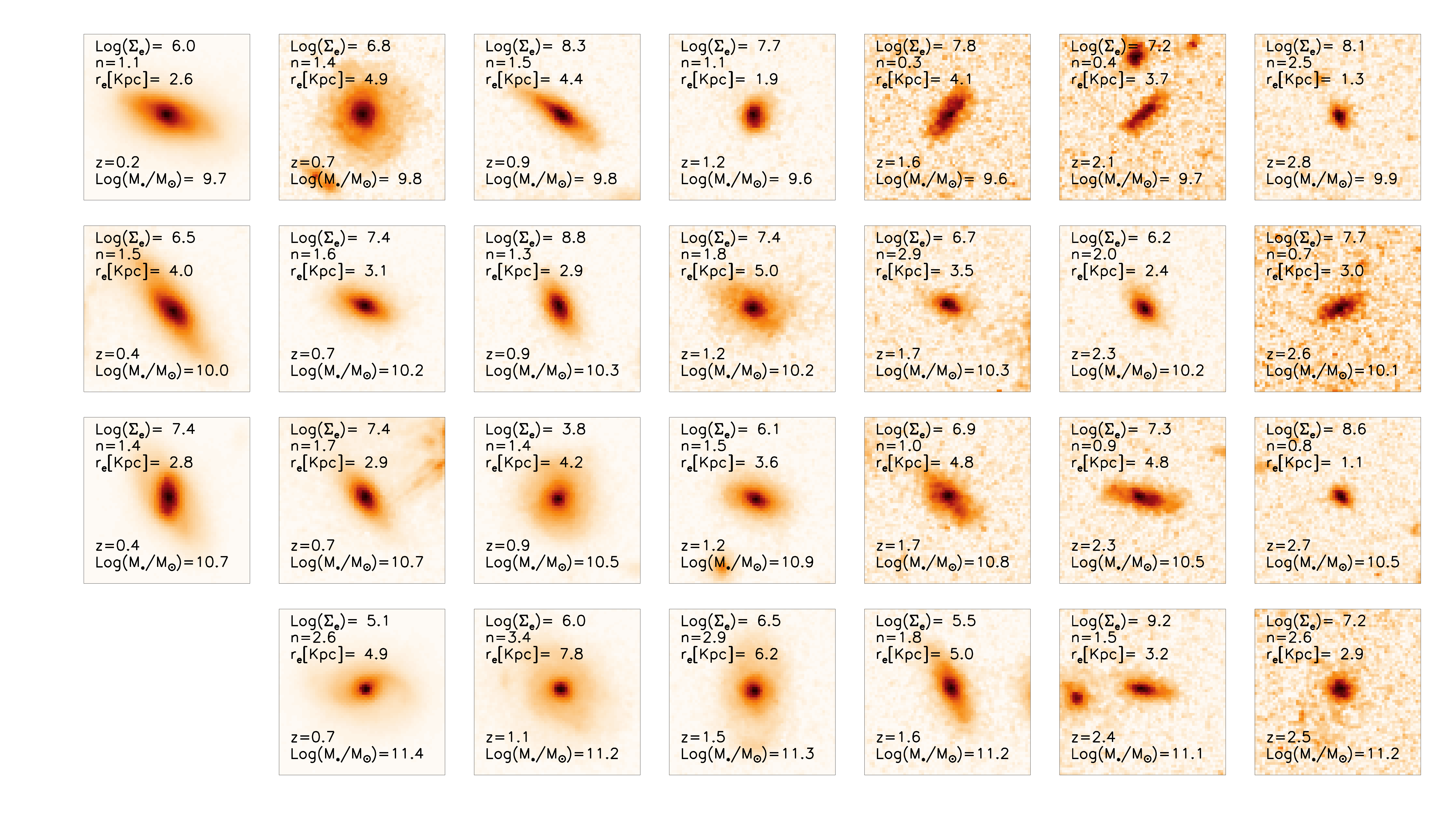}

\caption{Postage stamps of galaxies classified as disks ($\sim$late-type spirals) sorted by increasing stellar mass (vertical direction) and redshift (horizontal direction). } 
\label{fig:disk_stamps}
\end{center}
\end{figure*}

\begin{figure*}
\begin{center}

\includegraphics[width=\textwidth]{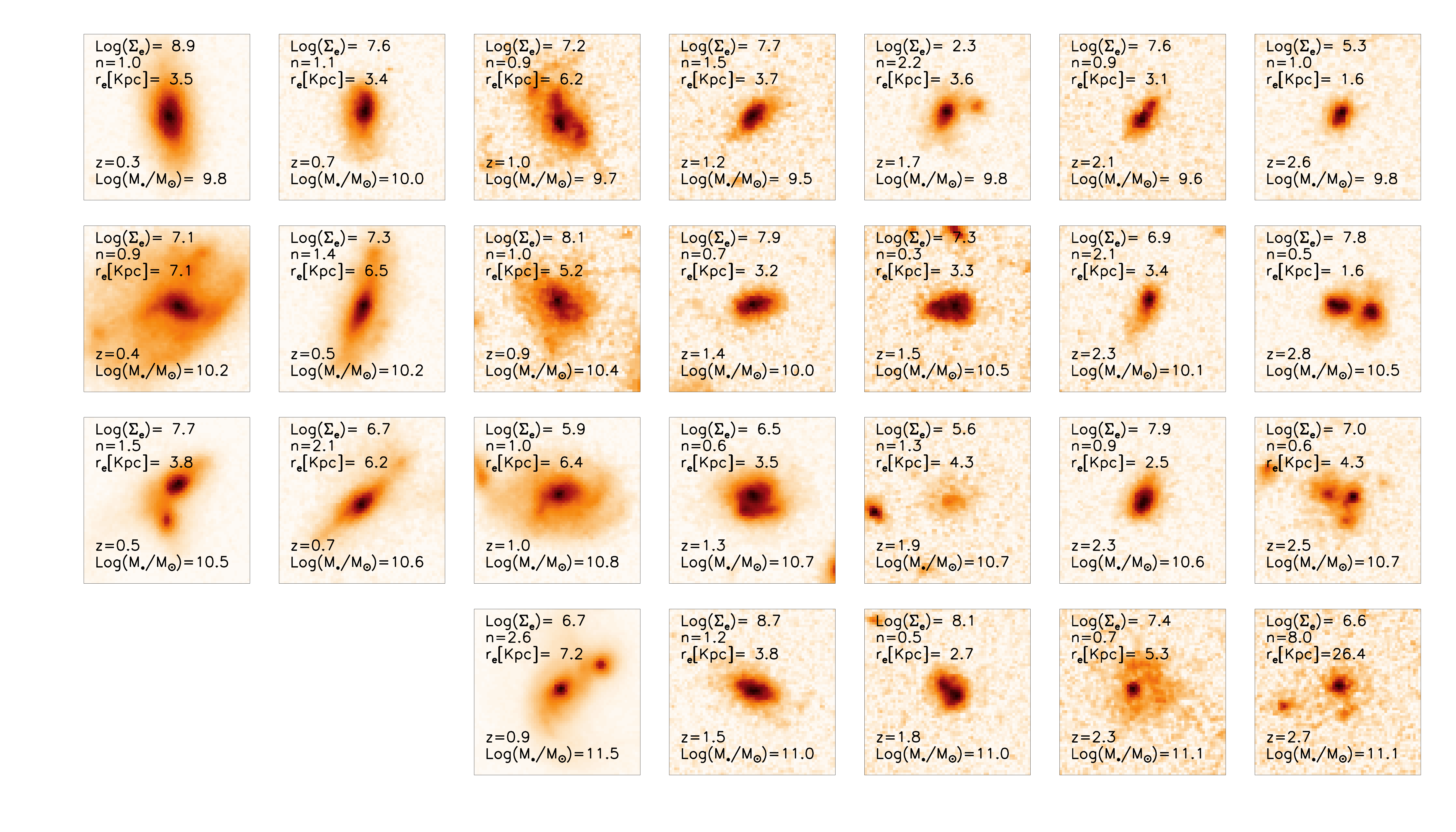}

\caption{Postage stamps of galaxies classified as irregulars sorted by increasing stellar mass (vertical direction) and redshift (horizontal direction). } 
\label{fig:irr_stamps}
\end{center}
\end{figure*}

Figures~\ref{fig:sph_stamps} to~\ref{fig:irr_stamps} show postage stamps of a random subset set of galaxies in each of the 4 main morphological classes used in this work, over a range of different stellar masses and redshifts. 

\section{Bulge-to-total ratios of different morphologies}
\label{app:struct_props_morph}

\begin{figure}
\begin{center}
\includegraphics[width=0.45\textwidth]{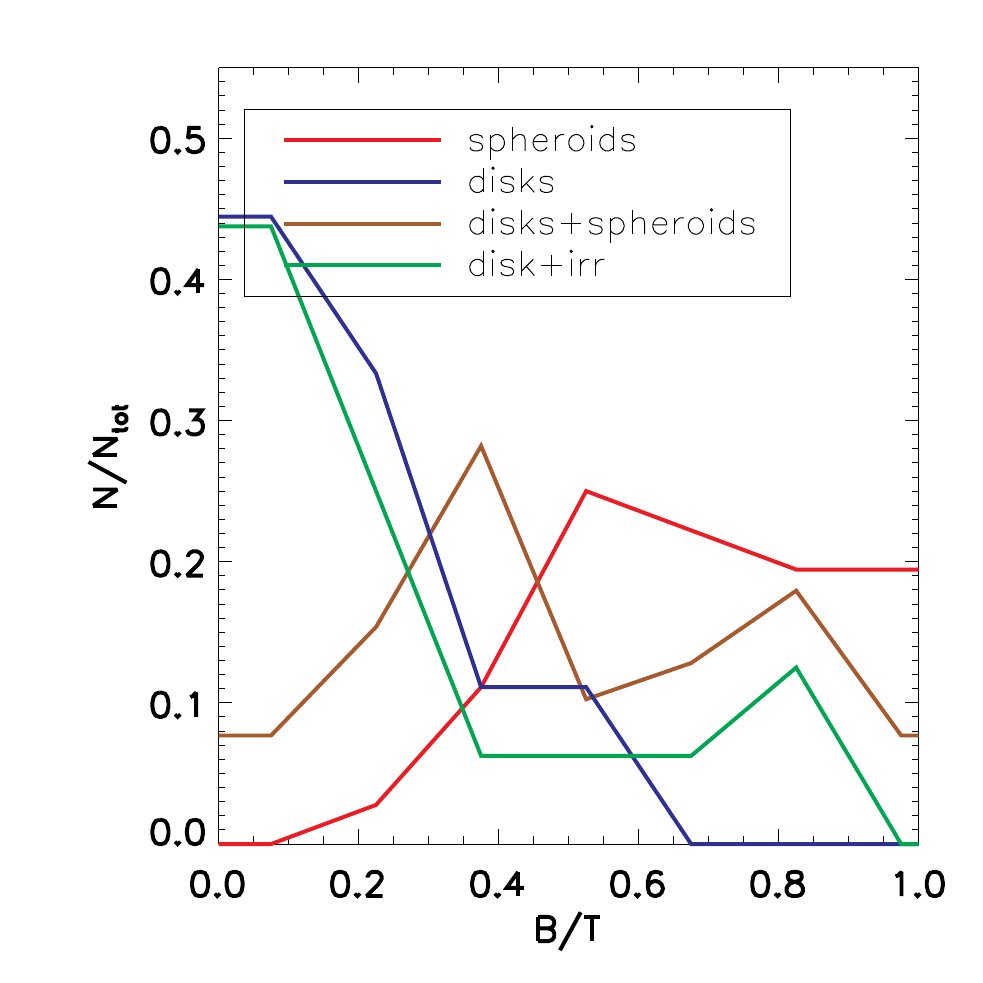}

\caption{Stellar mass bulge-to-total ratios for the different morphologies. The visual morphologies defined in this work are compared for some galaxies to the distribution of stellar mass bulge-to-total ratios. The expected trends are observed.} 
\label{fig:BT_visual}
\end{center}
\end{figure}

As an additional sanity check and given that many models use the stellar-mass disk to bulge ratio as a proxy for morphology, figure~\ref{fig:BT_visual} shows the stellar mass bulge-to-total (B/Ts) ratios for a subsample of galaxies from our dataset. Bulge fractions are obtained by fitting a 2-component Sersic+exponential model on 7 HST filters (from near UV to NIR) simultaneously using Megamorph \citep{2013MNRAS.430..330H}. Sizes of both components and Sersic indices of the bulges are allowed to change with wavelength following a polynomial of order 2. We then fit the 7 point SEDs of bulges and disks separately with BC03 templates and estimate the stellar masses of the two components separately.  While a detailed discussion of the procedure is beyond the scope of this paper (details are provided in Dimauro et al., in preparation), here we simply want to highlight the fact that the morphologies estimated independently with deep-learning do match the expected distribution of B/Ts reasonably well.  I.e., DISKs and IRRs tend to have bulge fractions smaller than 0.2 whereas SPHs have B/T greater than $\sim0.6$. DISKSPHs have a broader distribution of B/T values.  Note, however, that bulge/disk decompositions do not capture the irregularities in the light profile which are an important elements in this work.

\clearpage


\begin{thebibliography}{}





\bibitem[Baldry et al.(2008)]{2008MNRAS.388..945B} Baldry, I.~K., 
Glazebrook, K., \& Driver, S.~P.\ 2008, \mnras, 388, 945 

\bibitem[Barro et al.(2015)]{2015arXiv150900469B} Barro, G., Faber, S.~M., 
Koo, D.~C., et al.\ 2015, arXiv:1509.00469


\bibitem[Barro et al.(2014)]{2014ApJ...791...52B} Barro, G., Faber, S.~M., 
P{\'e}rez-Gonz{\'a}lez, P.~G., et al.\ 2014, \apj, 791, 52

\bibitem[Barro et al.(2013)]{2013ApJ...765..104B} Barro, G., Faber, S.~M., 
P{\'e}rez-Gonz{\'a}lez, P.~G., et al.\ 2013, \apj, 765, 104

\bibitem[Barro et al.(2011b)]{2011ApJS..193...30B} Barro, G., 
P{\'e}rez-Gonz{\'a}lez, P.~G., Gallego, J., et al.\ 2011, \apjs, 193, 30 

\bibitem[Barro et al.(2011a)]{2011ApJS..193...13B} Barro, G., 
P{\'e}rez-Gonz{\'a}lez, P.~G., Gallego, J., et al.\ 2011, \apjs, 193, 13

\bibitem[Behroozi et al.(2013)]{2013ApJ...777L..10B} Behroozi, P.~S., 
Marchesini, D., Wechsler, R.~H., et al.\ 2013, \apjl, 777, LL10 

\bibitem[Bell et al.(2005)]{2005ApJ...625...23B} Bell, E.~F., Papovich, C., 
Wolf, C., et al.\ 2005, \apj, 625, 23

\bibitem[Bernardi et al.(2016)]{2016arXiv160401036B} Bernardi, M., Meert, A., Sheth, R.~K., et al.\ 2016, arXiv:1604.01036

\bibitem[Bernardi et al.(2014)]{2014MNRAS.443..874B} Bernardi, M., Meert, 
A., Vikram, V., et al.\ 2014, \mnras, 443, 874

\bibitem[Bernardi et al.(2013)]{2013MNRAS.436..697B} Bernardi, M., Meert, 
A., Sheth, R.~K., et al.\ 2013, \mnras, 436, 697



\bibitem[Bernardi et al.(2011)]{2011MNRAS.412L...6B} Bernardi, M., Roche, 
N., Shankar, F., \& Sheth, R.~K.\ 2011, \mnras, 412, L6 

\bibitem[Bernardi et al.(2010)]{2010MNRAS.404.2087B} Bernardi, M., Shankar, 
F., Hyde, J.~B., et al.\ 2010, \mnras, 404, 2087 

\bibitem[Bezanson et al.(2011)]{2011ApJ...737L..31B} Bezanson, R., van 
Dokkum, P.~G., Franx, M., et al.\ 2011, \apjl, 737, LL31 


\bibitem[Bocquet et al.(2016)]{2016MNRAS.456.2361B} Bocquet, S., Saro, A., Dolag, K., \& Mohr, J.~J.\ 2016, \mnras, 456, 2361 

\bibitem[Bournaud et al.(2014)]{2014ApJ...780...57B} Bournaud, F., Perret, 
V., Renaud, F., et al.\ 2014, \apj, 780, 57

\bibitem[Brammer et al.(2012)]{2012ApJS..200...13B} Brammer, G.~B., van 
Dokkum, P.~G., Franx, M., et al.\ 2012, \apjs, 200, 13

\bibitem[Brammer et al.(2008)]{2008ApJ...686.1503B} Brammer, G.~B., van 
Dokkum, P.~G., \& Coppi, P.\ 2008, \apj, 686, 1503

\bibitem[Bruce et al.(2012)]{2012MNRAS.427.1666B} Bruce, V.~A., Dunlop, 
J.~S., Cirasuolo, M., et al.\ 2012, \mnras, 427, 1666

\bibitem[Bruzual 
\& Charlot(2003)]{2003MNRAS.344.1000B} Bruzual, G., \& Charlot, S.\ 2003, \mnras, 344, 1000 

\bibitem[Buitrago et al.(2011)]{2011arXiv1111.6993B} Buitrago, F., 
Trujillo, I., Conselice, C.~J., \& Haeussler, B.\ 2011, arXiv:1111.6993 

\bibitem[Buitrago et al.(2008)]{2008ApJ...687L..61B} Buitrago, F., 
Trujillo, I., Conselice, C.~J., et al.\ 2008, \apjl, 687, L61 

\bibitem[Bundy et al.(2005)]{2005ApJ...625..621B} Bundy, K., Ellis, R.~S., 
\& Conselice, C.~J.\ 2005, \apj, 625, 621

\bibitem[Cappellari(2016)]{2016arXiv160204267C} Cappellari, M.\ 2016, 
arXiv:1602.04267 


\bibitem[Cassata et al.(2013)]{2013ApJ...775..106C} Cassata, P., 
Giavalisco, M., Williams, C.~C., et al.\ 2013, \apj, 775, 106


\bibitem[Calzetti et al.(2000)]{2000ApJ...533..682C} Calzetti, D., Armus, 
L., Bohlin, R.~C., et al.\ 2000, \apj, 533, 682

128

\bibitem[Carollo et al.(2014)]{2014arXiv1402.1172C} Carollo, C.~M., 
Cibinel, A., Lilly, S.~J., et al.\ 2014, arXiv:1402.1172

\bibitem[Carollo et al.(2013)]{2013ApJ...773..112C} Carollo, C.~M., 
Bschorr, T.~J., Renzini, A., et al.\ 2013, \apj, 773, 112


\bibitem[Chabrier(2003)]{2003PASP..115..763C} Chabrier, G.\ 2003, \pasp, 
115, 763 



\bibitem[Cimatti et al.(2012)]{2012MNRAS.422L..62C} Cimatti, A., Nipoti, 
C., \& Cassata, P.\ 2012, \mnras, 422, L62 

\bibitem[Conroy \& Wechsler(2009)]{2009ApJ...696..620C} Conroy, C., \& Wechsler, R.~H.\ 2009, \apj, 696, 620

\bibitem[Conselice et al.(2005)]{2005ApJ...620..564C} Conselice, C.~J., Blackburne, J.~A., \& Papovich, C.\ 2005, \apj, 620, 564 

\bibitem[Conselice et al.(2013)]{2013MNRAS.430.1051C} Conselice, C.~J., 
Mortlock, A., Bluck, A.~F.~L., Gr{\"u}tzbauch, R., 
\& Duncan, K.\ 2013, \mnras, 430, 1051 



\bibitem[Daddi et al.(2005)]{2005ApJ...626..680D} Daddi, E., Renzini, A., 
Pirzkal, N., et al.\ 2005, \apj, 626, 680 

\bibitem[Dahlen et al.(2013)]{2013ApJ...775...93D} Dahlen, T., Mobasher, 
B., Faber, S.~M., et al.\ 2013, \apj, 775, 93 

\bibitem[Damjanov et al.(2015)]{2015arXiv150104976D} Damjanov, I., Geller, 
M.~J., Zahid, H.~J., \& Hwang, H.~S.\ 2015, arXiv:1501.04976 

\bibitem[Damjanov et al.(2011)]{2011ApJ...739L..44D} Damjanov, I., Abraham, 
R.~G., Glazebrook, K., et al.\ 2011, \apjl, 739, L44 

\bibitem[Dekel 
\& Burkert(2014)]{2014MNRAS.438.1870D} Dekel, A., \& Burkert, A.\ 2014, \mnras, 438, 1870

\bibitem[Feldmann et al.(2016)]{2016MNRAS.458L..14F} Feldmann, R., Hopkins, 
P.~F., Quataert, E., Faucher-Gigu{\`e}re, C.-A., 
\& Kere{\v s}, D.\ 2016, \mnras, 458, L14

\bibitem[Ferrarese et al.(2006)]{2006ApJ...644L..21F} Ferrarese, L., 
C{\^o}t{\'e}, P., Dalla Bont{\`a}, E., et al.\ 2006, \apjl, 644, L21


\bibitem[Fioc \& Rocca-Volmerange(1999)]{1999astro.ph.12179F} Fioc, M., \& Rocca-Volmerange, B.\ 1999, arXiv:astro-ph/9912179 

\bibitem[Fukushima (1980)]{Fukushima80}Fukushima, K. (1980). Neocognitron: A self-organizing neural network model for a mechanism of pattern recognition unaffected by shift in position. Biological Cybernetics, 36, 193–202.

\bibitem[Galametz et al.(2013)]{2013ApJS..206...10G} Galametz, A., Grazian, 
A., Fontana, A., et al.\ 2013, \apjs, 206, 10

\bibitem[Gehrels(1986)]{1986ApJ...303..336G} Gehrels, N.\ 1986, \apj, 303, 
336 

\bibitem[Gonz{\'a}lez et al.(2009)]{2009MNRAS.397.1254G} Gonz{\'a}lez, 
J.~E., Lacey, C.~G., Baugh, C.~M., Frenk, C.~S., 
\& Benson, A.~J.\ 2009, \mnras, 397, 1254 

\bibitem[Granato et al.(2004)]{2004ApJ...600..580G} Granato, G.~L., De 
Zotti, G., Silva, L., Bressan, A., \& Danese, L.\ 2004, \apj, 600, 580

\bibitem[Grogin et al.(2011)]{2011ApJS..197...35G} Grogin, N.~A., Kocevski, 
D.~D., Faber, S.~M., et al.\ 2011, \apjs, 197, 35

\bibitem[Guo et al.(2013)]{2013ApJS..207...24G} Guo, Y., Ferguson, H.~C., 
Giavalisco, M., et al.\ 2013, \apjs, 207, 24

\bibitem[Guo et al.(2010)]{2010MNRAS.404.1111G} Guo, Q., White, S., Li, C., 
\& Boylan-Kolchin, M.\ 2010, \mnras, 404, 1111 

\bibitem[Hammer et 
al.(2009)]{2009A&A...507.1313H} Hammer, F., Flores, H., Puech, M., et al.\ 2009, \aap, 507, 1313

\bibitem[H{\"a}u{\ss}ler et al.(2013)]{2013MNRAS.430..330H} 
H{\"a}u{\ss}ler, B., Bamford, S.~P., Vika, M., et al.\ 2013, \mnras, 430, 
330 

\bibitem[Huertas-Company et al.(2015)]{2015ApJS..221....8H} 
Huertas-Company, M., Gravet, R., Cabrera-Vives, G., et al.\ 2015, \apjs, 
221, 8 

\bibitem[Huertas-Company et al.(2013)]{2013MNRAS.428.1715H} 
Huertas-Company, M., Mei, S., Shankar, F., et al.\ 2013, \mnras, 428, 1715 

\bibitem[Hopkins et al.(2009)]{2009ApJ...691.1168H} Hopkins, P.~F., Cox, 
T.~J., Younger, J.~D., \& Hernquist, L.\ 2009, \apj, 691, 1168


\bibitem[Ilbert et 
al.(2013)]{2013A&A...556A..55I} Ilbert, O., McCracken, H.~J., Le F{\`e}vre, O., et al.\ 2013, \aap, 556, AA55

\bibitem[Kartaltepe et al.(2014)]{2014arXiv1401.2455K} Kartaltepe, J.~S., 
Mozena, M., Kocevski, D., et al.\ 2014, arXiv:1401.2455


\bibitem[Kennicutt(1998)]{1998ApJ...498..541K} Kennicutt, R.~C., Jr.\ 1998, 
\apj, 498, 541 

\bibitem[Koekemoer et al.(2011)]{2011ApJS..197...36K} Koekemoer, A.~M., 
Faber, S.~M., Ferguson, H.~C., et al.\ 2011, \apjs, 197, 36



\bibitem[Kriek et al.(2009)]{2009ApJ...705L..71K} Kriek, M., van Dokkum, 
P.~G., Franx, M., Illingworth, G.~D., 
\& Magee, D.~K.\ 2009, \apjl, 705, L71


\bibitem[Lapi et al.(2011)]{2011ApJ...742...24L} Lapi, A., 
Gonz{\'a}lez-Nuevo, J., Fan, L., et al.\ 2011, \apj, 742, 24

\bibitem[Leja et al.(2013)]{2013ApJ...766...33L} Leja, J., van Dokkum, P., 
\& Franx, M.\ 2013, \apj, 766, 33 

\bibitem[Lilly \& Carollo(2016)]{2016arXiv160406459L} Lilly, S.~J., \& Carollo, C.~M.\ 2016, arXiv:1604.06459

\bibitem[Lilly et al.(2013)]{2013ApJ...772..119L} Lilly, S.~J., Carollo, 
C.~M., Pipino, A., Renzini, A., \& Peng, Y.\ 2013, \apj, 772, 119

\bibitem[Madau 
\& Dickinson(2014)]{2014ARA&A..52..415M} Madau, P., \& Dickinson, M.\ 2014, \araa, 52, 415

\bibitem[Marchesini et al.(2014)]{2014ApJ...794...65M} Marchesini, D., 
Muzzin, A., Stefanon, M., et al.\ 2014, \apj, 794, 65

\bibitem[Marchesini et al.(2010)]{2010ApJ...725.1277M} Marchesini, D., 
Whitaker, K.~E., Brammer, G., et al.\ 2010, \apj, 725, 1277

\bibitem[Marchesini et al.(2009)]{2009ApJ...701.1765M} Marchesini, D., van 
Dokkum, P.~G., F{\"o}rster Schreiber, N.~M., et al.\ 2009, \apj, 701, 1765

\bibitem[Martig et al.(2009)]{2009ApJ...707..250M} Martig, M., Bournaud, 
F., Teyssier, R., \& Dekel, A.\ 2009, \apj, 707, 250

\bibitem[Masters et al.(2010)]{2010MNRAS.405..783M} Masters, K.~L., Mosleh, 
M., Romer, A.~K., et al.\ 2010, \mnras, 405, 783 

\bibitem[McCracken et 
al.(2012)]{2012A&A...544A.156M} McCracken, H.~J., Milvang-Jensen, B., Dunlop, J., et al.\ 2012, \aap, 544, A156

\bibitem[Mei et al.(2014)]{2014arXiv1403.7524M} Mei, S., Scarlata, C., 
Pentericci, L., et al.\ 2014, arXiv:1403.7524

\bibitem[Mei et al.(2009)]{2009ApJ...690...42M} Mei, S., Holden, B.~P., 
Blakeslee, J.~P., et al.\ 2009, \apj, 690, 42

\bibitem[Moffett et al.(2016)]{2016MNRAS.457.1308M} Moffett, A.~J., 
Ingarfield, S.~A., Driver, S.~P., et al.\ 2016, \mnras, 457, 1308

\bibitem[Mortlock et al.(2015)]{2015MNRAS.447....2M} Mortlock, A., Conselice, C.~J., Hartley, W.~G., et al.\ 2015, \mnras, 447, 2 

\bibitem[Mortlock et al.(2013)]{2013MNRAS.433.1185M} Mortlock, A., 
Conselice, C.~J., Hartley, W.~G., et al.\ 2013, \mnras, 433, 1185 

\bibitem[Mortlock et al.(2011)]{2011MNRAS.413.2845M} Mortlock, A., 
Conselice, C.~J., Bluck, A.~F.~L., et al.\ 2011, \mnras, 413, 2845 



\bibitem[Moster et al.(2013)]{2013MNRAS.428.3121M} Moster, B.~P., Naab, T., 
\& White, S.~D.~M.\ 2013, \mnras, 428, 3121 

\bibitem[Moster et al.(2011)]{2011ApJ...731..113M} Moster, B.~P., 
Somerville, R.~S., Newman, J.~A., \& Rix, H.-W.\ 2011, \apj, 731, 113

\bibitem[Moustakas et al.(2013)]{2013ApJ...767...50M} Moustakas, J., Coil, 
A.~L., Aird, J., et al.\ 2013, \apj, 767, 50

\bibitem[Moutard et al.(2016)]{2016arXiv160205917M} Moutard, T., Arnouts, 
S., Ilbert, O., et al.\ 2016, arXiv:1602.05917

\bibitem[Muzzin et al.(2013)]{2013ApJ...777...18M} Muzzin, A., Marchesini, 
D., Stefanon, M., et al.\ 2013, \apj, 777, 18 

\bibitem[Pozzetti et 
al.(2010)]{2010A&A...523A..13P} Pozzetti, L., Bolzonella, M., Zucca, E., et al.\ 2010, \aap, 523, A13

\bibitem[Naab et al.(2009)]{2009ApJ...699L.178N} Naab, T., Johansson, 
P.~H., \& Ostriker, J.~P.\ 2009, \apjl, 699, L178 


\bibitem[Newman et al.(2013)]{2013ApJ...767..104N} Newman, S.~F., Genzel, 
R., F{\"o}rster Schreiber, N.~M., et al.\ 2013, \apj, 767, 104

\bibitem[Newman et al.(2012)]{2012ApJ...746..162N} Newman, A.~B., Ellis, 
R.~S., Bundy, K., \& Treu, T.\ 2012, \apj, 746, 162 

\bibitem[Nipoti et al.(2012)]{2012MNRAS.422.1714N} Nipoti, C., Treu, T., 
Leauthaud, A., et al.\ 2012, \mnras, 422, 1714

\bibitem[Oser et al.(2010)]{2010ApJ...725.2312O} Oser, L., Ostriker, J.~P., 
Naab, T., Johansson, P.~H., \& Burkert, A.\ 2010, \apj, 725, 2312

\bibitem[Papovich et al.(2014)]{2014arXiv1412.3806P} Papovich, C., 
Labb{\'e}, I., Quadri, R., et al.\ 2014, arXiv:1412.3806



\bibitem[Patel et al.(2013)]{2013ApJ...766...15P} Patel, S.~G., van Dokkum, 
P.~G., Franx, M., et al.\ 2013, \apj, 766, 15

\bibitem[Peng et al.(2015)]{2015Natur.521..192P} Peng, Y., Maiolino, R., 
\& Cochrane, R.\ 2015, \nat, 521, 192 

\bibitem[Peng et al.(2010)]{2010ApJ...721..193P} Peng, Y.-j., Lilly, S.~J., 
Kova{\v c}, K., et al.\ 2010, \apj, 721, 193

\bibitem[Peng et al.(2002)]{2002AJ....124..266P} Peng, C.~Y., Ho, L.~C., 
Impey, C.~D., \& Rix, H.-W.\ 2002, \aj, 124, 266 

\bibitem[P{\'e}rez-Gonz{\'a}lez et al.(2008)]{2008ApJ...675..234P} 
P{\'e}rez-Gonz{\'a}lez, P.~G., Rieke, G.~H., Villar, V., et al.\ 2008, 
\apj, 675, 234

\bibitem[Poggianti et al.(2013)]{2013ApJ...777..125P} Poggianti, B.~M., 
Moretti, A., Calvi, R., et al.\ 2013, \apj, 777, 125 

\bibitem[Ribeiro et al.(2016)]{2016arXiv160201840R} Ribeiro, B., Le 
F{\`e}vre, O., Tasca, L.~A.~M., et al.\ 2016, arXiv:1602.01840

\bibitem[Schmidt(1959)]{1959ApJ...129..243S} Schmidt, M.\ 1959, \apj, 129, 
243

\bibitem[Schreiber et al.(2016)]{2016arXiv160104226S} Schreiber, C., Elbaz, 
D., Pannella, M., et al.\ 2016, arXiv:1601.04226

\bibitem[Shamir 
\& Wallin(2014)]{2014MNRAS.443.3528S} Shamir, L., \& Wallin, J.\ 2014, \mnras, 443, 3528

\bibitem[Schmidt(1968)]{1968ApJ...151..393S} Schmidt, M.\ 1968, \apj, 151, 393 

\bibitem[Shankar et al.(2015)]{2015arXiv150102800S} Shankar, F., Buchan, 
S., Rettura, A., et al.\ 2015, arXiv:1501.02800

\bibitem[Shankar et al.(2014)]{2014ApJ...797L..27S} Shankar, F., Guo, H., 
Bouillot, V., et al.\ 2014, \apjl, 797, LL27

\bibitem[Shankar et al.(2013)]{2013MNRAS.428..109S} Shankar, F., Marulli, 
F., Bernardi, M., et al.\ 2013, \mnras, 428, 109

\bibitem[Shen et al.(2003)]{2003MNRAS.343..978S} Shen, S., Mo, H.~J., 
White, S.~D.~M., et al.\ 2003, \mnras, 343, 978

\bibitem[Shibuya et al.(2015)]{2015ApJS..219...15S} Shibuya, T., Ouchi, M., 
\& Harikane, Y.\ 2015, \apjs, 219, 15

\bibitem[Shibuya et al.(2015)]{2015arXiv151107054S} Shibuya, T., Ouchi, M., 
Kubo, M., \& Harikane, Y.\ 2015, arXiv:1511.07054 

\bibitem[Silk 
\& Rees(1998)]{1998A&A...331L...1S} Silk, J., \& Rees, M.~J.\ 1998, \aap, 331, L1

\bibitem[Skelton et al.(2014)]{2014ApJS..214...24S} Skelton, R.~E., 
Whitaker, K.~E., Momcheva, I.~G., et al.\ 2014, \apjs, 214, 24 

\bibitem[Sonnenfeld et al.(2014)]{2014ApJ...786...89S} Sonnenfeld, A., 
Nipoti, C., \& Treu, T.\ 2014, \apj, 786, 89

\bibitem[Stott et al.(2016)]{2016MNRAS.457.1888S} Stott, J.~P., Swinbank, 
A.~M., Johnson, H.~L., et al.\ 2016, \mnras, 457, 1888

\bibitem[Stringer et al.(2014)]{2014MNRAS.441.1570S} Stringer, M.~J., 
Shankar, F., Novak, G.~S., et al.\ 2014, \mnras, 441, 1570 

\bibitem[Tacchella et al.(2015)]{2015Sci...348..314T} Tacchella, S., 
Carollo, C.~M., Renzini, A., et al.\ 2015, Science, 348, 314 

\bibitem[Trayford et al.(2016)]{2016arXiv160107907T} Trayford, J.~W., 
Theuns, T., Bower, R.~G., et al.\ 2016, arXiv:1601.07907

\bibitem[Trujillo et al.(2011)]{2011MNRAS.415.3903T} Trujillo, I., 
Ferreras, I., \& de La Rosa, I.~G.\ 2011, \mnras, 415, 3903 



\bibitem[Trujillo et al.(2006)]{2006ApJ...650...18T} Trujillo, I., 
F{\"o}rster Schreiber, N.~M., Rudnick, G., et al.\ 2006, \apj, 650, 18 



\bibitem[van der Wel et al.(2014)]{2014ApJ...788...28V} van der Wel, A., 
Franx, M., van Dokkum, P.~G., et al.\ 2014, \apj, 788, 28

\bibitem[van der Wel et al.(2012)]{2012ApJS..203...24V} van der Wel, A., 
Bell, E.~F., H{\"a}ussler, B., et al.\ 2012, \apjs, 203, 24

 

\bibitem[van der Wel et al.(2011)]{2011ApJ...730...38V} van der Wel, A., 
Rix, H.-W., Wuyts, S., et al.\ 2011, \apj, 730, 38

\bibitem[van der Wel et al.(2008)]{2008ApJ...688...48V} van der Wel, A., 
Holden, B.~P., Zirm, A.~W., et al.\ 2008, \apj, 688, 48

\bibitem[van Dokkum et al.(2015)]{2015ApJ...813...23V} van Dokkum, P.~G., 
Nelson, E.~J., Franx, M., et al.\ 2015, \apj, 813, 23 

\bibitem[van Dokkum et al.(2014)]{2014ApJ...791...45V} van Dokkum, P.~G., 
Bezanson, R., van der Wel, A., et al.\ 2014, \apj, 791, 45

\bibitem[van Dokkum et al.(2008)]{2008ApJ...677L...5V} van Dokkum, P.~G., 
Franx, M., Kriek, M., et al.\ 2008, \apjl, 677, L5 

\bibitem[van Dokkum et al.(2010)]{2010ApJ...709.1018V} van Dokkum, P.~G., 
Whitaker, K.~E., Brammer, G., et al.\ 2010, \apj, 709, 1018 

\bibitem[van Dokkum et al.(2011)]{2011ApJ...743L..15V} van Dokkum, P.~G., 
Brammer, G., Fumagalli, M., et al.\ 2011, \apjl, 743, L15 



\bibitem[Vogelsberger et al.(2014)]{2014MNRAS.444.1518V} Vogelsberger, M., 
Genel, S., Springel, V., et al.\ 2014, \mnras, 444, 1518

\bibitem[Whitaker et al.(2011)]{2011ApJ...735...86W} Whitaker, K.~E., 
Labb{\'e}, I., van Dokkum, P.~G., et al.\ 2011, \apj, 735, 86 

\bibitem[Whitaker et al.(2012)]{2012ApJ...754L..29W} Whitaker, K.~E., van 
Dokkum, P.~G., Brammer, G., \& Franx, M.\ 2012, \apjl, 754, LL29 

\bibitem[Wilkins et al.(2008)]{2008MNRAS.385..687W} Wilkins, S.~M., 
Trentham, N., \& Hopkins, A.~M.\ 2008, \mnras, 385, 687

\bibitem[Wisnioski et al.(2015)]{2015ApJ...799..209W} Wisnioski, E., F{\"o}rster Schreiber, N.~M., Wuyts, S., et al.\ 2015, \apj, 799, 209

\bibitem[Wuyts et al.(2012)]{2012ApJ...753..114W} Wuyts, S., F{\"o}rster 
Schreiber, N.~M., Genzel, R., et al.\ 2012, \apj, 753, 114

\bibitem[Wuyts et al.(2011)]{2011ApJ...742...96W} Wuyts, S., F{\"o}rster 
Schreiber, N.~M., van der Wel, A., et al.\ 2011, \apj, 742, 96




\end{thebibliography}
\end{document}